\documentclass[11pt]{article}
\pdfoutput=1 
\usepackage{amssymb}
\usepackage{amsmath}
\usepackage{amstext}
\usepackage{mathrsfs}
\usepackage{graphicx,epsfig}
\usepackage{epsfig}
\usepackage{verbatim} 
\usepackage{fancybox}
\usepackage{color}
\usepackage{ulem}
\usepackage{enumitem}
\usepackage{subfigure}
\usepackage{bbm}
\usepackage{parskip}
\usepackage{dsfont}
\usepackage[numbers,sort&compress]{natbib}
\usepackage{framed}
\usepackage{anyfontsize}
\usepackage{braket}
\usepackage{cancel}

\makeatletter
\renewcommand\@dotsep{200}
\makeatother

\usepackage{tikz}
\usetikzlibrary{decorations}
\pgfdeclaredecoration{complete sines}{initial}
{
    \state{initial}[
        width=+0pt,
        next state=upsine,
        persistent precomputation={\pgfmathsetmacro\matchinglength{
            \pgfdecoratedinputsegmentlength / int(\pgfdecoratedinputsegmentlength/\pgfdecorationsegmentlength)}
            \setlength{\pgfdecorationsegmentlength}{\matchinglength pt}
        }] {}
    \state{upsine}[width=\pgfdecorationsegmentlength,next state=downsine]{
        \pgfpathsine{\pgfpoint{0.25\pgfdecorationsegmentlength}{0.5\pgfdecorationsegmentamplitude}}
        \pgfpathcosine{\pgfpoint{0.25\pgfdecorationsegmentlength}{-0.5\pgfdecorationsegmentamplitude}}
    }
    \state{downsine}[width=\pgfdecorationsegmentlength,next state=upsine]{
        \pgfpathsine{\pgfpoint{0.25\pgfdecorationsegmentlength}{-0.5\pgfdecorationsegmentamplitude}}
        \pgfpathcosine{\pgfpoint{0.25\pgfdecorationsegmentlength}{0.5\pgfdecorationsegmentamplitude}}
}
    \state{final}{}
}

\newcommand{\vev}[1]{\bigl\langle#1\bigr\rangle}

\newcommand{\Comment}[1]{{}}
\definecolor{darkblue}{rgb}{0.15,0.35,0.55}
\definecolor{comment}{rgb}{1,0.4,0.4}
\definecolor{reddish}{rgb}{0.65, 0.2, 0.2}
\usepackage[linktocpage=true]{hyperref}
\hypersetup{
colorlinks=true,
citecolor=darkblue,
linkcolor=reddish,
urlcolor=darkblue,
pdfauthor={},
pdftitle={},
pdfsubject={}
}

\newcommand{\be}{\begin{equation}}
\newcommand{\ee}{\end{equation}}
\newcommand{\bea}{\begin{eqnarray}}
\newcommand{\eea}{\end{eqnarray}}
\newcommand{\beas}{\begin{eqnarray*}}
\newcommand{\eeas}{\end{eqnarray*}}

\def\({\left(}
\def\){\right)}

\newcommand{\rd}{{\rm d}}
\newcommand{\vp}{\varphi}

\DeclareMathOperator{\Exp}{e}

\newcommand{\D}{{\rm d}}

\newcommand{\Mpl}{M_{\rm Pl}}
\newcommand{\bfx}{{\vec{x}}}

\newcommand{\bfk}{{\vec{k}}}

\newcommand{\bfq}{{\vec{q}}}

\newcommand{\bfnabla}{{\vec{\nabla}}}

\newcommand{\lb}{\left[}
\newcommand{\rb}{\right]}

\def\gsim{ \lower .75ex \hbox{$\sim$} \llap{\raise .27ex \hbox{$>$}} }
\def\lsim{ \lower .75ex \hbox{$\sim$} \llap{\raise .27ex \hbox{$<$}} }

\newcommand{\beq}{\begin{equation}}
\newcommand{\eeq}{\end{equation}}
\newcommand{\beqs}{\begin{equation*}}
\newcommand{\eeqs}{\end{equation*}}
\newcommand{\beqar}{\begin{eqnarray}}
\newcommand{\eeqar}{\end{eqnarray}}
\newcommand{\bal}{\begin{aligned}}
\newcommand{\eal}{\end{aligned}}

\def\lagr{\hbox{$\cal L$}}
\def\ham{\hbox{$\cal H$}}

\def\dalam{\hbox
{\vrule\vbox{\hrule\hbox to 1ex{ \hfill}\kern 1 ex\hrule}\vrule}}

\def\1/2{\hbox{$ {1 \over 2}$ }}

\def\prl{\partial}

\def\h{\hbar}
\def\i/h{{i \over \h}}

\def\v{\vec}

\def\t{\tau}

\def\z{\zeta}

\def\a{\alpha} 
\def\b{\beta}

\def\d{\delta}
\def\l{\lambda} 
\def\e{\epsilon} \def\E{\hbox{$\cal E $}}

\def\xyma{\xymatrix@M.7em}
\def\xymas{\xymatrix@M.1em}
\newcommand{\ba}{\begin{eqnarray}}
\newcommand{\ea}{\end{eqnarray}}

\definecolor{darkred}{rgb}{0.7,0.3,0.3}
\definecolor{darkgreen}{rgb}{0.2,0.7,0.3}
\definecolor{greyish}{rgb}{.90,.90,.90}
\definecolor{greyish2}{rgb}{.96,.96,.96}
\definecolor{darkblue2}{rgb}{0.3,0.4,0.9}
\usepackage{pifont}
\usepackage{xcolor,colortbl}

\usepackage{tcolorbox}

\title{}
\author{}

\numberwithin{equation}{section}

\begin{document}
%
\setcounter{page}{1}
\renewcommand{\thefootnote}{\fnsymbol{footnote}}
%
%


\begin{center}
{\fontsize{21}{18} \bf Soft theorems for boosts}\\[7pt]
{\fontsize{21}{18} \bf and other time symmetries}
\end{center}

\vspace{1cm}

\begin{center}
{\fontsize{13}{18}\selectfont
Lam Hui,${}^{\rm a}$\footnote{\href{mailto:lh399@columbia.edu}{\texttt{lh399@columbia.edu}}}
Austin Joyce,${}^{\rm b}$\footnote{\href{mailto:austinjoyce@uchicago.edu}{\texttt{austinjoyce@uchicago.edu}}}
Ilia Komissarov,${}^{\rm a}$\footnote{\href{mailto:i.komissarov@columbia.edu}{\texttt{i.komissarov@columbia.edu}}}
Klaas Parmentier,${}^{\rm a}$\footnote{\href{mailto:k.parmentier@columbia.edu}{\texttt{k.parmentier@columbia.edu}}}
\\[5pt]
Luca Santoni,${}^{\rm c}$\footnote{\href{mailto:lsantoni@ictp.it}{\texttt{lsantoni@ictp.it}}}
and
Sam S. C. Wong${}^{\rm d}$\footnote{\href{mailto:scswong@sas.upenn.edu}{\texttt{scswong@sas.upenn.edu}}}
}
\end{center}
\vspace{0.8cm}

 \centerline{{\it ${}^{\rm a}$Center for Theoretical Physics, Department of Physics,}}
 \centerline{{\it Columbia University, New York, NY 10027, USA}} 
 
  \vspace{.3cm}

 \centerline{{\it ${}^{\rm b}$Kavli Institute for Cosmological Physics, Department of Astronomy and Astrophysics,}}
 \centerline{{\it The University of Chicago, Chicago, IL 60637, USA}}
 
  \vspace{.3cm}
  
  \centerline{{\it ${}^{\rm c}$ICTP, International Centre for Theoretical Physics,}}
 \centerline{{\it Strada Costiera 11, 34151, Trieste, Italy}}
 
  \vspace{.3cm}

\centerline{{\it ${}^{\rm d}$Center for Particle Cosmology, Department of Physics and Astronomy,}}
\centerline{{\it University of Pennsylvania 209 S. 33rd St., Philadelphia, PA 19104, USA}}
 \vspace{.25cm}

 \vspace{1cm}
\begin{abstract}
\noindent
We derive soft theorems for theories in which time symmetries---symmetries that involve the transformation of time, an example of which are Lorentz boosts---are spontaneously broken. 
The soft theorems involve unequal-time correlation functions with the 
insertion of a soft Goldstone in the far past. Explicit checks are
provided for several examples, including the effective theory of a relativistic superfluid and the effective field theory of inflation.
We discuss how in certain cases these unequal-time identities capture information at the level of observables that cannot be seen purely in terms of equal-time correlators of the field alone.
We also discuss when it is possible to phrase these soft theorems as identities involving equal-time correlators.
\end{abstract}

\newpage

\setcounter{tocdepth}{2}
\tableofcontents

\newpage
\renewcommand*{\thefootnote}{\arabic{footnote}}
\setcounter{footnote}{0}

\section{Introduction}

One of the most useful aspects of symmetry is that it tells us when things are impossible. Concretely, at the level of observable quantities, the most common manifestation of symmetries is through selection rules, which state that the probability for certain processes to happen is precisely zero. There are, of course, many examples where symmetries are spontaneously broken by dynamics. In these situations, rather than becoming useless, the utility of symmetry merely manifests differently.
Instead of enforcing selection rules, the on-shell avatars of spontaneously broken symmetries are soft theorems that relate observables to each other in particular kinematic limits. The most famous such statement is the Adler zero---which states that pion scattering amplitudes vanish when one of the external pion momenta is scaled to zero~\cite{Adler:1964um,Adler:1965ga}.

Soft theorems provide powerful kinematic constraints on the dynamics of systems. For example, Weinberg's soft photon and graviton theorems~\cite{Weinberg:1964ew,Weinberg:1965nx,Strominger:2013jfa, He:2014laa, He:2014cra,Strominger:2017zoo, Mirbabayi:2016xvc} tell us that particles can only interact with electromagnetism in ways that conserve total charge, and that gravity must couple democratically to all particles, respectively. Soft theorems also provide a useful organizing principle by which to classify effective field theories from the on-shell perspective. Field theories with enhanced symmetries lead to enhanced Adler zeroes of scattering amplitudes~\cite{Cheung:2014dqa,Cachazo:2014xea,Hinterbichler:2015pqa,Cheung:2016drk}, and their corresponding wavefunction coefficients satisfy particular Ward identities~\cite{Bittermann:2022nfh}.

In the cosmological context, Maldacena's soft theorem~\cite{Maldacena:2002vr,Creminelli:2004yq} serves as a constraint that all (attractor) models of single-clock inflation must satisfy (subject to some technical requirements).\footnote{The fact that there are technical requirements is in itself interesting---it implies that the soft theorems can be violated, indicating that they are not vacuous.} As such, cosmological soft theorems serve as powerful probes of new physics, with violations generically indicating the presence of new degrees of freedom. The promise of cosmological soft theorems as tests of new physics has motivated numerous generalizations in the inflationary context~\cite{Creminelli:2012ed,Assassi:2012zq,Hinterbichler:2012nm,Flauger:2013hra, Pimentel:2013gza,Goldberger:2013rsa,Hinterbichler:2013dpa,Berezhiani:2013ewa, Mirbabayi:2014zpa,Joyce:2014aqa,Kundu:2015xta,Pajer:2017hmb, Finelli:2017fml,Bravo:2017wyw,Bordin:2017ozj,Jazayeri:2019nbi,Avis:2019eav,Hui:2018cag}, and applications to observables in large-scale structure~\cite{Kehagias:2013yd,Peloso:2013zw, Creminelli:2013mca,Horn:2014rta,Esposito:2019jkb,Goldstein:2022hgr}. Interestingly, there are known symmetry consequences at the level of the action of inflationary perturbations that have thus far resisted a translation into observable quantities. For example certain cubic interactions in the effective description of inflation are fixed in terms of the sound speed of the inflaton~\cite{Cheung:2007st}. One of our main conceptual results is to understand how to see this relation at the level of correlation functions.

In this paper, we study the soft theorems associated with a particular kind of nonlinearly realized symmetry: those that involve the transformation of time. 
These symmetries, by their very nature, distinguish between space and time and are often associated with symmetries of special relativity (or its cosmological analogue, de Sitter symmetry). As such, they are ubiquitous in both cosmology and condensed matter physics. Indeed, a prime example is provided by Lorentz boosts, which serve to mix time and space coordinates as $\tau \mapsto \tau - \vec b \cdot \vec x \,;\, \vec
  x \mapsto \vec x - \vec b \, \tau$. In systems where boosts are nonlinearly realized, the transformation of the Goldstone mode therefore involves time. 
Symmetries of this type, involving shifts of time by spatial coordinates (or vice versa) are quite subtle. The conceptual reason is that these transformations disturb the time slices on which we define Hilbert spaces in canonical quantization. Deriving their consequences for observable quantities, therefore, takes some care. We derive these soft theorems satisfied by theories with nonlinearly realized time symmetries abstractly and apply this formalism to several examples of interest.
  
One motivation for the study of soft theorems for boost-like symmetries comes from condensed matter physics.
There is a sense in which gapless phases of matter can be classified according to how they nonlinearly realize spacetime symmetries~\cite{Dubovsky:2005xd,Dubovsky:2011sj,Nicolis:2013lma,Nicolis:2015sra,Alberte:2020eil, Gaiotto:2014kfa,Lake:2018dqm,Hofman:2018lfz,Delacretaz:2019brr,Benedetti:2021lxj, Hinterbichler:2022agn,McGreevy:2022oyu}. At low energies, one expects much of the dynamics of these systems to be controlled by the Goldstone modes that nonlinearly realize these symmetries.
It is therefore useful to understand how observables involving these Goldstone modes manifest symmetries in the form of soft theorems. As a particular example, we consider the effective field theory of a relativistic superfluid and explore how the nonlinearly realized boost symmetry constraints (unequal time) correlation functions of the superfluid phonon. Using these relations, we are able to reproduce relations between various operators in the effective field theory, purely by considering the properties of correlation functions. The philosophy is more general and can be applied to more complex condensed matter systems to derive the consequences of symmetry. It would be interesting to understand what general properties of these systems can be deduced from this universal soft dynamics.\footnote{
For example, one might imagine that the famous fluid Stokes drift effect \cite{Stokes} could be derived from soft theorems in the same way that the gravitational memory effect follows from soft theorems for graviton scattering.}

Our other motivation comes from cosmology, another context where a background condensate spontaneously breaks boost-like symmetries. In this case, the relevant Goldstone mode is the fluctuation of the inflaton (or the curvature perturbation on constant density hypersurfaces), which nonlinearly realizes the symmetries of de Sitter space in the limit where we decouple gravity. Since we only have observational access to these fluctuations at late times via their imprint in the cosmic microwave background, it is natural to try to reconstruct these correlation functions directly on the future boundary of de Sitter from physical consistency~\cite{Maldacena:2011nz,Creminelli:2011mw,Mata:2012bx,Bzowski:2013sza,Ghosh:2014kba,Arkani-Hamed:2015bza,Pajer:2016ieg,Arkani-Hamed:2017fdk,Arkani-Hamed:2018kmz,Goon:2018fyu,Benincasa:2019vqr,Bzowski:2019kwd,Hillman:2019wgh,Baumann:2019oyu, Baumann:2020dch,Goodhew:2020hob,Melville:2021lst,Cabass:2021fnw,Goodhew:2021oqg,Cabass:2022jda, Benincasa:2022gtd,Baumann:2022jpr}.
There has recently been substantial progress in this bootstrap approach in de Sitter space, including correlators of the inflaton itself~\cite{Pajer:2020wxk,Jazayeri:2021fvk,Baumann:2021fxj,Hillman:2021bnk,Bonifacio:2021azc, Pimentel:2022fsc,Jazayeri:2022kjy}.
One aspect of these constructions that remains somewhat mysterious is the precise consequences of the nonlinearly realized symmetries of the inflaton itself. Such an understanding would allow a definition of the effective theory of inflation at the level of observables and would help illuminate the possible parametric limits of the EFT, where certain operators dominate the dynamics, at the level of observables.
In this paper, we make progress toward such a description. Interestingly, however, we find that the full consequences of the inflaton's symmetries are only visible in unequal-time correlation functions. Concretely, we find that the relations between cubic couplings and the normalization of the inflaton's kinetic term can only be recovered from a soft theorem by considering correlators involving either fields evaluated at different times, or involving conjugate momenta. In contrast, at higher points, we can recover relations known from the EFT lagrangian from equal-time correlators.
On the cosmology side, our results will also clarify the assumptions underlying the formulation and validity of soft theorems in non-attractor models of inflation, such as ultra-slow-roll,  derived in~\cite{Finelli:2017fml, Bravo:2017wyw}. 

Aside from the examples that we consider, there are numerous systems where either relativistic symmetries are nonlinearly realized, or other time symmetries are present. We expect that the general techniques developed here will be more widely applicable to these systems and will help enable the study of these systems directly in terms of observables.

\vspace{-10pt}
\paragraph{Outline:} The paper is organized as follows.
In Section~\ref{sec:derivation}, we derive a general Ward identity valid for all types of symmetries (including time symmetries). The main novel feature is a careful accounting of boundary term contributions to the charges that generate these symmetries. The general Ward identity is a statement involving unequal-time correlation functions, we further discuss the requirements to recast it as a statement about equal-time correlators. In Section~\ref{sec:applications}, we apply the general theorem to specific examples and verify that it is satisfied by explicit calculation. We begin by considering two toy-model systems that are time-dependent field theories engineered to mimic slow-roll and ultra-slow-roll models of inflation.
We then consider the EFT of an ideal superfluid, and finally the effective field theory of inflation in the decoupling limit. We draw some lessons from these examples in Section~\ref{sec:conclusions}. We include a number of technical appendices that are somewhat peripheral to the main line of development. In Appendix~\ref{appendix:A0} we discuss the path-integral derivation of the soft theorem of Section~\ref{sec:derivation}. 
In Appendix~\ref{1ptconstr}, we derive an identity obeyed by two-point functions in these theories.
In Appendices~\ref{sec:superfluid} and \ref{EFToIapp}, we provide additional information about the superfluid and EFT of inflation examples, along with additional checks of the soft theorems in these models, and in generalizations---in particular in~\ref{sec:drivensuperfluid} we consider the symmetries of the EFT of a superfluid driven by an external source. In Appendix~\ref{app:timedependentsoundspeed} we consider the symmetries and soft theorems of ultra-slow-roll inflation with a time-dependent speed of sound. Finally, in Appendix~\ref{sec:softampl} we briefly comment on the manifestation of spontaneously broken boots at the level of scattering amplitudes of superfluid phonons.

\vspace{-10pt}
\paragraph{Notation and conventions:}   We work in mostly plus metric signature and use the curvature conventions that $R^{\rho}{}_{\sigma\mu\nu}=\partial_{\mu}\Gamma^{\rho}_{\nu\sigma}+\cdots$ and $R_{\mu\nu}=R^{\rho}{}_{\mu\rho\nu}$. We denote the reduced Planck mass by $\Mpl=(8\pi G)^{-1/2}$.
 We adopt the Fourier convention
 \be
 \phi(\bfx,\tau)=
 \int\frac{\D^3k}{(2\pi)^3}\,
 \Exp^{-i\bfk\cdot\bfx}\phi(\bfk,\tau)\,.
 \ee
 In the following, we will
occasionally omit explicit time dependence of field operators,
 e.g.~$\phi(\bfk, \tau) \equiv \phi(\bfk)$, when we  write equal-time
 correlation functions (see, e.g., Eq.~\eqref{toyidsrff} below). In
 unequal-time correlators, the time dependence will often be replaced
 by a label, e.g., $\phi_f (\vec k) \equiv \phi(\vec k, \tau_f)$ for
 the final time, or
$\phi_i (\vec k) \equiv \phi(\vec k, \tau_i)$ for initial time 
(see, e.g., Eq.~\eqref{softthd1}). 
We use two different phi symbols: $\phi(\v x, \t)$ to denote an operator-valued quantum field and $\vp(\v x)$ for classical field profiles.
We also use two different symbols for time in different contexts: $t$ and $\tau$.
Their difference is only relevant for cosmology: $t$ stands for the proper time
and $\tau$ for conformal time. Correspondingly, $\dot{}$ (overdot) denotes
$\partial_t$ and ${}'$ (prime) denotes $\partial_\tau$. For the superfluid
example---which takes place in flat space---$t$ and $\tau$ are
interchangeable though we mostly use $t$.
Lastly, we occasionally refer to soft theorems as consistency relations, which is a terminology inherited from the cosmology literature.

\newpage
\section{Soft theorems for spacetime symmetries}
\label{sec:derivation}
Spontaneously broken symmetries leave a nontrivial imprint in correlation functions of local operators in the guise of soft theorems obeyed by correlators involving the Nambu--Goldstone mode associated with the broken symmetry. Here we derive a general expression for this soft theorem as a Ward identity following~\cite{Hinterbichler:2013dpa}. 
The main novel ingredient is a careful accounting of the effects of (temporal) boundary terms on the Noether charges responsible for the Ward identities. As we will see, such boundary terms lead to important contributions to the soft theorems.

\subsection{Soft theorem derivation}
\label{derivation}

We are interested in correlation functions of local operators evaluated in the interacting vacuum of the theory, $\lvert 0_{\rm in}\rangle$.  It is convenient to consider the correlations of operators that are built out of some set of fundamental fields that we denote as $\phi(\vec x, \tau)$, where $\vec x$ is a spatial position and $\tau$ is the time coordinate.
 We then consider
correlation functions of operators built out of the field $\phi$, of the schematic form (in Fourier space)
\be
\langle 0_{\rm in} | {\cal O}  ({\vec k_1}, \cdots , {\vec k_N}) | 0_{\rm in} \rangle \, ,
\label{eq:corrdef}
\ee
where ${\cal O}$ abstractly denotes any composite built out of $\phi$. We will be mostly interested in the case where it is a product of the field operators at some fixed time, for example, ${\cal O}  ({\vec k_1}, \cdots , {\vec k_N})=\phi(\vec k_1, \tau_f) \cdots \phi(\vec k_N, \tau_f)$, where $\tau_f$ is the time of interest.\footnote{Here the label ``{\it f\,}'' denotes the ``final" time, where we evaluate the correlations. We will sometimes use the shorthand notation
$\phi_f ({\v k})$ to denote $\phi (\vec k, \tau_f)$.}
In principle the operator ${\cal O}$ can involve products of $\phi$'s at
different times:  we will consider such an example in Section~\ref{sec:EFTofI}.

Our goal is to understand the consequence of a nonlinearly realized symmetry for the correlation function~\eqref{eq:corrdef}. A convenient starting point is to consider the action of the charge, $Q$, that generates such a symmetry:\footnote{Strictly speaking, for a nonlinearly realized symmetry the charge $Q$ is ill-defined because it is IR divergent. Nevertheless, commutators of this charge with local operators make sense~\cite{Itzykson:1980rh,Brauner:2010wm}.}
\be
\label{Ward0}
i [Q, {\cal O}] = \delta {\cal O} \, .
\ee
Evaluating this in the in-vacuum, we have
\be
\label{WardC}
i \langle 0_{\rm in} | [Q,  {\cal O}] | 0_{\rm in} \rangle = \langle 0_{\rm in} | \delta {\cal O} | 0_{\rm in} \rangle \, .
\ee
In the cases of interest, the symmetry generated by $Q$ is nonlinearly realized, so that its action on ${\cal O}$ can be split as
\be
\delta {\cal O} = \delta_{\rm NL} {\cal O}  + \delta_{\rm L} {\cal O}\,,
\label{eq:NLLsplit}
\ee
where $\delta_{\rm NL} {\cal O}$ denotes the part of the transformation that is nonlinear (independent of the fields in the theory) and $\delta_{\rm L} {\cal O}$ denotes the part of the transformation that is linear in the fields. Importantly, the contribution to the right-hand side of~\eqref{WardC} coming from $\delta_{\rm NL}$ is localized at zero momentum and so does not contribute to connected correlation functions~\cite{Hinterbichler:2013dpa}. We can therefore drop this part of the transformation.

\begin{framed}
{\small
\vspace{-.15cm}
\noindent
{\bf\small Superfluid example:}
To make the discussion less abstract, it is useful to have a motivating example in mind. A particularly simple and interesting example is provided by the effective field theory of a relativistic superfluid~\cite{Son:2002zn}. The effective description is rather simple, at the lowest order in derivatives, it is given by a lagrangian of the form
\be
S = \Lambda^4\int\rd^4x P(X)\,,
\ee
where $X\equiv -(\partial\phi)^2/\Lambda^4$. The superfluid phonons, $\pi$, describe the fluctuations around the finite-density state $\phi = \mu t+\pi$. This state spontaneously breaks both time translations and the shift symmetry of $\phi$ but preserves their diagonal combination. Importantly, this state also spontaneously breaks Lorentz boosts, which therefore act on the phonons nonlinearly. Concretely we can write the action of infinitesimal boosts on $\pi$ as
\be
\delta_{\rm NL} \pi = \mu \,\vec b \cdot \vec x \, , \qquad \qquad
  \delta_{\rm L} \pi =  \vec b \cdot
(\tau \vec \nabla + \vec x \, \partial_\tau) \pi \, ,
\label{eq:sfluidtransf}
\ee
where we have made the same split as in~\eqref{eq:NLLsplit} and where $\vec b$ is the parameter of the boost transformation. As advertised, the transformation~\eqref{eq:sfluidtransf} contains both a part that is independent of $\pi$, which generates a field profile with a constant gradient and a piece linear in $\pi$. On the right-hand side of~\eqref{WardC}, only the piece linear in $\pi$ will appear.
}
\end{framed}

Let us now focus on the left-hand side of~\eqref{WardC}. We can evaluate this matrix element by inserting a complete set of field eigenstates at some early time:
\be
\label{WardCLHS}
i \langle 0_{\rm in} | [Q,  {\cal O}] | 0_{\rm in} \rangle = i \int {\cal D} \varphi_i \, 
\Big( \langle 0_{\rm in} | Q |\varphi_i \rangle
\langle \varphi_i | {\cal O}  | 0_{\rm in} \rangle - \langle 0_{\rm in} |  {\cal O} |\varphi_i \rangle
\langle \varphi_i | Q | 0_{\rm in} \rangle \Big) \, .
\ee
Here the $|\varphi_i \rangle$ are eigenstates of the Heisenberg-picture field operator:
$\phi (\vec k , \tau_i) |\varphi_i \rangle = \varphi_i (\vec k) |\varphi_i \rangle$,
where $\varphi_i (\vec k)$ is a (spatial) field profile.\footnote{Note that the state $|\varphi_i \rangle$ implicitly depends on time by virtue of its definition as an eigenstate of 
a Heisenberg operator at time $\tau_i$.}
The label ``{\it i\,}'' denotes the ``initial" time, meaning the far past. 
An implicit---and critical---assumption is that $Q$ is time independent, so that
it can  equally well be used to effect the transformation of ${\cal O}$ (which is evaluated at time $\tau_f$),
or to transform the initial wave function $\langle \varphi_i | 0_{\rm in} \rangle$. (This is a reasonable assumption because the charge is conserved in time.) Said another way,
the point is that $\langle \varphi_i | Q | 0_{\rm in} \rangle$ is relatively straightforward to compute because in the far past the in-vacuum approaches the free vacuum.

In order to evaluate the matrix elements in~\eqref{WardCLHS}, we need to parameterize the form of the charge $Q$. We envision that it arises from integrating a Noether current and that we can write it schematically as
\begin{equation}
Q = \int \D^3  x \, j^0 
= \int \D^3  x \Big[ \alpha(\vec x, \tau_i) \phi(\vec x, \tau_i) + \beta(\vec x, \tau_i) \Pi_\phi (\vec x, \tau_i) + \cdots \Big] \, ,
\label{ChargeAlphaBeta}
\end{equation}
where $\alpha$ and $\beta$ are arbitrary functions of space and time, and $\Pi_\phi$ is the canonical momentum conjugate to $\phi$ (which is just proportional to $\dot\phi$ in perturbation theory). The ellipsis represents terms that are either $\phi$ independent or which are quadratic or higher order in $\phi$, which can be ignored in the asymptotic past.\footnote{\label{morephis}In the interaction picture, terms with several $\phi$'s would
oscillate rapidly in the far past, leading to cancellations. See \cite{Hui:2018cag} for more details.
}
The term involving the canonical momentum $\Pi_\phi$ is expected, it is this term that effects the transformation of $\phi$ when we commute $Q$ with $\phi$, as a consequence of the canonical commutation relation between the field and its conjugate momentum. Since we are focusing on the parts of the charge linear in $\phi$, it is easy to see that the relevant $\beta$ is field-independent. That is, it is just the nonlinear part of the field transformation:
\be
\beta (\vec x, \tau) = \delta_{\rm NL} \phi (\vec x, \tau) \, .
\ee

The term linear in $\phi$ involving $\alpha$ is perhaps less familiar, but we can trace its origin to boundary terms in the Noether procedure. Suppose that under the symmetry transformation of interest, the lagrangian transforms as $\delta {\cal
  L} = \partial_\mu K^\mu$, then the Noether current is
\be
\label{j0F0}
j^\mu = \delta \phi {\partial {\cal L} \over \partial (\partial_\mu
  \phi)} - K^\mu \, .
\ee
Comparing this with~\eqref{ChargeAlphaBeta}, we see
that the $\alpha\phi$ contribution to $j^0$ arises from
$-K^0$, which comes from a boundary term in the variation of
the action, associated with the {\it time} boundary. 
The possible presence of these temporal boundary-term contributions was ignored in previous
discussions of cosmological soft theorems, and can be important in certain cases.\footnote{In view of the fact that we are primarily interested in correlation functions of the fields $\phi$ themselves,
one might reasonably ask why a linear $\phi$
contribution to $j^0$ would matter in the relevant commutator. 
The point is that we will take advantage of the conservation of $Q$ to
evaluate its effect in the far past, so the relevant commutator is then
between a $\phi$ in the far past and a string of $\phi$'s in
the far future, which need not vanish.} 
It can further be shown that the linear (in fields) contribution to $-K^0$ depends
on $\phi$ but not on its time derivatives. 
We will provide examples where $\alpha\neq 0$, and the contribution to the soft theorem must be tracked.

Given the form of the charge~\eqref{ChargeAlphaBeta}, we now turn to evaluate the matrix element~\eqref{WardCLHS}. It will be convenient to work in Fourier space.
Anticipating that the spatial Fourier transforms of $\alpha$ and
$\beta$ are localized at zero momentum, we define $\alpha(\vec q,
\t_i) \equiv  (2\pi)^3 \delta (\vec q) \alpha_0 (-\v q , \t_i)$ and
$\beta(\vec q, \t_i) \equiv (2\pi)^3 \delta (\vec q)
D_{-\v q} (\t_i)$.\footnote{In some examples, $D_{\v q}$ can
involve derivatives with respect to $\vec q$. As a concrete example, one can consider the superfluid~\eqref{eq:sfluidtransf}, where the nonlinear part of the symmetry is $\b = \mu
  \vec \epsilon  \cdot \vec x$.
In this case, we have $\b(\v q) = (2 \pi)^3 \d(\v
  q) \, i \mu \vec \epsilon \cdot \vec \nabla_{\vec q}$, and thus
$D_{- \v q} = i \mu \vec \epsilon \cdot \vec \nabla_{\vec q}$. 
Here, we are interpreting the derivatives distributionally such that the derivative does not act on the delta function.
The same understanding applies to $\alpha$ and $\alpha_0$,
though we will not encounter examples where $\alpha_0$
depends on $\v q$. Note that both $\alpha_0$ and $D_{\vec q}$ are
in general functions of time $\t_i$, which we mostly keep implicit.
}  Thus, we can write the charge as
\be
Q = \lim_{\vec q \rightarrow 0} \Big( \alpha_0\, \phi(\vec q, \tau_i) + D_{\v q} \Pi_\phi (\vec q, \tau_i) + \cdots \Big) \, .
\label{Charge_a_0}
\ee
From this, we can evaluate the matrix element in the field basis
\be
\begin{aligned}
\label{Qacts}
\langle \varphi_i | Q | 0_{\rm in} \rangle & = \lim_{\vec q \rightarrow 0} \left(
\alpha_0 \langle \varphi_i | \phi(\vec q, \tau_i)  | 0_{\rm in} \rangle
-i D_{\v q} {\delta \over \delta \varphi_i (\vec q)} \langle \varphi_i | 0_{\rm in} \rangle
\right)
 \\[2pt]  
 & = \lim_{\vec q \rightarrow 0}
\bigg(
\alpha_0 \langle \varphi_i | \phi(\vec q, \tau_i)  | 0_{\rm in} \rangle
+ i D_{\v q} \,{\cal E}_i (q) \langle \varphi_i | \phi(\vec q, \tau_i) | 0_{\rm in} \rangle \bigg)\, ,
\end{aligned}
\ee
where we have used that $\Pi_\phi = -i\delta/\delta\varphi$ and have assumed the initial wave function is Gaussian:
\be
\langle \varphi_i | 0_{\rm in} \rangle \propto {\,\rm exp\,}
\left(-{1\over 2} \int {\D^3  k \over (2\pi)^3} {\cal E}_i (k) \varphi_i (\vec k) \varphi_i (-\vec k)  \right) \, ,
\ee
where ${\cal E}_i$ is a kernel that can be parameterized as ${\cal E}_i (k) = i {\,\rm Im\,} {\cal E}_i (k) +
1/(2 P_i (k))$, with $P_i(k)$  the initial power spectrum of $\phi$ fluctuations.\footnote{Both $\E(k)$ and the power spectrum $P(k)$ depend on the momentum only via its magnitude $k \equiv |\v k|$. This is a consequence of translation and rotation invariance, which we always assume to be symmetries.}

Substituting~\eqref{Qacts} into~\eqref{WardCLHS} and combining the result
with~\eqref{WardC}, we find
\be
\begin{aligned}
\label{softthd0}
\lim_{\vec q \rightarrow 0} \bigg( i  (\alpha_0 - i D_{\v q}
{\cal E}_i^* (q)) \langle 0_{\rm in} | \phi(\vec q, \tau_i) {\cal O}  | 0_{\rm in} \rangle_c' &- 
i (\alpha_0 + i D_{\v q}
{\cal E}_i (q)) \langle 0_{\rm in} |  {\cal O} \phi(\vec q, \tau_i) |0_{\rm in}  \rangle_c' \bigg)
\\ &\hspace{5cm}= \langle 0_{\rm in} |  \delta_{\rm L} {\cal O} | 0_{\rm in} \rangle_c'
\, ,
\end{aligned}
\ee
where we have reduced the correlation functions to their connected
parts, (that this is possible as can be proved by induction---see~\cite{Hinterbichler:2013dpa})
and we have introduced the symbol $\langle \cdots \rangle_c'$ to denote a
connected correlation function with the overall momentum conserving
delta function removed.\footnote{\label{deltafuncRemove}Removing the momentum-conserving delta function is not
trivial, but that it can be done  follows from the hierarchical nature of
the relevant symmetries. For instance, for the superfluid, the
hierarchy of symmetries includes both time translations and boosts. The
former involves no momentum derivative while the latter involves one
such derivative. One finds that the delta function in the boost soft theorem can be removed by using the time translation soft theorem. See the explicit examples in the next section and~\cite{Hinterbichler:2013dpa} for details.}

The formula~\eqref{softthd0} is general---the operator ${\cal O}$ could even involve a product of $\phi$s at different times. 
For the case of primary interest, we have ${\cal O} = \phi_f (\v k_1) \cdots \phi_f (\v k_N)$, 
where we use  $\phi_f (\v k_1)$ as shorthand for $\phi ({\vec k_1}, \tau_f)$
(and likewise we use $\phi_i (\v q)$ as shorthand for $\phi ({\vec q}, \tau_i)$).
We can then write\footnote{Note that we have assumed (\textit {i}) $\phi(\vec x)$ is Hermitian such that
$\phi^\dagger (\vec k) = \phi(-\vec k)$, (\textit {ii}) parity invariance so that $\langle \phi(-\vec q) \phi(-\vec k_N) \cdots \phi (-\vec k_1) \rangle = \langle \phi(\vec q) \phi(\vec k_N) \cdots \phi (\vec k_1) \rangle$, and
(\textit {iii}) $\phi$'s at equal time commute. In particular, if ${\cal O}$ involves $\phi$ at
different times, one must revert to the more general statement
(\ref{softthd0}).
It is also worth noting---though we do not need it---that for real $\alpha(\v x)$ and $\beta(\v
x)$, their Fourier space counterparts obey $\alpha_0 (\v q) =
\alpha_0^* (-\v q)$ and $D_{\v q} = D^*_{-\v q}$. 
}
%
\begin{tcolorbox}[colframe=white,arc=0pt,colback=greyish2]
\be
\begin{aligned}
&\lim_{\vec q \rightarrow 0}\bigg[ D_{\v q} \left( {\cal E}_i (q)
 \langle 0_{\rm in} |  \phi_f (\v k_1) \cdots \phi_f
(\v k_N) \phi_i (\v q) |0_{\rm in}  \rangle_c' +  {\rm c. \, c.}
\right)
\\
& \qquad \quad 
+ \alpha_0 (\v q) \left(-i \langle 0_{\rm in} |  \phi_f (\v k_1) \cdots \phi_f
(\v k_N) \phi_i (\v q) |0_{\rm in}  \rangle_c' +  {\rm c. \, c.}
\right) \bigg]
\\ 
&\hspace{6cm}
=
\sum_{a=1}^N\langle 0_{\rm in} \rvert    \phi_f (\v k_1) \cdots\delta_L \phi_f (\v k_a)\cdots \phi_f (\v k_N) \lvert 0_{\rm in}  \rangle_c' \, ,
\label{softthd1}
\end{aligned}
\ee
\end{tcolorbox}
\noindent
where the differential operator $D_{\vec q}$ acts on everything to its right.
In the following, we will evaluate this general formula in specific examples. We first consider in what circumstances this identity can be phrased as a relation between equal-time correlation functions.


\subsection{Physical mode condition} 
\label{sec:physmodcond}

The soft theorem~\eqref{softthd1} involves the insertion of a soft mode at early times. At least observationally, we are typically concerned with correlations between operators all evaluated at the same time. We, therefore, want to understand when and how the expression~\eqref{softthd1} can be phrased as an identity involving equal-time correlators. (This same question was considered in a similar context in~\cite{Hui:2018cag}. Here we follow that discussion, with some emphasis on slightly different points.)

We begin from the assumption that the soft mode has
a certain time dependence $g(\tau)$ in the long-wavelength limit:
\be
\label{growth}
\lim_{\vec q\to 0}\phi_i (\v q) |0_{\rm in}\rangle = \lim_{\vec q\to 0}\phi_f
  (\v q) |0_{\rm in} \rangle \,{g (\tau_i) \over  g(\tau_f)} \Big( 1 +
 O(q) \Big)\, ,
\ee
where $ O(q)$ indicates that this formula is corrected by terms 
suppressed by (positive) powers
of $q$. 
This is merely a statement about the behavior of the soft mode at leading order in perturbation theory. That is to say, the linearized $\phi$ can be decomposed
into creation and annihilation operators, and $g(\tau)$ reflects the
time-dependence of the corresponding mode function at the vanishing
$q$--- this limit picks 
out a dominant mode (often called the growing mode).\footnote{One can make stronger, nonlinear, statements about the long-wavelength time dependence in certain cases. For example, the curvature perturbation in cosmology freezes outside the horizon nonlinearly~\cite{Maldacena:2002vr,Assassi:2012et,Pimentel:2012tw}. However, it is unclear if analogous results hold in the other examples we study in this paper, though they are not strictly needed for the statements that we make. This is an interesting subject to explore.
Since our soft mode is a Goldstone mode, there is a sense in
which the constant mode is always a solution of the fully nonlinear 
equation of motion.}
Since we are interested in the ratio of the mode
  function at two different times,  the mode function
  normalization (possibly $q$-dependent) does not matter.

Given~\eqref{growth}, we can write a correlation function involving a single early-time mode in terms of an equal-time correlation function with all modes at the late time as
\be
\label{ffi}
\langle 0_{\rm in} | \phi_f (\v k_1) \cdots \phi_f (\v k_N) \phi_i (\v q)
  | 0_{\rm in} \rangle'_c = \langle 0_{\rm in} | \phi_f (\v k_1) \cdots \phi_f (\v k_N) \phi_f (\v q)
  | 0_{\rm in} \rangle'_c {g(\tau_i) \over g(\tau_f)} \Big(1 + O(q)\Big) \, .
\ee
Note that in general, the unequal-time correlator on the left-hand side is complex, while the equal-time correlator on the right-hand side is real. The difference is made up by the correction terms, which are also in general complex.
Another implication of~\eqref{growth} is that the power spectrum at different times can be expressed as
\be
\label{Pif}
P_i(q) = P_f (q) \left( {g(\tau_i) \over
  g(\tau_f)} \right)^2 \Big(1 + O(q)\Big)\, ,
\ee
where $P_i$ and $P_f$ are the $\phi$ power spectrum at
initial and final time respectively. 

The above statements apply regardless of the form of the symmetry transformation. However, only for certain symmetries can these facts be used to phrase the soft theorems as equal time statements.
 We now introduce the assumption that makes the promotion to a final-time Ward identity possible---the physical mode condition that
\begin{tcolorbox}[colframe=white,arc=0pt,colback=greyish2]
\be
\label{physicalmodecond}
D_{\v q} (\tau_i) = D_{\v q} (\tau_f) {g(\tau_i) \over g(\tau_f)} \, .
\ee
\end{tcolorbox}
\noindent
In other words, the physical mode condition states that the
nonlinear part of the field transformation has the same time
dependence as the dominant (growing) mode solution to the equation of motion.\footnote{For a discussion of how this differs from the adiabatic mode
condition, relevant for soft theorems originating from gauge
symmetries, see~\cite{Hui:2018cag}.}

Applying these statements to~\eqref{softthd1}, we have
\be
\label{softequaltime}
D_{\v q} (\tau_f) \left({1 \over P_f (q)} 
\langle 0_{\rm in} | \phi_f (\v k_1) ... \phi_f (\v k_N) 
\phi_f (\v q) | 0_{\rm in} \rangle'_c\right) = 
\sum_{a=1}^N\langle 0_{\rm in} |    \phi_f (\v k_1) \cdots \delta_{\rm L}\phi_f(\v k_a)\cdots\phi_f (\v k_N) ] | 0_{\rm in}  \rangle_c' \, ,
\ee
which is the equal-time Ward identity we were looking for. 

Aside from ~\eqref{physicalmodecond}, we had to make some additional assumptions to obtain this expression: (1) The $\alpha_0$ term 
in~\eqref{softthd1} multiplies 
$\langle 0_{\rm in} | \phi_f (\v k_1) \cdots \phi_f (\v k_N) 
\phi_i (\v q) | 0_{\rm in} \rangle'_c - {\,\rm c. c.}$, and the latter
vanishes by virtue of~\eqref{ffi} assuming $g$ (or the ratio thereof) is real. Note that
$\alpha_0$ is in principle a function of $q$, and it is important that the
product of $\alpha_0$ and the corrections in~\eqref{ffi} remain
subleading in powers of $q$. (2) The kernel ${\cal E}_i$ has an imaginary part; it is important its
product with (the imaginary part of) corrections in~\eqref{ffi} is subleading in
powers of $q$ compared to what is kept in the equal-time
consistency relation~\eqref{softequaltime}.
(3) In cases where $D_{\v q} (\tau_i)$ or $D_{\v q} (\tau_f)$ does not
contain derivatives in $q$, there is no need to be careful
about how fast the corrections in~\eqref{ffi} vanish in the $q
\rightarrow 0$ limit. However, caution is needed in cases where $D_{\v
  q}$ has derivatives. 
In the superfluid example,
$D_{\v q}$ for the boost symmetry carries a $ \v \nabla_{\v q}$.
It is important the corrections start at order $q^n$ with $n > 1$
so that they do not contaminate the boost consistency relation once
it is promoted to equal (final) time.
A possible source of corrections in~\eqref{ffi}
comes from the mode function. It could happen
that the mode function at small $\vec q$ behaves in
such a way that~\eqref{growth} has $O(q)$ corrections, which
would imply the same in~\eqref{ffi}. The presence of these
corrections (once acted on by $\v \nabla_{\v q}$) would spoil the promotion of the boost consistency
relation to equal time.
We will see that this is exactly what happens in the superfluid case.\footnote{In cases
where the consistency relations come in a hierarchy 
such as with time-translations and boosts (see footnote~\ref{deltafuncRemove}), it is the {\it same} $g(\t)$ that
shows up in the physical mode condition for both. 
}
Henceforth, we implicitly assume these additional assumptions are
satisfied when we use the term physical mode condition.

\newpage
\section{Applications}
\label{sec:applications}

To demonstrate the  scope of the soft theorem~\eqref{softthd1}, in this section we discuss a variety of  models with
space/time-dependent nonlinearly realized symmetries and write the explicit form of the soft theorems satisfied by these theories. 
We will also compute explicit correlators for these models, 
which serve as (perturbative) checks of the soft theorem.

We start by considering (in Section~\ref{sec:toymodels}) two simple toy examples with time-dependent interactions that mimic features of the dynamics of perturbations in slow-roll and ultra-slow-roll inflation in cosmology (see also Ref.~\cite{Hui:2018cag}). These models are time-dependent field theories of the form
\be
S = \int\rd^3 x\rd\tau f(\tau) \left(\Exp^{3 \zeta}  \zeta'^2 - \Exp^\zeta (\partial_i \zeta)^2 \right) \,,
\label{eq:toyEFT}
\ee
where $f(\tau)$ is a function of (conformal) time, which can be chosen to capture different features of cosmological models.
The advantage of these toy examples, with respect to the actual inflationary models, is that they are simple enough that they allow us to compute explicitly all the correlation functions of interest while containing all the necessary conceptual ingredients to illustrate and check the soft theorem~\eqref{softthd1}.

We then discuss in Section~\ref{sec:superfluidmain} the low-energy effective theory of a relativistic superfluid. In particular, we derive the consequences of the spontaneously broken time translations and Lorentz boosts.
Finally, we consider cosmology and translate the general soft theorem into the effective field theory of inflation in the decoupling limit in Section~\ref{sec:EFTofI}.
An interesting feature of the superfluid and EFT of inflation examples is that the soft theorems are somewhat trivial if the hard modes involved in the correlator are evaluated at the same time, but become nontrivial statements if the hard modes are evaluated at different times.

While the initial-final soft theorem \eqref{softthd1} is general and just follows from symmetry, the
existence of a soft theorem at late times is not guaranteed
and requires---as discussed in~\cite{Hui:2018cag} and reviewed in
Section~\ref{sec:physmodcond}---the physical mode condition to hold. In the
examples presented below, we will check when this is the case, and also verify the equal-time soft theorems whenever these can be written.

\subsection{Time-dependent toy models}
\label{sec:toymodels}
As a simple warm-up, we consider the action~\eqref{eq:toyEFT} with two different choices of $f(\tau)$ meant to mimic slow-roll and ultra-slow-roll inflation, respectively. These are time-dependent field theories that have symmetries---and corresponding soft theorems obeyed by their correlation functions---but which are simple enough that computing correlators explicitly is straightforward.

\subsubsection{Slow-roll toy model}
\label{sec:Slow-Roll}

We  start by considering the following action, 
introduced by~\cite{Hui:2018cag} to replicate 
certain features of
slow-roll inflation:
\be
S=  \int \D\tau \, \D^3x\,\frac{1}{\tau^2 } \left(\Exp^{3 \zeta}  \zeta'^2 - \Exp^\zeta (\partial_i \zeta)^2 \right) \, ,
\label{actiontoysr}
\ee
where $\tau$ denotes the time coordinate and  $\zeta'
\equiv \partial_\tau \zeta$.\footnote{This is a flat-space model designed to mimic the evolution of the curvature perturbation in slow-roll inflation. We use $\tau$, instead of $t$, to denote the time coordinate as it plays the same role as conformal time in cosmology.} The scalar field $\zeta$ in \eqref{actiontoysr} plays the same role as $\phi$ in the previous sections. 
Besides spatial dilation and special conformal symmetries, which were studied in~\cite{Hui:2018cag},
the action \eqref{actiontoysr}  is invariant under the time-dependent symmetry transformation 
\be
\delta \zeta = 1 - \tau \partial_\tau \zeta \, ,
\label{srsymm}
\ee
up to a boundary term  $\delta S =  \int \D\tau \,
\D^3x~\partial_\tau(-\tau {\cal L})$.
%
Applying the standard Noether procedure by deforming the parameter of the symmetry transformation to a spacetime-depending function, we can extract the Noether current, whose temporal component is
\begin{equation}
j^0 =   \frac{2}{\tau^2}\zeta'+ \ldots
\label{eq:j0SR}
\end{equation}
up to corrections that are higher order in the field $\zeta$.\footnote{To read off the terms in $j^0$ that are
  linear in the field, it is sufficient to focus on the nonlinear part
  of the symmetry and work out $\delta S$ for the quadratic part of
  the action.}
In the notation of equation~\eqref{Charge_a_0}, we then have
\begin{equation}
\alpha_0= 0\, ,
\qquad\qquad
D_{\v q}= 1\, ,
\end{equation}
and the unequal-time soft theorem~\eqref{softthd1} reads
\be
\lim_{\vec q \rightarrow 0}
 {\cal E}_i (\vec q)  \langle \zeta(\v k_1,\t_f) \zeta(\v k_2,\t_f) \zeta(\v q,\t_i) \rangle' + {\rm c. c.} = -\tau_f
\left[ \langle \zeta'(\v k_1,\t_f) \zeta(\v k_2,\t_f) \rangle' + 
\langle \zeta(\v k_1,\t_f) \zeta'(\v k_2,\t_f) \rangle' \right] \, ,
\label{toyidsrif}
\ee
where we have suppressed the appearance of the in-vacuum.

It is straightforward to check that the relation~\eqref{toyidsrif} is
satisfied by plugging in the explicit expressions for the correlators
(at tree level). The two-point function and wavefunction kernel are
\be
\langle\zeta(\v k,\t_f)\zeta( - \v k,\t_f) \rangle'   = \frac{1}{4  k^3}(1+k^2 \tau_f^2)\, , 
\qquad
\mathcal{E}_i(q)  = \frac{2 q^2 M^2}{i H^2 \tau_i (1-i q \tau_i)}\, , 
\ee
while the unequal-time three-point function in the soft limit is given by
\be \lim_{\v q\rightarrow 0} \mathcal{E}_i(q)   \langle \zeta(\v
k,\t_f) \zeta(- \v k - \v q, \t_f) \zeta(\v q,\t_i) \rangle '  +\text{c.c.}  = - \frac{ \tau_f^2}{2 k} \, .
\ee
In addition, one can check that the physical mode condition is
satisfied for the toy model~\eqref{actiontoysr} (see also~\cite{Hui:2018cag}). This allows us to promote the initial-final
soft theorem \eqref{toyidsrif} to late times as follows:
\be
\lim_{\bfq\rightarrow 0} \frac{ \langle \zeta(\vec k) \zeta(-{\vec k}-{\vec q}) \zeta({\vec q}) \rangle ' }{\langle\zeta ({\vec q})\zeta({-\vec q}) \rangle' }
= - \tau \partial_{\tau}  \langle\zeta(\vec k)\zeta(-\vec k)\rangle'    \, ,
\label{toyidsrff}
\ee
where we have suppressed the time-dependence of $\zeta$, assuming that
all the fields are evaluated at the same (late) time $\tau$. A
perturbative computation of the relevant correlators also confirms
this equal-time version of the soft theorem.

\subsubsection{Ultra-slow-roll toy model}
\label{sec:Ultra-Slow-Roll}

Let's now consider a similar-looking scalar theory but with a different 
time-dependent
factor designed to mimic the time evolution of the curvature perturbation in ultra-slow-roll inflation~\cite{Tsamis:2003px,Kinney:2005vj}:
\be
S = \int \D\tau \, \D^3  x \, \tau^4 \left( \Exp^{3\zeta} \zeta'^2 - \Exp^\zeta (\partial_i \zeta)^2 \right) \, .
\label{eq:USRaction}
\ee
The action \eqref{eq:USRaction} is invariant under the following
nonlinearly realized symmetry:
\be
\delta\zeta = {1\over \tau^3} + {1\over 2 \tau^2} \zeta' \, .
\label{usrsymm}
\ee
We see that the nonlinear part of the transformation is given by
$\delta_{\rm NL} \zeta = 1/\tau^3$ and the linear part is $\delta_{\rm L} \zeta = (1 /2 \tau^2) \zeta'$. 
We can again extract the time component  of the conserved Noether current $j^\mu$ associated with \eqref{usrsymm} in the usual way
\be
j^0 = 2 + 6  \zeta  + 2 \tau \zeta' + \cdots = 
2+ 6 \zeta + {1 \over \tau^3} \Pi_\zeta + \cdots \, ,
\label{eq:j0USR}
\ee
where the ellipsis represents terms involving two or more $\zeta$s.\footnote{The constant part of  $j^0$ gives rise to a divergent contribution to the charge $Q
= \int \D^3  x \, j^0$, but it can be ignored because it gets
canceled in the commutator combination \eqref{WardC}.}
Translating into the notation of Section~\ref{sec:derivation}, we have
\be
\alpha_0 = 6\, , \qquad\qquad D_{\v q} = \frac{1}{\tau^3}\,,
\label{eq:usrdefs}
\ee
and, using $\delta_{\rm L} \zeta_f  = (1 /2 \tau_f^2) \zeta'_f$, we arrive at the following 
 unequal-time soft theorem relating the three and two-point functions:
\be
\label{usrcc}
\lim_{\vec q \rightarrow 0}
\left( {1\over \tau_i^3} {\cal E}_i (\vec q) - 6 i \right) \langle \zeta(\v k_1,\t_f) \zeta(\v k_2,\t_f) \zeta(\v q,\t_i) \rangle' + {\rm \, c.  c.} = {1\over 2 \tau_f^2} \frac{\D}{\D \t_f}
 \langle \zeta(\v k_1,\t_f) \zeta(\v k_2,\t_f) \rangle' \, .
\ee
This relation can be explicitly checked  using the expression for the two-point function
\be
\begin{aligned}
\langle \zeta(\v k,\t_f) \zeta(-\v k,\t_f) \rangle'    = \frac{1+k^2 \t_f^2}{4  k^3 \t_f^6} \, , \qquad 
\qquad \mathcal{E}_i(q)  = i \left( 6 \t_i^3 - \frac{2 \t_i^5 q^2}{1-i q \t_i} \right)\, , 
\end{aligned}
\ee
along with the following limits of the three-point function
\begin{align}
 \lim\limits_{\bfq\rightarrow 0} \frac{1}{\t_i^3} \mathcal{E}_i(q)  \langle \zeta(\v k,\t_f) \zeta(-\v k - \v q,\t_f) \zeta(\v q,\t_i) \rangle '  +\text{c.c.}   &= 0\, , 
\\
 - 6 i  \lim\limits_{\bfq\rightarrow 0} \langle \zeta(\v k,\t_f) \zeta(- \v k - \v q,\t_f) \zeta(\v q,\t_i) \rangle '  + \text{c.c.}  &= - \frac{3 + 2 k^2  \t_f^2}{4  k^3  \t_f^9} \, .
 \label{eq:USRcheck}
\end{align}

It is furthermore straightforward to check that the physical mode condition
holds for
the model~\eqref{eq:USRaction}, which allows us to promote the initial-final soft theorem~\eqref{usrcc} to late times, where we obtain
\be
\lim_{\vec q \rightarrow 0}
 \frac{\langle \zeta(\v k) \zeta(-\vec k - \v q) \zeta (\vec q) \rangle' }{\braket{\z(\v q) \z (-\v q)}'}   = \frac{\t}{2} \frac{\D}{\D \t}
\langle \zeta (\vec k) \zeta (-\vec k) \rangle' \, .
\ee
This equal-time consistency relation can also be verified by explicit computation of the relevant
correlators.

An interesting fact about this model is that
\beq
\lim \limits_{\v q \to 0} \frac{i}{\t_i^3} {\rm Im}\, \E_i(\v q) = 6i \, .
\eeq
In other words, only the real part of
the combination $({\cal E}_i(\v q)/\tau_i^3 - 6i)$ survives in the
soft limit, which implies a relation between the boundary term $\alpha$ and the imaginary part of the wave function. In some circumstances, this observation generalizes to other examples (see Appendix~\ref{1ptconstr} for a discussion).

A point worth stressing is that the crucial difference between the case discussed here and the one in Section~\ref{sec:Slow-Roll} is the presence of a non-vanishing term linear in the field $\zeta$
in the   Noether current \eqref{eq:j0USR}, i.e., $\alpha\neq0$ (as
defined in~\eqref{eq:usrdefs}).   
Such a nonzero $\alpha$ follows from (temporal) boundary terms in the symmetry
variation of the action (see \eqref{j0F0}), 
and is required in order for \eqref{Ward0} with ${\cal O}\equiv \Pi_\z$ to hold. In fact, one can directly compute the commutator of the Noether charge $Q = \int \D^3  x\, j^0(\v x)$ with the conjugate momentum $\Pi_\z$ as
\beq
i \lb Q , \, \Pi_\z(\v y)  \rb =i  \int \D^3  x \, \lb 6  \z(\v x)
+ \cdots \, , \, \Pi_\z(\v y) \rb = - 6  + \cdots\, ,
\label{commQpi}
\eeq
where the ellipsis denotes linear and higher order terms in the field $\z$.
It is easy to check that \eqref{commQpi} is consistent with the variation of  $\Pi_\z$ under the symmetry \eqref{usrsymm}:
\beq
\d_{\rm} \Pi_\z = \d_{\rm} \(2 \t^4 e^{3 \z} \z' \) = 2 \t^4 \partial_\tau( \d_{\rm NL}\z) + \cdots = - 6  + \cdots \, .
\eeq
In cases where $\delta\Pi_\z$ does not contain any field-independent piece  (like in the  example of  Section~\ref{sec:Slow-Roll} where $\delta \Pi_\z \propto \d_{\rm NL}\z' = 0$),
then $\alpha=0$ and the generalized soft theorem \eqref{softthd0} (as well as its late-time version, if the physical mode condition is satisfied) reduces to the previous results of, e.g., \cite{Hinterbichler:2013dpa,Hui:2018cag}.

\subsection{Superfluid}
\label{sec:superfluidmain}

The next example we consider is the effective field theory of a relativistic superfluid.
We can take as the starting point the action~\cite{Son:2002zn}
\begin{equation}
S = \int {\rm d}^4x \, \Lambda^4P(X) \, ,
\label{Lsuperfluid}
\end{equation}
where $P$ is a generic function of $X$, defined by
\begin{equation}
X\equiv -\frac{1}{\Lambda^4}\partial_\mu\phi \partial^\mu\phi \, .
\end{equation}  
The superfluid ground state $\bar \phi = \mu t$ (where $\mu$ is the chemical potential) spontaneously breaks time translations and boosts.
Expanding the action in perturbations
around this state:
\begin{equation}
\phi(\bfx, t) = \mu t + \pi(\bfx, t) \, ,
\label{phidef}
\end{equation}
we have a theory of
the low-energy dynamics of an ideal superfluid at zero temperature and finite density (or finite chemical potential $\mu$), where the function $P(\mu)$ is the equation of state of the superfluid, giving the pressure as a function of chemical potential~\cite{Son:2002zn,Nicolis:2011pv,Nicolis:2013lma,Nicolis:2015sra} (see also Appendix~\ref{sec:superfluid} below for more details).

The field fluctuations  $\pi$ are the (derivatively coupled) Goldstone bosons that parametrize the phonon excitations in the superfluid.
In addition to constant shifts ($\pi\mapsto\,\pi+c$), the field $\pi$ also nonlinearly realizes time translations and Lorentz boosts, which are spontaneously broken by the time-dependent background $\bar \phi $. (There is, however, a diagonal combination of a shift of $\pi$ and a time translation that remains linearly realized.) Our goal in this section is to derive the soft theorems that result from the spontaneous breaking of these symmetries at the level of in-in correlation functions of $\pi$. 
(We will comment briefly  on soft theorems for scattering amplitudes
 in Appendix~\ref{sec:softampl}, but see~\cite{Green:2022slj} for an extensive discussion.)

By expanding the action~\eqref{Lsuperfluid} around the background~\eqref{phidef}, one can find the lagrangian for $\pi$ at all orders, which depends on the functional form of $P(X)$. For instance, the quadratic action for $\pi$ is
\be
S^{(2)} = \int {\rm d}^4x  \, {P'(\mu^2) \over c_s^2} \Big( \dot\pi^2 - c_s^2 (\partial_i \pi)^2\Big) \, ,
\label{S2phnon}
\ee
where $P'\equiv \D P/\D X$, $\dot \pi \equiv \partial_t \pi$, and the sound speed (squared) is defined by
\be
c_s^2 \equiv {P'(\mu^2) \over P' (\mu^2) + 2 \mu^2 P''(\mu^2)} \, .
\ee
The expressions for the cubic and quartic lagrangians, which we will need later on for the explicit checks of the soft theorems, are reported in~\eqref{superfluidL} in Appendix \ref{sec:superfluid}. As it is clear from \eqref{S2phnon} and \eqref{superfluidL}, the effective couplings are (constant) functions of $P$ and its derivatives, computed on the background $\bar \phi$.\footnote{The fact that the couplings in the low-energy effective theory for the Goldstone $\pi$ are time-independent is a consequence of the existence of the diagonal combination of time translations and constant shifts that is linearly realized on $ \phi$ \cite{Nicolis:2011pv}.}
Note that, at any fixed order in perturbation theory, there are some operators whose couplings are fixed in terms of lower-order couplings. This is the case, for instance, for the operators $\dot \pi(\partial_i\pi)^2$ and $(\partial_i\pi)^4$ (see Eqs.~\eqref{superfluidL}), whose coefficients, apart from a common overall normalization factor, are functions of the sound speed $c_s$ only. This is a result of the specific symmetry breaking pattern realized in the system: the Poincar\'e group is broken down to spatial translations and rotation.
As we will show, this pattern gets reflected at the level of soft theorems.

\subsubsection{Spontaneously broken time translations}

Let us start by considering time translations.
Time translations act nonlinearly on $\pi$ as 
\be
\delta\pi = \mu + \dot\pi \, .
\label{superflshift}
\ee
The corresponding Noether current is 
\be
j^0 =  {2\mu P'(\mu^2) \over c_s^2} \dot\pi + \cdots =  \mu \Pi_\pi + \cdots \, .
\ee
Thus, we have, in the language of Section~\ref{sec:derivation}
\be
\alpha_0 = 0\, , \qquad D_{\v q} = \mu \, .
\ee
Recalling that
$\delta_{\rm L} \pi (t_f) = \dot\pi (t_f)$, we conclude that
the consistency condition~\eqref{softthd1} takes the form 
\be
\lim_{\vec q \rightarrow 0} \mu {\cal E}_i (q) \langle \pi (\vec k_1,t_f) \pi(\vec k_2,t_f) \pi(\vec q,t_i) \rangle' + {\,\rm c.  c.} = \langle \dot\pi(\vec k_1,t_f) \pi(\vec k_2,t_f) \rangle'
+ \langle \pi(\vec k_1,t_f) \dot \pi(\vec k_2,t_f) \rangle' \, .
\label{shiftccsuperfl}
\ee
From the action for the field $\pi $ (see~\eqref{superfluidL}) one can compute explicitly the correlators in \eqref{shiftccsuperfl} and check in fact that \eqref{shiftccsuperfl} is satisfied, but in a trivial way, namely, both sides of the equation are zero. Essentially this is a consequence of the shift symmetry of $\pi$,  which implies that the left-hand side of \eqref{shiftccsuperfl} must vanish.\footnote{The fact that $\pi$ has a shift symmetry can be thought of as a consequence of the fact that the background for $\phi$ is linear in time~\cite{Finelli:2017fml,Finelli:2018upr,Floess:2018ths}. In Appendix~\ref{sec:drivensuperfluid} we comment on a slight deformation of  \eqref{Lsuperfluid} that is still shift symmetric but that admits a background solution for $\phi$ that is not linear in time. We will see in that case, the left-hand side of the soft theorem \eqref{3to2drivtimetransl} does not vanish.}  Instead of time translations, we could have instead considered this soft theorem associated with the shift symmetry. The difference between this soft theorem and~\eqref{shiftccsuperfl} is the Ward identity for the linearly realized time translations.

Note that the physical mode condition is satisfied for the symmetry~\eqref{superflshift}. In fact,  the mode function for $\pi$, which is just a  plane wave, is constant at  leading order in the soft limit,
\beq
\lim_{\vec q\to 0} \Exp^{-i c_s qt} \simeq 1 - i c_s q t + \cdots \, ,
\eeq
like the nonlinear part of the transformation \eqref{superflshift}.
As a result, the soft theorem \eqref{shiftccsuperfl} can be promoted to late times, as can be explicitly checked.

\subsubsection{Soft theorems from spontaneously broken boosts}
\label{subsec:SuperBoost}

In addition to time translations, Lorentz boosts are also spontaneously broken by the superfluid ground state. The action of such a boost on $\pi$ is
\be
\delta\pi = b^j \Big[ \mu x_j +  (t \partial_j + x_j \partial_t) \pi  \Big] \, ,
\label{superflboost}
\ee
where $\vec b$ is an infinitesimal constant vector.
It is straightforward to derive the corresponding Noether current and extract
\be
j^0 = b_j {2\mu P'(\mu^2) \over c_s^2} x^j \dot\pi + \cdots = 
b_j\mu x^j \Pi_\pi + \cdots \, .
\ee
Translating this into the notation of Section~\ref{sec:derivation}, we find
\beq
\alpha_0 = 0\, , \qquad\quad D_{\v q} = -i \epsilon^j\mu \partial_{q^j} \, ,
\eeq
with the understanding that the action of the operator $D_{\v q}$ also involves sending $\vec q \rightarrow 0$.\footnote{The fact that the soft mode generated by a symmetry transformation is localized around $\vec q = 0$ follows from the Fourier transform
$\int \D^3  x \,  x^j \, \Exp^{-i {\vec q} \cdot {\vec x}} = i(2\pi)^3  \partial_{q^j} 
 \delta ({\vec q})$.}
 
With these ingredients, we can adapt the formula~\eqref{softthd1} to this model. We first consider the soft theorem that relates the three and two-point functions:
\be
\lim_{\vec q \rightarrow 0} \mu \partial_{q^j} \left( {\cal E}_i (q) 
\langle \pi (\vec k_1, t_f) \pi (\vec k_2, t_f) \pi (\vec q, t_i) \rangle' 
+ {\,\rm c. c.} \right) =   
\partial_{t_f} \partial_{k^j_1} \langle \pi_f (\vec k_1, t_f) \pi_f (-\vec
k_1, t_f) \rangle' \, 
\label{superflccboost} 
\ee
If one starts from the version of \eqref{softthd1} before the
momentum-conserving delta function removal, to arrive at the above
expression requires being careful about how to treat 
momentum derivatives under the delta function constraint.
This will be discussed below.

The identity~\eqref{superflccboost}  is indeed satisfied by correlation functions of a superfluid, but in a somewhat trivial way---both sides vanish identically, similar to the soft theorem~\eqref{shiftccsuperfl} for time translations considered above.
However, we can consider a slightly more general situation where the identity is nontrivial. That is, we consider the case where the operators carrying the hard momenta are inserted at different times. Effectively this will lead to a soft theorem involving operators at three different times.

Let us go back to the original formulation
(\ref{softthd0}) and choose ${\cal O} = \pi(\vec{k}_1,t_1)
\pi(\vec{k}_2,t_2)$,
where $t_1$ and $t_2$ are two different times. 
Moreover, let us start from the formulation before delta function removal.
We thus have:
\be
\begin{aligned}
 \lim_{\vec q \rightarrow 0} \mu \partial_{q^j} \Big({\cal E}_i (q) \Big[
\langle \pi_{\vec k_1}(t_1) \pi_{\vec k_2}(t_2) \pi_{\vec q}(t_i) \rangle &+ 
\langle \pi_{\vec q}(t_i)  \pi_{\vec k_1}(t_1) \pi_{\vec k_2}(t_2)
\rangle \Big] \Big)
\\
&=  \left(\partial_{k_1^j} \partial_{t_1}+\partial_{k_2^j}\partial_{t_2}+ t_1 k_{1}^j +t_2 k_{2}^j \right) \braket{ \pi_{\v k_1}(t_1) \pi_{\v k_2}(t_2)}
\, .
\label{superfl3to2uneqtime}
\end{aligned}
\ee
Notice that the $t \partial_j \pi$ part of the boost symmetry now acts nontrivially on the two-point function on the right-hand side of this relation, since we have separated the final times. Removing the momentum-conserving delta function from this expression is rather subtle, commuting the $\partial_k$ derivatives past the delta function eventually leads to the identity
\begin{tcolorbox}[colframe=white,arc=0pt,colback=greyish2]
\begin{align}
\label{softtht1t2}
&\lim\limits_{q \rightarrow 0} \frac{\mu}{2} \prl_{q^j}\left(\frac{\braket{\pi_{\v k_1}(t_1) \pi_{\v k_2}(t_2) \pi_{\v q}(t_i) }' + \braket{\pi_{\v q}(t_i) \pi_{\v k_1}(t_1) \pi_{\v k_2}(t_2)  }'}{\braket{ \pi_{-\v q}(t_i) \pi_{\v q}(t_i)}'} \right) \\\nonumber
&\hspace{7.5cm}
= \left(\prl_{t_1}\prl_{k_{1}^j}+k_{1}^j (t_1-t_2)\right) \braket{  \pi_{\v k_1}(t_1) \pi_{-\v k_1}(t_2)}' \, ,
\end{align}
\end{tcolorbox}
\noindent
which we derive explicitly in the following inset.

\begin{framed}
{\small
\vspace{-.15cm}
\noindent
{\bf\small Commuting through the delta function:} Here we describe how
to get from~\eqref{superfl3to2uneqtime} to~\eqref{softtht1t2}.
Let us consider the first two terms on the right-hand side of the relation \eqref{superfl3to2uneqtime}. First, we separate the contributions where the momentum derivative acts on the delta function. Introducing the primed correlator with a delta function removed, i.e.,
\beq
\langle \pi_{\vec k_1}(t_1) \pi_{\vec k_2}(t_2) \rangle \equiv  \d(\v k_1 + \v k_2) \langle \pi_{\vec k_1} (t_1)\pi_{\vec k_2}(t_2) \rangle'\, ,
\eeq
we can write
\be
\begin{aligned}
 &\left( \partial_{k_1^j} \d(\v k_1 + \v k_2) \right) \langle \dot \pi_{\vec k_1}(t_1) \pi_{\vec k_2}(t_2) \rangle'
+  \left( \partial_{k_2^j} \d(\v k_1 + \v k_2)  \right) \langle \pi_{\vec k_1}(t_1)  \dot \pi_{\vec k_2}(t_2) \rangle' \\
 &\hspace{1.25cm}= \left[ \left( \partial_{k_1^j} \d(\v k_1 + \v k_2) \right) \langle \dot \pi_{\vec k_1}(t_1) \pi_{-\vec k_1}(t_2) \rangle'
+  \left( \partial_{k_1^j} \d(\v k_1 + \v k_2)  \right) \langle \pi_{\vec k_1}(t_1)  \dot \pi_{- \vec k_1} (t_2)\rangle'  \right] \\
&\hspace{1.4cm}~~~~
-   \d(\v k_1 + \v k_2) \langle  (\partial_{k_1^j} \dot \pi_{\vec k_1}(t_1) )\pi_{-\vec k_1}(t_2) \rangle'
-   \d(\v k_1 + \v k_2)   \langle (\partial_{k_1^j}  \pi_{\vec k_1}(t_1) ) \dot \pi_{-\vec k_1}(t_2) \rangle' \, ,
\label{superfl3to2deltaderiv}
\end{aligned}
\ee
where we have used the distributional identities
\beq
\begin{aligned}
f(\v k_1, \v k_2) \d(\v k_1 + \v k_2) &  = f(\v k_1, -\v k_1) \d(\v k_1 + \v k_2) \, , \\
f(\v k_1, \v k_2) \prl_{k_1^j}\d(\v k_1 + \v k_2) & = f(\v k_1, -\v k_1) \prl_{k_1^j}\d(\v k_1 + \v k_2) - \left. \left( \prl_{k_1^j} f(\v k_1, \v k_2) \right) \right|_{\v k_2 = - \v k_1} \d(\v k_1 + \v k_2)\, .
\end{aligned}
\eeq
Of the terms on the right hand side of (\ref{superfl3to2deltaderiv}), 
the ones where the delta function comes with a derivative will cancel
analgous terms on the left hand side of 
(\ref{superfl3to2uneqtime}), making use of the (lower order) time
translation soft theorem (essentially \ref{shiftccsuperfl} but with unequal times on the hard modes). 
The remaining terms on 
the right hand side of (\ref{superfl3to2deltaderiv}) can be combined
with analogous terms
on the right hand side of (\ref{superfl3to2uneqtime}) 
--- those without delta function derivative ---
to obtain
\eqref{softtht1t2}.
Note that the two-point function is overall time translation invariant and
is thus a function of $t_1 - t_2$.

}
\end{framed}

We now would like to verify that the identity~\eqref{softtht1t2} is satisfied in a nontrivial way. We can compute the relevant correlators explicitly using the lagrangian~\eqref{superfluidL}. The two-point function is given by
\be
\braket{ \pi_{\v k_1}(t_1) \pi_{-\v k_1}(t_2)}'= \frac{c_s}{4 P'(\mu^2) k_1} \Exp^{-i c_s k_1 ( t_1 - t_2)}\,,
\ee
while the soft limit of the three-point function is
\be \label{eqn:3ptsuperflsqzed}
\lim\limits_{\vec q \rightarrow 0} \frac{\mu}{2} \prl_{q^j}\left(\frac{\braket{\pi_{\v k_1}(t_1) \pi_{\v k_2}(t_2) \pi_{\v q}(t_i) }' + \braket{\pi_{\v q}(t_i) \pi_{\v k_1}(t_1) \pi_{\v k_2}(t_2)  }'}{\braket{ \pi_{-\v q}(t_i) \pi_{\v q}(t_i)}'} \right) = \frac{c_s(1-c_s^2) k_{1}^j}{4 P'(\mu^2) k_1} (t_1-t_2)\Exp^{-i c_s k_1 ( t_1 - t_2)} .
\ee
Using these expressions, it is straightforward to check that the relation~\eqref{softtht1t2} is indeed satisfied.

\noindent
{\bf Constraints on the action:}
We can think of the identity~\eqref{softtht1t2} as the manifestation of the spontaneously broken boost symmetry of a relativistic superfluid at the level of observables. In order to make this more explicit, it is useful to take a slightly different perspective.
Consider the cubic action for a superfluid~\eqref{superfluidL}, but allow the couplings to be arbitrary:
\be
S  =\int\rd^4x\frac{P'(\mu^2)}{c_s^2}\bigg(\dot{\pi}^2 -  c_s^2(\partial_i\pi)^2 
+
c_{33} \dot \pi^3
+
c_{31}\dot \pi(\partial_i\pi)^2 
 \bigg) \, .
 \label{eq:freefluid}
\ee
In the actual superfluid EFT $c_{33}$ and $c_{31}$ are known and are given by
\begin{equation}
\label{c33c31}
c_{33} = \frac{(1-c_s^2)(3 c_s^2 + 2 c_3)}{3 c_s^2 \mu}\,, \qquad ~~~~~ c_{31}  =- \frac{(1-c_s^2)}{ \mu} \, .
\end{equation}
However, we can think of them as free parameters and ask how they are constrained by the soft theorem~\eqref{softtht1t2}. If one computes the correlators following from the action~\eqref{eq:freefluid} and then enforces~\eqref{softtht1t2} as a constraint, this uniquely fixes $c_{31}$ to take  the value~\eqref{c33c31}, while $c_{33}$ is unconstrained. This is not in itself surprising, because we know that the form of the coupling $c_{31}$ is fixed by the spontaneously broken Lorentz symmetries, as can be checked directly---the action~\eqref{eq:freefluid} is only invariant under~\eqref{superflboost} when $c_{31}$ is given by~\eqref{c33c31}.

If one considers the boost soft theorem that relates the three and two-point functions, and where the two hard modes are evaluated at the same time, the identity is satisfied in a trivial way---both sides are zero. This is perhaps slightly puzzling because we know that the boost symmetry of the action relates some of the cubic interactions to the quadratic action: where has this information gone?
The nontrivial feature is that to recover this information at the
level of observables we need to consider the hard modes at unequal
times.\footnote{A philosophically similar situation arises in the
  subleading consistency condition for the large scale structure
  bispectrum \cite{Horn:2014rta, Kehagias:2013yd, Peloso:2013zw, Creminelli:2013mca}, and for the EFT of inflation, as we will describe.}

\noindent
{\bf Four points:}
The previous discussion can be
straightforwardly extended to higher-point correlation functions. We can, for instance, consider the four-point function.
The quartic Hamiltonian in the interaction picture can be written generically as
\beq
\label{HamiltoniansPX}
\bal
\ham^{(4)} & = c_{44} \dot \pi^4 + c_{42} \dot \pi^2 (\prl_i \pi)^2 + c_{40} (\prl_i \pi)^4\, ,
\eal
\eeq
for some (known) constants $c_{44}$, $ c_{42}$ and $ c_{40}$.
The unequal-time soft theorem relating the four and three-point functions is given by
\beq
\lim \limits_ {\v q \rightarrow 0} \mu \prl_{q^j} \left({\cal{E}}_i(q) \braket{\pi_{\v k_1}(t_f) \pi_{\v k_2}(t_f) \pi_{\v k_3}(t_f)  \pi_{\v q}(t_i)}' \right) + {\rm{c.c.}} =  \left( \prl_{t_1}\prl_{k_{1}^j} + \prl_{t_2}\prl_{k_{2}^j} \right) \braket{\pi_{\v k_1}(t_1) \pi_{\v k_2}(t_2) \pi_{\v k_3}(t_3)}' \,,
\label{superfl4to3uneqtime}
\eeq
where all the fields on the right-hand side should be evaluated at $t_f$ after taking the derivatives.

Using \eqref{HamiltoniansPX}, the left-hand side of \eqref{superfl4to3uneqtime} can be evaluated as\footnote{Note that exchange diagrams do not contribute in the soft limit.} 
\be
\begin{aligned}
\lim \limits_ {\v q \rightarrow 0} \mu \prl_{q_j} \left({\cal{E}}_i(q) \braket{\pi_{\v k_1}(t_f) \pi_{\v k_2}(t_f) \pi_{\v k_3}(t_f)  \pi_{\v q}(t_i)}' \right) + {\rm{c.c.}}  &= \frac{ c_s^2 \mu k_1^j(k_1-k_3)(c_s^2 c_{42} k_2 - 2 c_{40}(k_1+k_3))}{8 P'^3(\mu) k_1 k_2 k_3 (k_1 + k_2 + k_3)} \\
&\hspace{.25cm}+ \frac{ c_s^2 \mu k_2^j (k_2 - k_3)(c_s^2 c_{42} k_1 - 2 c_{40} (k_2+k_3))} {8 P'^3(\mu) k_1 k_2 k_3 (k_1 + k_2 + k_3)} \, ,
\end{aligned}
\ee
while the right-hand side of \eqref{superfl4to3uneqtime} is
\be
\begin{aligned}
\left. \left( \prl_{t_1}\prl_{k_{1j}} + \prl_{t_2}\prl_{k_{2j}} \right) \braket{\pi_{\v k_1}(t_1) \pi_{\v k_2}(t_2) \pi_{-\v k_1-\v k_2}(t_3)}' \right|_{t_f} = 
\frac{c_s^4 k_1^j (k_1-k_3)(c_{31}(k_1-2 k_2 +k_3) - 3 c_s^2 c_{33} k_2)}{16 P'^3(\mu) k_1 k_2 k_3 (k_1+k_2+k_3)}\\
+ \frac{c_s^4 k_2^j (k_2-k_3)(c_{31}(k_2-2 k_1 +k_3) - 3 c_s^2 c_{33} k_1)}{16 P'^3(\mu) k_1 k_2 k_3 (k_1+k_2+k_3)} \, ,
\end{aligned}
\ee
where $k_3=\sqrt{k_1^2+k_2^2+2(\v k_1 \cdot \v k_2)}$. We can then equate the coefficients of the different independent combinations of momenta to obtain the following constraints on the coefficients appearing in~\eqref{HamiltoniansPX}:
\beq
\label{relsHcoeffs}
c_s^4 c_{31} + 4 c_s^2 \mu c_{40} = 0 \, , \qquad
2 c_{31} + 3 c_s^2 c_{33} + 2 \mu c_{42} = 0 \, .
\eeq
It is straightforward to check that the relations \eqref{relsHcoeffs} are indeed satisfied
by the explicit  expressions for the couplings $c_{44}$, $c_{42}$ and $c_{40}$ given in~\eqref{eq:sfluidH} and summarized here: 
\beq
\bal
c_{44} & = -\frac{(1-c_s^2)[c_s^2(c_s^2+4 c_3 +2 c_4) - (1 - c_s^2)(3 c_s^2 + 2 c_3)^2] P'(\mu^2)}{4 c_s^6 \mu^2} \, , \\
c_{42} & = \frac{(1-c_s^2)(-2+3 c_s^2 +2 c_3)P'(\mu^2)}{2 c_s^2 \mu^2}\, , \qquad c_{40}  = - \frac{(1-c_s^2) P'(\mu^2)}{4 \mu^2}\, .
\eal
\label{superflhamcoeff}
\eeq
It is worth stressing the result~\eqref{relsHcoeffs} follows simply
from knowing that boosts are nonlinearly realized in the theory. Note also that the soft theorem relating the four and three-point functions is satisfied nontrivially even when all of the hard modes are evaluated at the same time, in contrast to the three-point case.

The relation~\eqref{superfl4to3uneqtime} cannot be 
promoted to equal time
for all (soft and hard) modes.
As we stressed in Section~\ref{sec:physmodcond}, 
it is important for subleading identities like the boost one that the mode functions do not contain a linear in $q$ piece with a different time dependence from the symmetry. However, in the limit of small $q$, the mode function for the superfluid phonon becomes
$\Exp^{-i c_s q t} \simeq 1 - i c_s q t + \cdots \, $. Note that the nonlinear
part of the boost symmetry is constant in time, while the $O(q)$ part
of the mode function is linear in time, so the physical mode condition
is not satisfied. Consequently, the equal-time consistency relation is
not satisfied, as can be verified by an explicit check. 

\subsubsection{Identities involving correlators of $\dot{\pi}$ }

We have seen that the identities lead to a nontrivial statement when all $\pi$s are evaluated at different times in a correlator. One may still wonder if there exist nontrivial statements associated with correlators with all late-time operators evaluated at the same time. Heuristically, $\pi_k(t_2)$ can be expressed in terms of expansions around $\pi_k(t_1)$ by
\be
  \pi_k(t_2) = \pi_k(t_1) + (t_2-t_1) \dot{\pi}(t_1) + {\cal O}\left((t_2-t_1)^2\right),
\ee
therefore, one expects that a non-trivial soft theorem involving the squeezed limit of the correlator $\langle\pi(t_1) \dot{\pi}(t_1) \pi(t_i) \rangle$.\footnote{We thank Enrico Pajer for raising this point.} Explicitly, the corresponding identity is 
\be
\begin{aligned}
 & \lim_{\vec q \rightarrow 0} \mu \partial_{q^j} \Big({\cal E}_i (q) \Big[
\vev{\pi_{\vec k_1}(t) \dot{\pi}_{\vec k_2}(t) \pi_{\vec q}(t_i) } + 
\vev{ \pi_{\vec q}(t_i)  \pi_{\vec k_1}(t) \dot{\pi}_{\vec k_2}(t) }\Big] \Big)  \\ 
&\hspace{3cm }= t k_1^j \vev{\pi_{\vec k_1}(t) \dot{\pi}_{\vec k_2}(t)} + \partial_{k_1^j} \vev{\dot{\pi}_{\vec k_1}(t) \dot{\pi}_{\vec k_2}(t)}  \nonumber \\
& \hspace{3cm }\quad + k_2^j \vev{\pi_{\vec k_1}(t) \pi_{\vec k_2}(t)} + t k_2^j \vev{\pi_{\vec k_1}(t) \dot{\pi}_{\vec k_2}(t)}  + \partial_{k_2^j} \vev{\dot{\pi}_{\vec k_1}(t) \ddot{\pi}_{\vec k_2}(t)},
 \end{aligned}
 \ee
 where the second line on the right hand side comes from
\be
    \delta \dot{\pi} = b^j(\partial_j \pi + t \partial_j \dot{\pi} + x_j \ddot{\pi}). 
\ee
After carefully removing the momentum conserving delta functions, it reads
\be
\begin{aligned}
& \lim_{\vec q \rightarrow 0} \mu \partial_{q^j} \Big({\cal E}_i (q) \Big[
\vev{\pi_{\vec k_1}(t) \dot{\pi}_{\vec k_2}(t) \pi_{\vec q}(t_i) }' + 
\vev{ \pi_{\vec q}(t_i)  \pi_{\vec k_1}(t) \dot{\pi}_{\vec k_2}(t) }' \Big] \Big) \\ 
&\qquad = \frac{1}{2} \partial_{k_1^j} \vev{\dot{\pi}_{\vec k_1}(t) \dot{\pi}_{-\vec k_1}(t)}' - k_1^j \vev{\pi_{\vec k_1}(t) \pi_{-\vec k_1}(t)}' - \frac{1}{2}\partial_{k_1^j} \vev{ \pi_{\vec k_1}(t) \ddot{\pi}_{-\vec k_1}(t)}' \\
&\qquad =  \partial_{k_1^j} \vev{\dot{\pi}_{\vec k_1}(t) \dot{\pi}_{-\vec k_1}(t)}' - k_1^j \vev{\pi_{\vec k_1}(t) \pi_{-\vec k_1}(t)}',
\end{aligned}
\ee
where in the last line we have used the fact that correlators are time-translation invariant. Another way to obtain this identity is to take derivative of $t_2$ on \eqref{softtht1t2} and then setting $t_2 = t_1$. From an explicit computation, both sides of the identity are evaluated to be\footnote{In the computation of correlators associated with $\dot{\pi}$, say $\vev{\pi_{\vec k_1}(t) \dot{\pi}_{\vec k_2}(t) \pi_{\vec q}(t_i) }$, one has to carefully use the interaction picture field $\dot{\pi}_I$ since $\dot{\pi} = {\cal U}^{\dagger} (\dot{\pi}_I - i [\dot{\pi}_I,\, H_I] ) {\cal U}   $.}
\be
 -\frac{c_s(1-c_s^2)}{4P'(\mu^2)} \frac{k_1^j}{k_1}\, .
\ee
Notice that this is the same as what one would get by taking  $\partial_{t_2} $ of either the right or left-hand side of the unequal-time identity involving only $\pi$ and then setting
$t_2 =t_1$.
Therefore, this is a non-trivial identity where all the late time
operators are at equal time. In other words, like the all un-equal
time $\pi$ correlators, correlators involving $\dot{\pi}$ (or more
formally, the canonical momentum conjugate to $\pi$) also have
information about the relation between the coefficient of $c_{31}
\dot{\pi} (\partial_i \pi)^2 $ and the quadratic action.
However, it should be stressed that the soft mode still has to be at the
initial time, different from the time of the hard modes, because
the physical mode condition is not satisfied for boost in superfluid.

\subsection{Effective field theory of inflation}
\label{sec:EFTofI}

As a final example, we consider symmetries in the effective field theory of inflation~\cite{Creminelli:2006xe,Cheung:2007st}. This effective field theory describes adiabatic perturbations around a homogeneous and isotropic background driven by a single degree of freedom acting like a clock.\footnote{As such, its domain of applicability is somewhat broader than just inflation, applying to any single-clock FLRW cosmology~\cite{Creminelli:2006xe,Cheung:2007st,Creminelli:2016zwa,Finelli:2018upr}.}

To  leading order in the derivative expansion, in unitary gauge, the EFT takes the form
\begin{equation}
S = \int\D t \, \D^3  x \, \sqrt{-g} \left[ \frac{\Mpl^2}{2}R - \Mpl^2 \left(3H^2 +2 \dot{H} \right) 
+ \Mpl^2\dot{H}\delta g^{00} 
+ \sum_{n\geq2}\frac{M_n^4}{n!}(\delta g^{00})^n
+ \cdots \right] \, ,
\label{EFTofI}
\end{equation}
where $\delta g_{\mu\nu}\equiv g_{\mu\nu}-\bar{g}_{\mu\nu}$, with $\bar{g}_{\mu\nu}=\text{diag}(-1,a^2(t),a^2(t),a^2(t))$, denotes the perturbation of the metric, $H\equiv {\dot a}/{a}$ is the Hubble function, and $M_n$ are generic time-dependent couplings.
Under the assumption that these do not vary significantly in one Hubble time, the action inherits an approximate time-translational invariance. After reintroducing the Goldstone via the St\"uckelberg trick
i.e.,~$t\mapsto t+\pi$, we can take the decoupling limit where $\dot H\to 0$ and $M_{\rm Pl}\to \infty$, with their product held fixed. In this limit, we isolate the dynamics of the Goldstone mode $\pi$ in an exact de Sitter background. The action for $\pi$ up to the fourth order in the field is~\cite{Cheung:2007st}
\begin{equation}
S_\pi  =  \int \D t \,  \D^3x~ a^3\left( \lagr^{(2)}+ \lagr^{(3)} +  \lagr^{(4)} \right) \, ,
\label{eftaction4}
\end{equation}
where
\be
\begin{aligned}
\label{eftlagr234}
\lagr^{(2)}&= -\frac{\Mpl^2 \dot{H}}{c_s^2} \left( \dot{\pi}^2-\frac{c_s^2}{a^2}(\partial_i \pi)^2 \right)\, ,\\
\lagr^{(3)}&=  \Mpl^2 \dot{H} \left(1-\frac{1}{c_s^2} \right) \left( \dot{\pi}^3-\dot{\pi} \frac{(\partial_i \pi)^2}{a^2} \right) - \frac{4}{3}M_3^4\dot{\pi}^3 \, ,\\
\lagr^{(4)}&=   \left( \frac{M_2^4}{2} - 2M_3^4 + \frac{2}{3}M_4^4 \right)\dot \pi^4 - \left( M_2^4 - 2 M_3^4\right)\dot \pi^2 \frac{(\prl_i \pi)^2}{a^2} + \frac{M_2^4}{2} \frac{((\prl_i \pi)^2)^2}{a^4} \, ,
\end{aligned}
\ee
with the sound speed defined by
\begin{equation}
c_s^{-2}  =  1- \frac{2 M_2^4}{\Mpl^2 \dot H}\, .
\end{equation}
The action \eqref{eftaction4} is invariant under the following set of global symmetries (in addition to the obvious spatial translation and rotation symmetries):
\begin{itemize}

\item Shifts of the Goldstone field,  $\delta \pi = 1$. We can think of this as an accidental symmetry that is a consequence of the near time-translation invariance of the background, which becomes exact in the decoupling limit.

\item Dilations, acting on $\pi$ as
\begin{equation}
\delta_D \pi = -\frac{1}{H}(1+\dot \pi) +  \bfx \cdot  \bfnabla \pi 	\, ;
\label{dilpi}
\end{equation}

\item Special conformal transformations (SCTs), which act as
\beq
\delta_{\text{SCT}} \pi =   - \frac{2}{H}\vec{b}\cdot \vec{x} +b^j\left(  - \frac{2}{H}x_j \dot \pi + 2 x_j  \bfx \cdot  \bfnabla \pi - x^2 \prl_j \pi +\frac{1}{a^2 H^2} \prl_j \pi  \right) \, ,
\label{SCTsymm}
\eeq
parametrized by the infinitesimal vector $\vec{b}$.

\end{itemize}
Note that much like in the superfluid case, a linear combination of the shift and dilation symmetries is realized linearly, causing the correlators in this theory to be scale invariant.\footnote{There is, therefore, a precise sense in which the EFT of inflation in the decoupling limit is like a superfluid in de Sitter space.}

The soft theorems  resulting from the nonlinearly realized symmetries~\eqref{dilpi} and~\eqref{SCTsymm} for the late-time cosmological correlators are well known, and have been studied in many contexts~\cite{Maldacena:2002vr,Creminelli:2004yq,Creminelli:2012ed,Kehagias:2012pd, Assassi:2012zq,Hinterbichler:2012nm, Goldberger:2013rsa,Pimentel:2013gza}. Here we wish to focus on two (related) facets. The first is more conceptual: given recent progress in the abstract bootstrap construction of inflationary correlators, it is natural to ask how the nonlinearly realized symmetries further constrain these objects. Related to this, the second thing we wish to elucidate is a small puzzle involving the three-point function in the EFT of inflation. Much like in the superfluid case, the known equal-time soft theorems are satisfied by $\pi$ correlators in a trivial way: both sides of the relation happen to vanish. This is somewhat strange because the coefficient of the $\dot\pi(\partial\pi)^2$ interaction is completely fixed by symmetry. It is then natural to ask: how does one recover this fact at the level of observables? We will see that---like the superfluid---one must consider unequal-time soft theorems in order to see this relation at the level of correlators.

In this Section, we derive and check the unequal-time soft theorem that relates the three and two-point functions in the EFT of inflation. (We additionally discuss equal-time soft theorems for $\pi$ in Appendix~\ref{EFToIapp}.)

\noindent
{\bf Unequal-time soft theorem for SCTs:}
We want to understand the consequences of the nonlinearly realized SCT (boost) symmetry of the EFT of inflation Goldstone mode~\eqref{SCTsymm}. As a first step, we consider the relation between the three and two-point functions implied by this symmetry. It turns out that the soft theorems where the hard modes are evaluated at the same time are all trivially satisfied (in the sense that both sides are zero). This includes the case where all of the operators (including the soft mode) are at the same time. To obtain a nontrivial identity, we, therefore, concentrate on 
correlators involving two Goldstone fields $\pi_f$ taken at different final times $\t_1$ and $\t_2$. As in the superfluid case, removing the momentum-conserving delta function is somewhat subtle, but after doing this, the resulting Ward identity takes the form:
\begin{tcolorbox}[colframe=white,arc=0pt,colback=greyish2]
\vspace{-.25cm}
\begin{align}
\label{3to2EFToI}
\lim\limits_{\v q \rightarrow 0}& \frac{1}{H} \frac{\prl}{\prl q_j} \left[ {\cal E}_i(q) \left( \braket{\pi_{\v k_1}(\t_1)\pi_{\v k_2}(\t_2) \pi_{\v q}(\t_i)}' +  \braket{\pi_{\v q}(\t_i) \pi_{\v k_1}(\t_1)\pi_{\v k_2}(\t_2)}'\right) \right] \\ \nonumber
 &\hspace{3cm} =\frac{k_1^j}{k_1} \left(- \t_1 \prl_{k_1}\prl_{\t_1}  + \frac{1}{2}\left( 4\prl_{k_1} + k_1 \prl_{k_1}^2  - k_1(\t_1^2 - \t_2^2) \right) \right) \braket{\pi_{\v k_1}(\t_1)\pi_{-\v k_1}(\t_2)}' \, .
\end{align}
\end{tcolorbox}
\noindent
Note that the three-point function involves a soft mode evaluated in the initial state, along with two hard modes evaluated at (late) times $\t_1$ and $\t_2$.

We can verify that the soft theorem~\eqref{3to2EFToI} is satisfied in
the EFT of inflation using~\eqref{eftlagr234}. We find that the unequal-time two-point function is given by
\be
\braket{\pi_{\v k_1}(\t_1)\pi_{-\v k_1}(\t_2)}' = \frac{H^2}{4 \Mpl^2 |\dot H| c_s k_1^3}(-i + c_s \t_1 k_1) (i + c_s \t_2 k_1)e^{-i c_s k_1(\t_1 - \t_2)}\, ,
\ee
while the soft limit of the three-point function is given by
\begin{align}
&\lim\limits_{\v q \rightarrow 0}  \frac{\prl}{\prl q_j} \left[  {\cal E}_i(q) \left( \braket{\pi_{\v k_1}(\t_1)\pi_{\v k_2}(\t_2) \pi_{\v q}(\t_i)}' +  \braket{\pi_{\v q}(\t_i) \pi_{\v k_1}(\t_1)\pi_{\v k_2}(\t_2)}'\right) \right] \\\nonumber
&\hspace{7cm}
= \frac{k_1^j}{2}(1-c_s^2) (\t_2^2-\t_1^2 )\braket{\pi_{\v k_1}(\t_1)\pi_{\v k_2}(\t_2)}' \,.
\end{align}
Putting these together, it is straightforward to check
that~\eqref{3to2EFToI} is satisfied. Note also that both sides are
nonzero, so there is a nontrivial relationship between the soft limit
of the three-point function and the unequal-time two-point
function. Much as in the superfluid case, this relation can also be
seen in correlators involving $\dot\pi$. And different from the
superfluid case, the initial time of the soft mode can be promoted to
late times, because for $\pi$ in the EFT of inflation, the physical mode
condition is satisfied for SCT (as well as dilation).

It is somewhat interesting that we have to consider correlators beyond just equal time correlators of $\pi$ in order to
recover the relation between cubic and quadratic terms in the action that is a consequence of the nonlinearly realized SCT symmetry.\footnote{One way to think about why this is necessary is that the normalization of the three-point function is degenerate with rescaling the sound speed of fluctuations. The unequal time correlator is sensitive to the speed of propagation of fluctuations, so it provides a way of breaking this degeneracy and relating the normalization of the three and two-point functions.} Much as in the superfluid case, it is only the coefficient of the $\dot\pi(\partial\pi)^2$ term that is constrained by the soft theorem; the correlator coming from the $\dot\pi^3$ interaction vanishes in the soft limit so its coefficient is unconstrained. In particular, the action up to cubic order in conformal time is schematically
\beq
S = -\frac{M_{\rm Pl}^2 \dot H}{c_s^2} \int \D \t \D^3 x\, \( \frac{1}{H^2 \t^2} \( \pi'^2 - c_s^2 (\prl_i \pi)^2 \) + \frac{1}{H \t} (c_{33}  \pi'^3 + c_{31}  \pi ' (\prl_i \pi)^2 ) \) \, .
\eeq
It is only the coefficient $c_{31}$ in the cubic action that is fixed entirely in terms of the speed of sound
\beq
c_{31} = 1 - c_s^2 \, ,
\eeq
and this is precisely the physical content captured by the relation
(\ref{3to2EFToI}).
It is also worth pointing out that the physical mode condition is satisfied 
for the EFT of inflation. Thus, in Eq. (\ref{3to2EFToI}), 
the initial time $\tau_i$ for the soft mode 
can be promoted to match $\tau_1$ or $\tau_2$ for the hard modes,
unlike its counterpart in the superfluid example.

From the perspective of bootstrapping inflationary correlators, it is
somewhat interesting that the relation between cubic couplings and the
sound speed seems to be invisible at the level of equal-time $\pi$
correlation functions. At higher points, it does seem possible to
reconstruct the full set of relations following from the nonlinearly
realized symmetries by looking at equal-time correlators, as we
discuss in Appendix~\ref{EFToIapp}. Typically in cosmology we are only
able to measure equal-time correlation functions, so it would be
somewhat more desirable to find a way of extracting this information
on-shell from these correlators.
Perhaps the most natural approach is to consider bootstrapping correlators of the momentum conjugate to $\pi$, which are sensitive to phases of the cosmological wavefunction. Alternatively, perhaps these symmetry considerations should be taken as an indication that we should consider bootstrapping unequal time correlators. In any case, it would be very interesting to further systematize the application of symmetry constraints to the EFT of inflation at the level of observables.

\newpage
\section{Conclusions}
\label{sec:conclusions}

A wide variety of systems---all the way from condensed matter to cosmology---spontaneously break spacetime symmetries in their ground state. These spacetime symmetries often involve transforming the time coordinate. The corresponding nonlinear realization of these ``time symmetries" is somewhat subtle because of the way that they transform the foliation of spacetime used to quantize the system. Nevertheless, many important examples of symmetries, for instance, Lorentz boosts, fall into this class, and so it is necessary to understand the consequences of these symmetries for observables.

In this paper, we have derived a soft theorem satisfied by unequal-time correlation functions~\eqref{softthd1} in theories with time symmetries. The primary technical difference with previous studies is carefully tracking the effects of (temporal) boundary terms on the charges that generate the spacetime symmetries of interest. In many cases, the unequal-time relation~\eqref{softthd1} can be promoted to an equal-time soft theorem, provided that the system satisfies the so-called physical mode condition---that the nonlinear part of the symmetry transformation has the same time dependence as the soft limit of the growing-mode solution to the equations of motion.

We have applied this technology to several examples in this paper. In addition to two instructive examples provided by time-dependent field theories designed to mimic features of inflation, we have considered the EFT of a relativistic superfluid and the EFT of inflation in the decoupling limit. An interesting feature of the superfluid and EFT of inflation cases is that the previously known soft theorems were unable to reproduce the relation between cubic couplings and the speed of sound of perturbations that is clear from the action perspective. What we find is that these relations can be reproduced by considering a particular unequal-time three-point function in the soft limit. Or, alternatively a three-point function involving momentum.

One lesson that we can abstract from these studies is that it seems that for time symmetries, unequal-time correlation functions (or ones involving canonical momenta) are somewhat more natural observables than equal-time ones. Given recent progress in the bootstrap construction of equal-time correlators, it is natural to consider bootstrapping these objects. An additional intriguing feature of unequal-time objects is that they may encode more directly consequences of causality, the imprints of which are somewhat mysterious at the level of equal-time correlators.

We expect that the tools developed in this paper will serve useful to study the consequences of symmetries for observable quantities of phases of matter and in the cosmological setting.

\vspace{-10pt}
\paragraph{Acknowledgements.} We would like to thank James Bonifacio,  Paolo Creminelli, Daniel Green, Hayden Lee,  Alberto Nicolis,  Chia-Hsien Shen, and Marko Simonovi\'c
for useful discussions. We thank in particular Enrico Pajer for useful comments on the draft. LH, IK, and KP are supported in part by the DOE
DE-SC011941. LH acknowledges support by a Simons Fellowship in
Theoretical Physics. The work of AJ is supported in part by DOE (HEP) Award DE-SC0009924. The work of SW is supported in part by the DOE (HEP) Award DE-SC0013528.

\newpage
\appendix

\section{Path integral derivation of the soft theorem}
\label{appendix:A0}

It is interesting to compare the derivation of the soft theorem of Section~\ref{sec:derivation}---which is based on the operator formalism in canonical quantization~\cite{Hinterbichler:2013dpa,Kehagias:2012pd, Finelli:2017fml}---with the derivation from the path integral~\cite{Hui:2018cag,Goldberger:2013rsa, Berezhiani:2014tda, Berezhiani:2013ewa, Avery:2015rga}. In this Appendix, we derive the same identity using the path integral. There are two related ways to think about the path integral derivation, depending on whether one treats the symmetry transformation in the active or passive picture.

We first review the basic elements of the wavefunctional/path integral definition of correlation functions. Following~\cite{Hui:2018cag}, let $\phi(\bfx, \tau_a)$ be the field
operator at position $\bfx$ and  time $\tau_a$, and let $\lvert\vp_a\rangle$ be field eigenstates satisfying $\phi(\bfx ,\tau_a)\lvert\vp_a\rangle = \vp_a(\bfx)\lvert\vp_a\rangle$.
The expectation value of a general operator $\mathcal{O}$,   built out of  the fields $\vp_f(\vec x) =\phi(\vec x,\tau_f) $   at some final time $\tau=\tau_f$ and generic spatial positions $\bfx_a$, can be computed according to the definition
\begin{equation}
\langle{\cal O}(\bfx_1,\ldots, \bfx_N)\rangle = \int{\cal D}\vp_f \,  {\cal O}(\bfx_1,\ldots, \bfx_N)\,\vert\Psi[\vp_f,\tau_f]\vert^2 \, ,
\label{eq:schrodingercorrs}
\end{equation}
where $\vert\Psi[\vp_f, \tau_f]\vert^2$ is the normalized probability distribution for spatial field profiles at time $\tau_f$. 
This wavefunctional $\Psi[\vp_f,\tau_f]$  can  be constructed via a path integral from some initial vacuum state, $\vert 0_\text{in}\rangle$, by inserting a complete set of field eigenstates
\be
\Psi[\vp_f,\tau_f] =
\int{\cal D}\vp_i\, \langle\vp_f, \tau_f\rvert\vp_i,\tau_i\rangle\langle\vp_i,\tau_i\rvert 0_{\rm in}\rangle = 
\int {\cal D}\vp_i
\raisebox{.09cm}{$\underset{\substack{{\scriptscriptstyle \varphi(\tau_f)=\varphi_f} \\ {\scriptscriptstyle\varphi(\tau_i) =\varphi_i}}}{\displaystyle\int}$}   \hspace{-.3cm}
{\cal D}\varphi
 \, \Exp^{i \int_{\tau_i}^{\tau_f}\D \tau \int \D^3x \, \mathcal{L}[\varphi(\tau)]}\Psi_{0}[\vp_i, \tau_i] \, ,
\label{eq:wavefunctional}
\ee
where we have written the transition  amplitude $\langle\vp_f,
\tau_f\rvert\vp_i,\tau_i\rangle$ as\footnote{It is worth emphasizing that in the
  expression \eqref{eq:wavefunctionalta}, as well as in other
  analogous equations, $\varphi$ does not denote a field operator but is merely an integration variable. }
\be
\langle\vp_f, \tau_f\rvert\vp_i,\tau_i\rangle =
\raisebox{.09cm}{$\underset{\substack{{\scriptscriptstyle \varphi(\tau_f)=\varphi_f} \\ {\scriptscriptstyle\varphi(\tau_i) =\varphi_i}}}{\displaystyle\int}$}   \hspace{-.3cm}
{\cal D}\varphi
 \, \Exp^{i \int_{\tau_i}^{\tau_f}\D \tau \int \D^3x \, \mathcal{L}[\varphi(\tau)]} \, .
\label{eq:wavefunctionalta}
\ee
The path integral in~\eqref{eq:wavefunctional} sums over all possible field configurations for $\phi$ that start from the vacuum and which satisfy the late-time boundary condition that  $\phi(\bfx, \tau_f) = \vp_f(\bfx)$.
We have  introduced the vacuum wavefunctional  $\Psi_0[\vp_i,\tau_i] \equiv \langle\vp_i,\tau_i\rvert 0_{\rm in}\rangle $, which captures the initial conditions for the path integral. In the limit $\tau_i\rightarrow-\infty$, we assume that it takes the Gaussian form
\be
\Psi_0[\vp_i,\tau_i]  \propto
\exp\left(-\frac{1}{2}\int\frac{\rd^3\bfk}{(2\pi)^3} \, 
{\cal E}_i (k) \varphi_i(\bfk)
\varphi_i(-\bfk)\right) \, .
\label{eq:gaussianwavef}
\ee
With these preliminaries out of the way, we would like to understand
the consequences of symmetries for the correlation function~\eqref{eq:schrodingercorrs} in both the active and passive formalisms.

\subsection{Active picture}
\label{appendix:A}

We first consider the active picture.
In the path integral derivation of Ward identities, it is important that the action of a symmetry transformation on the Heisenberg eigenstate can be expressed as
\be
\Exp^{-i Q} |\varphi\rangle = |\varphi + \delta\varphi\rangle \, .
\ee
Time symmetries (in the active sense) generally cannot be expressed
this way (i.e., typically the symmetries do not take an eigenstate of the field
operator into another eigenstate of the field operator), thus necessitating the operator approach taken in the main text. Nonetheless, in cases where the above condition holds, we can meaningfully compare the two derivations.

We start by considering~\eqref{eq:schrodingercorrs} with ${\cal O}(\bfx_1,\cdots, \bfx_N) = \phi (\v k_1 , \t_f) \cdots \phi (\v k_N , \t_f)$.
We can change variables on the right-hand side to relabel $\varphi_f$ to $\varphi_f + \delta\varphi_f$. Assuming the integral measure is invariant under this change, we conclude that to the first order in the variation,
\be
\begin{aligned}
\label{WardCb}
-\int {\cal D}\varphi_f \, \varphi_f ({\vec k_1}) \cdots \varphi_f ({\vec k_N}) &\Big(\Psi^* \lb \varphi_f, \t_f \rb \delta \Psi \lb \varphi_f, \t_f \rb + \Psi \lb \varphi_f, \t_f \rb \delta \Psi^* \lb \varphi_f, \t_f \rb\Big)
\\
&\hspace{2cm}
= \int {\cal D}\varphi_f \, \delta [ \varphi_f ({\vec k_1}) \cdots \varphi_f ({\vec k_N}) ] |\Psi \lb \varphi_f, \t_f \rb|^2 \, .
\end{aligned}
\ee
Recalling that 
\be
\delta \Psi \lb \varphi_f, \t_f \rb = \langle \varphi_f + \delta\varphi_f | 0_{\rm in} \rangle
- \langle \varphi_f | 0_{\rm in} \rangle = i \langle \varphi_f | Q | 0_{\rm in} \rangle \, ,
\ee
we recognize that~\eqref{WardCb} is equivalent to~\eqref{WardC}. 
This makes it clear neither statement is a statement of symmetry per se. In
the path integral derivation, the statement follows from a change of
the dummy variable of integration; in the operator derivation, it
follows from the existence of an operator that can effect a certain
transformation on the field.\footnote{That such an operator is topological (conserved) is, however, a non-trivial physical assumption.}

We next rewrite the left-hand side of~\eqref{WardCb} using
\be
\Psi \lb \varphi_f, \t_f \rb = \int {\cal D} \varphi_i \, \langle \varphi_f | \varphi_i \rangle \Psi_0 \lb \varphi_i, \t_i \rb \, ,
\ee
where $\Psi_0 \lb \varphi_i, \t_i \rb \equiv \langle \varphi_i | 0_{\rm in} \rangle$ is the initial wavefunctional, and
\be
\Psi\lb \varphi_f + \d\varphi_f, \t_f \rb = \int {\cal D} \varphi_i \, \langle \varphi_f + \delta\varphi_f | \varphi_i \rangle \Psi_0 \lb \varphi_i, \t_i \rb \, .
\ee
The trick is to again relabel the variable of integration $\varphi_i$ by $\varphi_i + \delta\varphi_i$,
assume the measure is invariant, and crucially, that the transition amplitude satisfies
\be
\label{symmphys}
\langle \varphi_f + \delta\varphi_f | \varphi_i + \delta\varphi_i\rangle = \langle \varphi_f | \varphi_i \rangle \, .
\ee
It then follows that
\be
\delta \Psi \lb \varphi_f, \t_f \rb = \int {\cal D} \varphi_i \, \langle \varphi_f | \varphi_i \rangle \delta \Psi_0 \lb \varphi_i, \t_i \rb \, ,
\ee
and the soft theorem can be derived by rewriting the left-hand side of~\eqref{WardCb},
noting how the initial wavefunctional transforms because it is a Gaussian.

Equation~\eqref{symmphys} is where the assumption of symmetry is used. 
The analog of this in the operator approach is when we go from considering how the charge operator acts on the final fields, to how it acts on the in-vacuum, assuming the charge is time-independent.

\subsection{Passive picture}
\label{passive}

We now want to consider the path integral derivation in the passive picture. It is a bit more general than the active picture derivation since it can deal with some time symmetries---those where time is transformed by a space-independent amount (for example time translations). The operator approach is still more general since it even applies to cases where the time variable shifts by a space-dependent function, such as it does for boosts.
Nonetheless, the passive picture path integral derivation provides a useful check of the results.
To be precise, it is useful to work in something of a mixed picture, where time symmetries are treated passively, while other symmetries (including spatial symmetries) are treated in the active picture.

Let us assume  that the theory has some time-dependent symmetry, acting nonlinearly on $\phi$, that can be written as
\begin{equation}
\phi (\tau) \mapsto \tilde{\phi}(\tilde{\tau}) =  \phi (\tau) + a(\phi,\tau) \, , 
\qquad
\tau \mapsto\tilde{\tau} = \tau +b (\tau) \, .
\label{symm}
\end{equation}
In the following, we will consider infinitesimal transformations of
the type \eqref{symm}, with $a(\tau,\varphi)$ and $b(\tau)$  arbitrary
functions of their arguments. In addition, $a$ can also be a function
of space, but $b$ is assumed to be space-independent---this last
assumption is crucial for the next step of the derivation to make
sense.\footnote{This is also why we utilized the operator formalism in the main text, to cover cases like boosts where
time is transformed by a space-dependent amount.}
Note that in this formulation of the symmetry, the time
coordinate is allowed to transform (and in this sense
the formulation is passive), but the spatial
coordinate, which is kept implicit, is not transformed 
(and in this sense the formulation is active).
In other words, let us express the above in its completely active
form:
\be
\delta\varphi = \varphi + a - b \partial_\tau \varphi
\ee
We can say $a = \delta_{\rm NL} \varphi + \tilde \delta_{\rm L}\varphi$
and $\delta_{\rm L} \varphi = \tilde \delta_{\rm L}\varphi -
b \partial_\tau \varphi$. That is to say, the full nonlinear part of
the field transformation is contained in $a$, while the full linear
part of the transformation is partly in $-b\partial_\tau \varphi$
associated with the time transformation, and partly in $a$
(the part we call $\tilde \delta_{\rm L}\varphi$, which is the linear
transformation that could come from a spatial coordinate
transformation).
We are interested in computing 
\be
\langle \vp_f + a_f, \tau_f  + b_f \rvert\vp_i+a_i,\tau_i+b_i\rangle \equiv
\langle \tilde{\vp}_f, \tilde{\tau}_f\rvert\tilde{\vp}_i,\tilde{\tau}_i\rangle =
\raisebox{.09cm}{$\underset{\substack{{\scriptscriptstyle \tilde \varphi(\tilde \tau_f)=\tilde\varphi_f} \\ {\scriptscriptstyle \tilde \varphi(\tilde \tau_i) =\tilde \varphi_i}}}{\displaystyle\int}$}   \hspace{-.3cm}
{\cal D}\tilde \varphi
 \, \Exp^{i \int_{\tilde \tau_i}^{\tilde \tau_f}\D \tau \int \D^3x \, \mathcal{L}[\tilde \varphi( \tau)]} \, .
\label{eq:wavefunctionaltashifted}
\ee
First, we assume that ${\cal D}\tilde \varphi = {\cal D}
\varphi$. Then, by the assumption that \eqref{symm} is a symmetry of the action,
\be
\int_{\tilde \tau_i}^{\tilde \tau_f}\D \tilde\tau \int \D^3x\,
\mathcal{L}[\tilde \varphi( \tilde \tau)]
= \int_{ \tau_i}^{ \tau_f}\D \tau \int \D^3x \, \mathcal{L}[\varphi(\tau)]  + \delta\theta[\varphi_f,\varphi_i; \tau_f,\tau_i] \, ,
\label{ThetaDef}
\ee
where $\delta \theta[\varphi_f,\varphi_i; \tau_f,\tau_i]\equiv \theta[\varphi_f;\tau_f]-\theta[\varphi_i;\tau_i]$ captures possible boundary terms.
Thus, 
\be
\langle \tilde{\vp}_f, \tilde{\tau}_f\rvert\tilde{\vp}_i,\tilde{\tau}_i\rangle =
\langle \vp_f, \tau_f\rvert   \vp_i, \tau_i\rangle \Exp^{i \delta \theta} \, ,
\label{eq:wavefunctionaltashifted-2}
\ee
where one should keep in mind that $\delta \theta$ is of the same order as the functions $a$ and $b$ that define the infinitesimal transformation \eqref{symm}.
From~\eqref{eq:wavefunctional}, we can write
\begin{equation}
\begin{split}
\Psi[\vp_f+a_f,\tau_f+b_f]
& =
\int{\cal D}(\vp_i+a_i)\, \langle \vp_f+a_f, \tau_f+b_f \rvert \vp_i+a_i,\tau_i+b_i \rangle \Psi_0[\vp_i+a_i,\tau_i+b_i] 
\\
& = 
\int{\cal D}\vp_i\, \langle \vp_f, \tau_f\rvert \vp_i,\tau_i \rangle \Exp^{i \delta \theta} \Psi_0[\vp_i+a_i,\tau_i+b_i] \, .
\end{split}
\label{eq:wavefunctionaltrans}
\end{equation}
Subtracting \eqref{eq:wavefunctional} from \eqref{eq:wavefunctionaltrans}, and keeping only  terms up to linear order in the symmetry transformation,
\be
\begin{aligned}
b_f \partial_{\tau_f} &\Psi[\vp_f,\tau_f] + \Psi[\vp_f+a_f,\tau_f] - \Psi[\vp_f,\tau_f] 
\\
&= \int{\cal D}\vp_i\, \langle \vp_f, \tau_f\rvert \vp_i,\tau_i \rangle
\Big[
b_i \partial_{\tau_i} \Psi_0[\vp_i,\tau_i] 
+ \Psi_0[\vp_i+a_i,\tau_i] - \Psi_0[\vp_i,\tau_i]
+ i  \delta \theta \Psi_0[\vp_i,\tau_i]
\Big] \, .
\label{eq:wavefunctionaldiff}
\end{aligned}
\ee
It is then straightforward to use~\eqref{eq:wavefunctionaldiff} along with the quantum mechanics formula~\eqref{eq:schrodingercorrs} expanded to linear order in the field variation to deduce the soft theorem.
It can also be shown that 
$\theta = \int d^3 {\vec x} \, (K^0 + b {\cal L})$ at both the initial
and final times,
where $K^\mu$ is defined by 
$\delta{\cal L} = \partial_\mu K^\mu$ in the active picture.
It can further be shown that $\theta$ can depend on $\varphi$
but not $\partial_\tau \varphi$ (and likewise for the linear
contribution to $K^0$). 
%

\section{Transformation of the one-point function}
\label{1ptconstr}

We now wish to discuss the special case where the operator
${\cal O}$ in the soft theorem is just a single field $\phi$.
In this case, the corresponding momentum
must be soft rather than hard in order to arrive at a non-trivial symmetry
statement. The right-hand side of the soft theorem will involve the nonlinear transformation of the fields involved.
In other words, we have
\begin{equation}
\label{softone}
\bal
\lim_{\vec q \rightarrow 0} D_{\v q}(
\t_i) \left(
{\cal E}_i (q) \langle 0_{\rm in} |  \phi_f (-\v q) \phi_i (\v q) |0_{\rm in}  \rangle'  +  {\rm c.c.}
\right) &
\\   - i \alpha_0(\v q , \tau_i) &
\left(\langle 0_{\rm in} |  \phi_f (-\v q) \phi_i (\v q) |0_{\rm in}  \rangle' -  {\rm c.c.}
\right) 
= \lim_{\vec q \rightarrow 0}  D_{\v q} (\tau_f) \, .
\eal
\end{equation}
We confine ourselves to cases where $D_{\v q}$ does not involve
derivatives with respect to $\v q$.
The above can then be rewritten as
\beq
\bal
 \lim_{\vec q \rightarrow 0} D_{\v q} & (
\t_i) \left( {1 \over P_i (q)}  {\,\rm Re\,} \langle 0_{\rm
  in} |  \phi_f (-\v q) \phi_i (\v q) |0_{\rm in}  \rangle' \right)  \\
& \qquad\quad + 2  \left( \alpha_0(\v q, \tau_i) - D_{\v q}(\tau_i) {\,\rm Im\,}{\cal E}_i
  (q)\right) {\,\rm Im\,}  \langle 0_{\rm
  in} |  \phi_f (-\v q) \phi_i (\v q) |0_{\rm in}  \rangle' = \lim_{\vec q \rightarrow 0} D_{\v q}
   (\tau_f) \, .
   \eal
\eeq
{\it If} the physical mode condition is satisfied, such that the first
term on the left equals the term on the right, up to corrections
suppressed by powers of $q$, {\it then} we have:
\be
\lim_{\vec q \rightarrow 0} \left( \alpha_0(\v q, \tau_i) - D_{\v q}(\tau_i) {\,\rm Im\,}{\cal E}_i
  (q)\right) {\,\rm Im\,}  \langle 0_{\rm
  in} |  \phi_f (-\v q) \phi_i (\v q) |0_{\rm in}  \rangle' = 0 \, .
\ee
It is interesting that in the cases we have performed perturbative checks
(and where the physical mode condition holds and $D_{\v q}$ does not contain derivatives with respect to $\v q$), the way that this identity is satisfied is that 
$\alpha_0(\v q, \tau_i) = D_{\v q}(\tau_i) {\,\rm Im\,}{\cal E}_i(q)$
as $\v q \to 0$.\footnote{This holds, for example, in the slow-roll and ultra-slow-roll inflation toy models (Sections \ref{sec:Slow-Roll} and \ref{sec:Ultra-Slow-Roll}), and for the shift symmetry identity for a driven superfluid (Appendix \ref{sec:drivensuperfluid}).}

The observation that this relation between the imaginary part of the wavefunction kernel and the boundary-term contribution to the Noether charge is valid non-perturbatively in $\v q$ is perhaps not that surprising. The quantity $\alpha_0(\v q, \tau_i)$ is related to the temporal boundary term that the action shifts by, $K^0$.
On the other hand, the initial wavefunctional kernel ${\cal E}_i (q)$ is equal, up to a coefficient, to the classical on-shell action at the initial time boundary $\t = \t_i$, and so, is also essentially a time boundary term.

\section{More on superfluids}
\label{sec:superfluid}

Here we elaborate on the details of the EFT of superfluids that we considered as an example in the main text.
A superfluid is a system with an internal $U(1)$ symmetry that is spontaneously broken in a state of finite density for the associated conserved $U(1)$ charge~\cite{Son:2002zn,Nicolis:2011pv,Nicolis:2013lma}.
The symmetry breaking pattern is conveniently parametrized in terms of a scalar field $\phi(\bfx,t)$ with a nonzero time-dependent expectation value, $\bar \phi = \mu t$, where $\mu$ is the chemical potential. The fluctuations around this background describe phonon excitations in the superfluid, $\phi(\bfx,t)= \mu t +\pi(\bfx,t)$.
The field fluctuation  $\pi(\bfx, t)$ nonlinearly  realizes not only the internal $U(1)$ symmetry, which acts on $\pi$ as a constant shift ($\pi\rightarrow \pi+\e$), but also time translations and Lorentz boosts, which are spontaneously broken by $\bar \phi$. There is a diagonal contribution of shifts and time translations that remains linearly realized~\cite{Nicolis:2011pv}.

At leading order in derivatives, the low-energy effective action is~\cite{Son:2002zn}
\begin{equation}
S = \int \D t \, \D^3 x \, P(X) \, ,
\label{Lsuperfluidapp}
\end{equation}
where $X\equiv -\partial_\mu\phi\partial^\mu\phi$. Expanding this action around the superfluid ground state, we find the following interactions for the phonon $\pi$, as an expansion in powers of the field:
\begin{subequations}
\label{superfluidL}
\begin{align} 
{\cal L}^{(2)} & = \frac{P'(\mu^2)}{c_s^2} \left[ \dot{\pi}^2 -  c_s^2(\partial_i\pi)^2 \right] \, ,
\\
\label{superfluidL3}
{\cal L}^{(3)} & = \frac{(1-c_s^2)P'(\mu^2)}{c_s^2\mu }
 \left[ \left(1+ \frac{2 c_3}{3c_s^2}\right)  \dot \pi^3
-
 \dot \pi(\partial_i\pi)^2 \right] \, ,
\\
{\cal L}^{(4)} & = \frac{(1-c_s^2)P'(\mu^2)}{4c_s^2\mu^2 }
\left[ \left(1+ \frac{4 c_3}{c_s^2}+\frac{2 c_4}{c_s^2}\right) 
\dot \pi^4
-2 \left(1+\frac{2 c_3}{c_s^2}\right)
\dot \pi^2(\partial_i\pi)^2
+ (\partial_i\pi)^4 \right]
 \, ,
\end{align}
\end{subequations}
where the speed of sound (squared) is
\begin{equation}
c_s^2 = \frac{P'(\mu^2)}{P'(\mu^2) + 2\mu^2 P''(\mu^2)} \, ,
\label{soundspeedsquared}
\end{equation}
and where we defined  the constants
\begin{equation}
c_3 \equiv \frac{2 c_s^4 \mu^4 P'''(\mu ^2)}{\left(1-c_s^2\right) P'(\mu ^2)} \, ,
\qquad\qquad
c_4 \equiv \frac{4 c_s^4 \mu^6 P^{(4)}(\mu ^2)}{3 \left(1-c_s^2\right) P'(\mu ^2)} \, ,
\end{equation}
in terms of the function $P(X)$ that defines the equation-of-state of the superfluid.
The fact that the sound speed \eqref{soundspeedsquared} is in general $\neq 1$ is a consequence of the spontaneous breaking of  Lorentz symmetries. The effective couplings in the lagrangian~\eqref{superfluidL} are all constant as a result of the presence of an unbroken combination of time translations and constant shifts on $\pi$~\cite{Nicolis:2011pv,Finelli:2018upr}.

\subsection{Symmetries}

The time-dependent expectation value for $\phi$ is responsible for the spontaneous breaking of time translations and Lorentz boosts that are symmetries of the superfluid action~\eqref{Lsuperfluidapp}. Here we briefly review how these symmetries are nonlinearly realized on $\pi$.
The field transformations \eqref{T12} below are the main ingredients for the derivation of the soft theorems presented in Section~\ref{sec:superfluidmain}.

Spacetime translations and Lorentz transformations are generated by the Killing vectors
\begin{equation}
P_\mu = -i \partial_\mu \, ,
\qquad\qquad
J_{\mu\nu} = i\left(x_\mu \partial_\nu - x_\nu \partial_\mu \right) \,.
\end{equation}
The action of these Poincar\'e transformations on a scalar is by the Lie derivative
\begin{subequations}
\label{T12}
\begin{align}
\delta_{P_\mu} \phi(x) & = -\partial_\mu \phi(x) \, ,
\label{t1}
\\
\delta_{J_{\mu\nu}}\phi (x) & = \left( x_\mu \partial_\nu - x_\nu \partial_\mu \right)\phi(x) \, .
\label{t2}
\end{align}
\end{subequations}
By expanding $\phi = \mu t+\pi$, we can obtain the action of Poincar\'e transformations on the superfluid phonon. Time translations and boosts are realized nonlinearly
\begin{subequations}
\begin{align}
\delta_{P_0} \pi(x) & = -\mu-\dot \pi(x) \, ,
\\
\delta_{J_{0i}}\phi (x) & =  -\mu x_i-\left( t\partial_i + x_i\partial_0 \right)\pi(x) \, ,
\end{align}
\end{subequations}
which is of course a consequence of the fact that we are expanding around a nontrivial background for $\phi$.

\paragraph{Interactions:}

The field's correlators can be computed for instance using the in-in formalism~\cite{Weinberg:2005vy}. In order to do this, we require the interaction-picture Hamiltonian. Using~\eqref{superfluidL}, we obtain
\begin{subequations}
\label{eq:sfluidH}
\begin{align}
\mathcal{H}_\text{free} &= 
	\frac{P'(\mu^2)}{c_s^2} \left[ \dot \pi_I^2 + c_s^2 (\partial_i\pi_I)^2 \right]
  \,  ,\\
\mathcal{H}^{(3)}_I &= \frac{(1-c_s^2)P'(\mu^2)}{c_s^2\mu }
 \left[ - \left(1+ \frac{2 c_3}{3c_s^2}\right)  \dot \pi_I^3
+
 \dot \pi_I(\partial_i\pi_I)^2 \right] \, ,\\
\mathcal{H}^{(4)}_I &= \frac{(1-c_s^2)P'(\mu^2)}{4c_s^2\mu^2 }
\Bigg[ \left( \frac{4 c_3^2 \left(c_s^2-1\right)}{c_s^4}+c_3 \left(12-\frac{8}{c_s^2}\right)+9 c_s^2+\frac{2 c_4}{c_s^2}-8 \right) 
\dot \pi^4_I \nonumber \\
&\hspace{2cm}+  \left( 6- \frac{4}{c_s^2}+\frac{4 c_3}{c_s^2} \right)
\dot \pi^2_I(\partial_i\pi_I)^2
- (\partial_i\pi_I)^4 \Bigg]  \, .
\end{align}
\end{subequations}
Recall that, as a consequence of the derivative interactions, the relation between the interaction Hamiltonian and the interaction lagrangian is nontrivial.

Equal-time correlation functions can be computed from these interaction Hamiltonian expressions using the in-in master formula~\cite{Weinberg:2005vy}:
\begin{equation}
\langle \mathcal{O}(t)\rangle = \langle 0 \vert \bar{T} \Exp^{i\int_{-\infty-i\epsilon}^{t} H_I(t')\D t' } \mathcal{O}_I(t) T \Exp^{-i\int_{-\infty+i\epsilon}^{t} H_I(t')\D t' }   \vert 0\rangle
 \, ,
\label{res0}
\end{equation}
where everything is in the interaction picture and where we introduced the $i\epsilon$-prescription to turn off the interactions at $t=-\infty$. The generalization to unequal-time correlators is straightforward and just involves time-evolving each operator separately.

\subsection{Driven superfluid}
\label{sec:drivensuperfluid}

It is instructive to consider a deformation of the theory~\eqref{Lsuperfluid}, which has a slightly different action of the symmetries. We imagine adding a term linear in $\phi$ to the action:
\begin{equation}
S = \int \D t \, \D^3 x \, \Big( P(X) + \lambda \phi\Big) \, ,
\label{Lsuperfluiddriven}
\end{equation}
which can be thought of as describing a superfluid coupled to an external constant source that provides a sort of  ``driving''~\cite{Pajer:2018egx,Finelli:2018upr}. The linear potential preserves the shift symmetry, but allows for more general, nonlinear time-dependent solutions for the background $\bar\phi$. The interesting features can be seen already in the case where $P(X) = \frac{1}{2}X$, so we will specialize to this case for simplicity. The background equation of motion now admits the quadratic solution $\bar\phi =\lambda t^2/2$. If we parametrize the field fluctuations as $\phi=\bar{\phi}+\delta\phi$, the theory for $\delta\phi$ clearly has no nontrivial interactions. Let's instead decompose $\pi$  as
\be
\phi = \frac{\l}{2} (t+ \pi)^2.
\ee
where $\pi$ plays the role of the Stueckelberg field for the spontaneously broken time translations.
The Goldstone action for $\pi$ is
\beq
S= -\frac{1}{2} \lambda^2  \int \D t \, \D^3  x~ (t+\pi)^2 (\partial_\mu\pi)^2 \, .
\label{gactionds}
\eeq
Though~\eqref{gactionds}  now appears to have interactions, the theory is of course still a (field-redefined) free theory. It is nevertheless instructive to check the soft theorems that result from the nonlinearly realized symmetries on $\pi$.
The action~\eqref{gactionds} has the following time-dependent symmetries, which are realized nonlinearly  on $\pi$:
\begin{itemize}
\item {\bf Shift symmetry:} $\delta \pi = 1/(t+\pi)$,

\item {\bf Time translations:} $\delta \pi = 1 + \dot \pi$,

\item {\bf Boosts:} $\delta_j \pi =  x_j +  (t \prl_j +x_j \prl_t) \pi$.

\end{itemize}
Note that this model is slightly different from a conventional superfluid in that there is no linearly realized combination of shifts and constant time translations.\footnote{There is however a linearly realized combination of shifts and time-dependent temporal diffeomorphisms~\cite{Finelli:2018upr}.} Consequently, the EFT couplings are time-dependent.
In the following, we will analyze each of the symmetries separately and discuss when they lead to equal-time soft theorems for the Goldstone's correlation functions.

\vspace{-10pt}
\paragraph{Shift symmetry.} We first consider the shift symmetry, $\phi\rightarrow\phi+\text{const}$, which acts on $\pi$ as
\begin{equation}
\delta \pi = \frac{1}{t+\pi}= \frac{1}{t}-\frac{\pi}{t^2}+\cdots.
\label{drivenshift}
\end{equation}
Under this transformation, the action~\eqref{gactionds} changes by a  total derivative:
\begin{equation}
\delta S = -\frac{1}{2} \lambda^2  \int \D t \, \D^3  x~ 2 \dot \pi \, .
\end{equation} 
The  Noether charge associated with the symmetry \eqref{drivenshift} is thus
\begin{equation}
Q = \lambda^2   \int \D^3 x ~ \left( t \dot{\pi } + \pi + \cdots \right) \, .
\end{equation}
Translating this into the notation of Section~\ref{sec:derivation}, we have
\begin{equation}
\alpha_0 = \lambda^2 \, , \qquad\qquad D_{\v q} = \frac{1}{t} \, .
\label{albet0drsupshift}
\end{equation}
The soft theorem~\eqref{softthd1} reads 
\beq
\label{softthdrivenunequalt}
\lim\limits_{\bfq \rightarrow 0} \left( \frac{1}{t_i} \mathcal{E}_i(q)  - i \lambda^2  \right) \braket{\pi_f(\bfk) \pi_f(-\bfk - \bfq) \pi_i(\bfq)}' + \text{c.c.} = -\frac{2}{t_f^2} \braket{\pi_f(\bfk) \pi_f(-\bfk)}'.
\eeq
This expression can be  checked upon using the explicit form for the two-point function
\be
\braket{\pi_f(\bfk) \pi_f(-\bfk)}'   =  \frac{1}{2\lambda^2 k t_f^2}\,,
\ee
along with the soft limit of the unequal time three-point function
\be
\label{eqcrrslts}
 \lim\limits_{\bfq \rightarrow 0} \left(\frac{1}{t_i} \mathcal{E}_i(q) - i\lambda^2 \right) \braket{\pi_f(\bfk) \pi_f(-\bfk-\bfq) \pi_i(\bfq)}' + \text{c.c.} = -\frac{1}{\lambda^2 k t_f^4}\,  .
\ee
It is also straightforward to check that the symmetry~\eqref{drivenshift} satisfies the physical mode condition---the nonlinear part of the transformation $D_{\v q}$~\eqref{albet0drsupshift} and the soft mode function in this model $u_q \sim 1/t -iq + \cdots$ exhibit the same time dependence in the leading order in soft momentum $q$. We can therefore extend~\eqref{softthdrivenunequalt} to late times by simply replacing the initial soft mode $\pi_i(\bfq)$ with $\pi_f(\bfq)$. Suppressing the dependence on the final time, one finds the following equal-time soft theorem:
\beq
\lim\limits_{\bfq \rightarrow 0} \frac{\braket{\pi(\bfk) \pi(-\bfk - \bfq)\pi(\bfq)}'}{t\braket{\pi(\bfq) \pi(-\bfq)}'} = -\frac{2}{t^2} \braket{\pi(\bfk) \pi(-\bfk)}' ,
\label{eq:dsflatetime}
\eeq
It is straightforward to compute the correlators
\beq
\braket{\pi(\bfk) \pi(-\bfk)}'   =  \frac{1}{2\lambda^2 k t^2}\, , \qquad 
\braket{\pi(\bfk) \pi(-\bfk-\bfq) \pi(\bfq)}' = - \frac{k + |\bfk + \bfq| + q}{4\lambda^4 t^5 k |\bfk + \bfq| q} \, ,
\label{drsup23}
\eeq
and verify that the relation~\eqref{eq:dsflatetime} is satisfied.

\vspace{-10pt}
\paragraph{Time translations.} Time translations are nonlinearly realized on $\pi$ in the standard way:
\begin{equation}
\delta\pi = 1+\dot\pi \, .
\end{equation}
After computing the corresponding conserved current, one finds $\alpha_0=0$ and $D_{\v q}=1$, leading to the following unequal-time soft theorem for $\pi$:
\beq
 \lim\limits_{\bfq \rightarrow 0} \mathcal{E}_i(q) \braket{\pi_f(\bfk) \pi_f(-\bfk - \bfq) \pi_i(\bfq)}' + \text{c.c.} = \prl_{t_f} \braket{\pi_f(\bfk) \pi_f(- \bfk)}'.
\label{3to2drivtimetransl}
\eeq
In contrast to~\eqref{drivenshift}, the physical mode condition associated with the nonlinearly realized time translations is not satisfied, since the nonlinear part of the symmetry $D_{\v q}$ is now constant in time. One is therefore not guaranteed to be able to find equal-time soft theorems for Goldstone's correlation functions. Incidentally, it just so happens that the soft theorem relating three and two-point functions at equal times, obtained by simply replacing $\pi_i$ with $\pi_f$ in~\eqref{3to2drivtimetransl}, 
\beq
 \lim\limits_{\bfq \rightarrow 0} \frac{\braket{\pi(\bfk) \pi(-\bfk - \bfq) \pi(\bfq)}'}{\braket{\pi(\bfq) \pi(-\bfq)}'}  = \partial_t \braket{\pi(\bfk) \pi(-\bfk)}' .
\eeq
is satisfied, as it can be explicitly checked upon using \eqref{drsup23}.
One can verify that this is however an accident of the $3$-to-$2$ soft theorem. The analogous $4$-to-$3$ equal-time consistency condition from time translations is, in fact, violated.

\vspace{-10pt}
\paragraph{Boosts.} Under an infinitesimal boost, the field $\pi$ transforms as 
\begin{equation}
\delta_j \pi =  x_j +  (t \prl_j +x_j \prl_t) \pi \, .
\end{equation}
Following the standard procedure, one finds the Noether current $j^0= b_j x^j \Pi_\pi$ and, therefore, $\alpha_0=0$ and $D_{\v q} = -i b_j \partial_{q^j}$, leading to the same soft theorem as in Eq.~\eqref{superflccboost} with $\mu=1$. What is different are the explicit expressions for the correlation functions, which can be read off from \eqref{eqcrrslts}. 
For boosts, the conditions required for promoting the identity (\ref{superflccboost}) to final times are also not met: similarly to the  superfluid in Section~\ref{sec:superfluidmain}, the mode function in the soft limit $u_q \sim 1/t -iq + \cdots$ contains a linear  piece in $q$, which violates the physical mode condition (see Section~\ref{sec:physmodcond}).

\section{More on the EFT of inflation}
\label{EFToIapp}

Here we elaborate on some of the details of the effective field theory of inflation~\cite{Creminelli:2006xe,Cheung:2007st}.
We first show how to derive symmetries~\eqref{dilpi} and~\eqref{SCTsymm} from the de Sitter isometries, and briefly review how they can be re-expressed in terms of the curvature perturbation $\zeta$. We then discuss some aspects of the soft theorems satisfied by $\pi$.

\subsection{Symmetries}
\label{sec:relpizeta}

Let's consider the symmetries of the EFT~\eqref{EFTofI}. For simplicity, we will work in the limit where
the couplings do not depend explicitly on time. This ensures that after a St\"uckelberg transformation, the action for the scalar fluctuation $\pi$ is manifestly invariant under constant shifts~\cite{Cheung:2007st}.\footnote{This is equivalent to assuming that, in the coordinate where the `clock' is linear in time, there is a shift symmetry acting on the scalar field.}
Everything we discuss here can be easily generalized, following \cite{Finelli:2017fml,Finelli:2018upr}, to shift-symmetric theories where the scalar clock is not a linear function of time.\footnote{It is of course always possible to redefine time in such a way to make the background of $\phi$ linear in the new temporal coordinate. However, in these coordinates, the field transformation is no longer a constant shift \cite{Finelli:2018upr}.}  We will discuss an example of this in Appendix~\ref{app:timedependentsoundspeed}.

Besides translations and rotations, de Sitter space possesses  $4$ additional isometries that can be written, in their infinitesimal form, as
\begin{equation}
t \mapsto t +\frac{\lambda}{H} \, ,
\qquad
x^i\mapsto (1-\lambda)x^i \, ,
\label{dils}
\end{equation}
and 
\begin{equation}
t \mapsto t +\frac{2}{H}\vec{b}\cdot\vec{x} \, ,
\qquad
x^i\mapsto x^i+ b^i \left(\vec{x}^2  - \frac{1}{a^2H^2} \right) - 2 x^i (\vec{b}\cdot\vec{x}) \, ,
\label{scts}
\end{equation}
where $a = \Exp^{Ht}$.
The transformations \eqref{dils} correspond to infinitesimal dilations, while \eqref{scts} at late times acts like a special conformal transformation on the future boundary.

In single-field inflation, in the limit in which the background metric can be approximated by de Sitter space, the transformation laws~\eqref{dilpi} and~\eqref{SCTsymm} can be derived as follows. Using an appropriate field redefinition, we can always choose to parametrize perturbations as
\begin{equation}
\phi (\vec{x}, t) = t +\pi(x) \, .
\end{equation}
Using the fact that $\phi (\vec{x}, t) $ transforms as a scalar, i.e., that~$\tilde{\phi}(\tilde{x})=\phi(x)$, it is straightforward to derive that $\pi$ transforms under dilations as
\begin{equation}
\delta_D \pi = -\frac{\lambda}{H}(1+\dot \pi) + \lambda \, \bfx \cdot  \bfnabla \pi 	\, ,
\label{app:dilpi}
\end{equation}
while for the isometries \eqref{scts} one finds
\beq
\delta_{\text{SCT}} \pi =   - \frac{2}{H}\vec{b}\cdot \vec{x} +b^j\left(  - \frac{2}{H}x_j \dot \pi + 2 x_j  \bfx \cdot  \bfnabla \pi - x^2 \prl_j \pi +\frac{1}{a^2 H^2} \prl_j \pi  \right) \, .
\label{app:SCTsymm}
\eeq

In   terms of the  curvature perturbation $\zeta$, defined by $g_{ij}=a^2(t){\rm e}^{2\zeta}\delta_{ij}$~\cite{Bardeen:1983qw,Salopek:1990jq} the transformation laws \eqref{app:dilpi}  and \eqref{app:SCTsymm}  take the form
\begin{equation}
\delta_D\zeta = \lambda(1+ \vec{x}\cdot\vec{\nabla} \zeta) 
\label{tlszeta1}
\end{equation}
 and
\begin{equation}
\delta_{\text{SCT}} \zeta = 2 \vec{b}\cdot \vec{x} + \left(\frac{b^j}{a^2H^2} + 2 \vec{b}\cdot \vec{x}  x^j - \vec{x}^2 b^j \right)\partial_j\zeta \, ,
\label{tlszeta2}
\end{equation}
respectively.
These are usually derived by searching for residual large gauge transformations that are the zero momentum limit of physical solutions of the Einstein equations~\cite{Weinberg:2003sw,Weinberg:2008zzc}. 
Alternatively,~\eqref{tlszeta1} and~\eqref{tlszeta2} can be  obtained also from~\eqref{app:dilpi}  and~\eqref{app:SCTsymm} by taking the variation of the nonlinear relation between $\zeta$  and  $\pi$  \cite{Maldacena:2002vr,Cheung:2007sv}:
\begin{equation}
\zeta = - H\pi +  H \pi\dot{\pi} +\frac{1}{2}\dot{H}\pi^2 + \frac{1}{4a^2}\left( -\partial_i\pi\partial_i\pi + \nabla^{-2}\partial_i\partial_j(\partial_i\pi\partial_j\pi) \right) + O(\pi^3) \, .
\label{zetapinonlinear}
\end{equation}

It is straightforward to check that \eqref{tlszeta1} and \eqref{tlszeta2} are  symmetries of the action for $\zeta$ after the non-dynamical components of the metric are integrated out  (see Ref.~\cite{Maldacena:2002vr} for details and \cite{Creminelli:2012ed} for an explicit check in the context of  slow-roll single-field inflation).\footnote{Note that the piece proportional to  $1/(a^2H^2)$  in \eqref{tlszeta2} originates from a (large) time-dependent spatial diffeomorphism $\xi^i$ whose form is fixed by the requirement that the Hamiltonian constraints are solved nontrivially in the zero momentum limit \cite{Creminelli:2012ed}. Concretely the diffeomorphism parameter that is an adiabatic mode is
\begin{equation}
\xi^i = \epsilon^i \vec{x}^2 - 2 x^i (\vec{\epsilon}\cdot \vec{x}) + \delta x^i(t) \, ,
\qquad
\qquad \delta x^i(t)\equiv -\frac{\epsilon^i}{a^2H^2} \, .
\label{xi}
\end{equation}
The form of $\delta x^i(t)$ in  \eqref{xi} is in agreement with the result of the integral in the general expression for $\xi^i$~\cite{Mirbabayi:2014zpa},
\begin{equation}
\xi^i(t) =\left( 1+ \int^t \frac{{\rm d} t'}{a(t')^3}\int^{t'}a(t''){\rm d}t''  \, \nabla^2\right)\bar{\xi}^i \, ,
\end{equation}
where one should substitute $\bar{\xi}^i= \epsilon^i \vec{x}^2 - 2 x^i (\vec{\epsilon}\cdot \vec{x})$ and work at the leading order in the exact de Sitter limit.}

\subsection{Soft theorems}
\label{app:EFTapp2}

We now want to discuss the soft theorems following from the symmetries of the $\pi$ action~\eqref{eftlagr234}. These soft theorems are well-studied (e.g.,~\cite{Creminelli:2012ed,Assassi:2012zq,Hinterbichler:2012nm,Flauger:2013hra,Pimentel:2013gza,Goldberger:2013rsa,Hinterbichler:2013dpa,Berezhiani:2013ewa,Kundu:2015xta,Hui:2018cag}), but for completeness we review these results in order to put them in context with our discussion in the main text.

\vspace{-10pt}
\paragraph{Shift symmetry:}
We first consider the shift symmetry of $\pi$, which acts as $\delta\pi= 1$. 
 The corresponding unequal-time consistency condition takes the form
\beq
\lim\limits_{\bfq \rightarrow 0} \mathcal{E}_i(q) \braket{\pi_{\bfk}(\t_f) \pi_{-\bfk - \bfq}(\t_f) \pi_{\bfq}(\t_i)}' + \text{c.c.} = 0 \, .
\label{utstttEFTofI}
\eeq
This can be explicitly checked using the expression for the soft limit of the three-point function:
 \begin{align}
 \left. \lim\limits_{\bfq \rightarrow 0} \mathcal{E}_i(q)  \braket{\pi_{\bfk}(\t_f) \pi_{-\bfk - \bfq}(\t_f) \pi_{\bfq}(\t_i)}' + \text{c.c.} \right|_{\t_f=0} & =\frac{5 H^3 (1-c_s^2) }{16 c_s^3  |\dot H|  \Mpl^2} \frac{(\v k \cdot \v q)^2}{k^7} \\ &~~ + \frac{H^3 (4 c_s^4 M_3^4 + (3 c_s^4 + 5 c_s^2-8) |\dot H| \Mpl^2)}{16 c_s^3 |\dot H|^2  \Mpl^4} \frac{q^2}{k^5}\, .
 \nonumber
\end{align}
which indeed vanishes in the soft limit, as expected.

One can check that the physical mode condition associated with constant shifts of $\pi$ in the theory \eqref{EFTofI} is satisfied. This ensures that in the soft theorem \eqref{utstttEFTofI} one can promote $\t_i$ to $\t_f$. The consistency condition at equal times thus reads
\beq
\lim\limits_{\bfq \rightarrow 0} \frac{\braket{\pi_{\bfk} \pi_{-\bfk - \bfq} \pi_{\bfq}}'}{\braket{\pi_{\bfq} \pi_{-\bfq} }'}=0  \, ,
\label{etccttEFTofI}
\eeq
As is well-known, the soft limit of the equal-time three-point function following from an explicit calculation vanishes as $q^2$ in the soft limit.

\vspace{-10pt}
\paragraph{Dilation soft theorem at three points:}

Dilations are nonlinearly realized on $\pi$ as in \eqref{app:dilpi}.
This can be used to derive the following unequal time soft theorem relating the three and two-point functions
\beq
-\lim\limits_{\bfq \rightarrow 0} \frac{1}{H} \mathcal{E}_i(q)  \braket{\pi_{\bfk}(\t_f) \pi_{-\bfk - \bfq}(\t_f) \pi_{\bfq}(\t_i)}' + \text{c.c.} = \left( \t_f \prl_{\t_f} -3 - k \prl_{k} \right) \braket{\pi_{\bfk}(\t_f) \pi_{-\bfk}(\t_f) }'.
\label{dilidEFTofI}
\eeq
The physical mode condition for the dilation symmetry \eqref{app:dilpi} is satisfied, which implies that \eqref{dilidEFTofI} can be promoted to final times as~\cite{Creminelli:2012ed}
\beq
-\lim\limits_{\bfq \rightarrow 0} \frac{1}{H} \frac{\braket{\pi(\bfk) \pi(-\bfk - \bfq) \pi(\bfq)}'}{\braket{\pi(\bfq) \pi(-\bfq)}'} = \left( \t \prl_{\t} -3 - k \prl_{k} \right) \braket{\pi(\bfk)\pi(-\bfk) }'.
\label{dilidEFTofIff}
\eeq
It is straightforward to check explicitly that both sides of~\eqref{dilidEFTofI}, as well as of~\eqref{dilidEFTofIff}, vanish identically.

\vspace{-10pt}
\paragraph{SCT soft theorem at three points:}

Special conformal transformations act nonlinearly on $\pi$ as in \eqref{app:SCTsymm}. 
The corresponding  soft theorem that relates a three-point function with one field evaluated at the initial time to a late-time two-point function is:
\beq
\lim\limits_{\bfq \rightarrow 0} \frac{1}{H} \frac{\prl}{\prl q_j}\left(\mathcal{E}_i(q) \braket{\pi_{\bfk}(\t_f) \pi_{-\bfk - \bfq}(\t_f) \pi_{\bfq}(\t_i)}'\right) + \text{c.c.} = \frac{k_j}{2 k}\left(- \t_f \frac{\prl^2}{\prl k \prl \t_f} +4 \prl_k + k \prl_k^2  \right)\braket{\pi_{\bfk}(\t_f) \pi_{-\bfk}(\t_f)}'.
\label{softthSCTEFTofI3p}
\eeq
Note that, in contrast to \eqref{3to2EFToI}, we are taking  the final times in \eqref{softthSCTEFTofI3p} to be equal, $\tau_1=\tau_2\equiv \tau_f$.
In this limit, each side of \eqref{softthSCTEFTofI3p} vanishes identically, as it can be checked explicitly:
\beq
\braket{\pi_{\bfk}(\t_f) \pi_{- \bfk}(\t_f) }' = \frac{H^2}{4 \Mpl^2 c_s |\dot{H}| k^3} (1+ c_s^2 k^2 \t_f^2), 
\qquad
 \lim\limits_{\bfq \rightarrow 0} \frac{1}{H} \frac{\prl}{\prl q_j}\left(\mathcal{E}_i(q) \braket{\pi_{\bfk}(\t_f) \pi_{-\bfk - \bfq}(\t_f) \pi_{\bfq}(\t_i)}'\right)=0.
\eeq
As was mentioned in Section~\ref{sec:EFTofI}, this means that we cannot reproduce the relation between the cubic couplings and the normalization of the two-point function. This instead requires the unequal-time soft theorem discussed in Section~\ref{sec:EFTofI}.

Slow-roll inflation satisfies the physical mode condition. Using this, along with the identity~\eqref{dilidEFTofIff} for dilations, we can promote the soft theorem~\eqref{softthSCTEFTofI3p} to equal times as
\beq
\lim\limits_{q \rightarrow 0} \frac{1}{H} \frac{\prl}{\prl q_j}\left(\frac{\braket{\pi_{\bfk} \pi_{-\bfk - \bfq} \pi_{\bfq}}'}{\braket{\pi_{- \bfq} \pi_{\bfq} }'}\right) = \frac{k_j}{2 k}\left(- \t \frac{\prl^2}{\prl k \prl \t} +4 \prl_k + k \prl_k^2  \right)\braket{\pi_{\bfk} \pi_{-\bfk}}'\, ,
\eeq
One can also verify by an explicit calculation that this relation is satisfied in the sense that both sides vanish~\cite{Creminelli:2012ed}.

\vspace{-10pt}
\paragraph{SCT soft theorem at four points:}
\label{subsec:sct43}
We now want to consider the consequences of the equal-time soft theorem following from SCTs at four points. We saw above that at three points, the corresponding SCT soft theorem is satisfied in a somewhat trivial way---both sides vanish. As such, the three-point relation does not place any constraints on the parameters of the EFT. At four points, the situation is more interesting.

The soft theorem was checked explicitly in~\cite{Creminelli:2012ed} for the correlators derived from the EFT of inflation. Here, we want to emphasize a slightly different viewpoint, spiritually related to the bootstrap construction of inflationary correlators. To that end, we imagine taking the interactions of the EFT of inflation~\eqref{eftlagr234} and allowing all the couplings to be arbitrary, even the ones that are in reality fixed in terms of the sound speed:
\be
\begin{aligned}
S
=\int \D t\,\D^3x\, a^3 \bigg[  \frac{1}{2}\left(\dot{\pi}^2 - \frac{c_s^2}{a^2}(\partial_i \pi)^2 \right)   + \alpha_1 \dot{\pi}^3 &+ \alpha_2 \dot{\pi} \frac{(\partial_i \pi)^2}{a^2} 
\\
&+ \beta_1 \dot{\pi}^4 + \beta_2 \dot{\pi}^2 \frac{(\partial_i \pi)^2}{a^2} + \beta_3 \frac{(\partial_i \pi)^4}{a^4} + \cdots \bigg] \, .
\end{aligned}
\label{eq:eft4pt}
\ee
Conceptually, we can think of~\eqref{eq:eft4pt} as a way of generating four-point correlators that are consistent with locality, unitarity, etc. in de Sitter space, but one could alternatively imagine directly bootstrapping these shapes following the approach of~\cite{Bonifacio:2021azc}. (We will see that only contact four-point functions actually enter the soft theorem.) We now want to see how the nonlinearly realized symmetries of $\pi$ additionally constrain the shapes generated by~\eqref{eq:eft4pt}.

All of the shapes generated by the quartic operators vanish in the soft limit and satisfy the dilation identity trivially, so the only nontrivial constraint comes from the SCT symmetry of $\pi$. The equal-time soft theorem following from this symmetry reads 
\begin{equation}
\label{4to3EFToI}
 \lim_{ q \to 0} \frac{\partial}{\partial q^i} \left(\frac{1}{P(q)} \vev{ \pi_{{\vec k}_1} \pi_{{\vec k}_2} \pi_{{\vec k}_3} \pi_{\vec q} } \right) =  \frac{H}{2} \hat{{\cal K}}^i  \vev{\pi_{{\vec k}_1} \pi_{{\vec k}_2} \pi_{{\vec k}_3} } \, ,
\end{equation}
where we have defined 
\be
\hat{{\cal K}}^i \equiv \sum_{a =1}^3\hat{K}^i_{ {\vec k}_{a}}   \, ,~~{\rm with}~~\, \hat{K}^i_{ {\vec k}_{a} }  = 6 \frac{\partial}{\partial k_{a} ^i} +2 k^a \frac{\partial^2}{\partial k_{a} ^i\partial k_{a} ^a} -k^i \frac{\partial^2}{\partial k_{a} ^a\partial k_{\ell} ^a}\, .
\ee
Using the action~\eqref{eq:eft4pt}, we can explicitly compute the late-time three-point function
 \begin{equation}
\vev{\pi_{{\vec k}_1} \pi_{{\vec k}_2} \pi_{{\vec k}_3} }' 
= \frac{12 ( \alpha_1 c_s^2+\alpha_2)K_3^2 +\alpha_2 K_1^6-3 \alpha_2 K_1^4 K_2+11 \alpha_2  K_1^3 K_3-4 \alpha _2  K_1^2K_2^2-4 \alpha_2 K_1 K_2 K_3 }{4 K_3^3 K_1^3 c_s^8}\, ,
 \end{equation}
where $K_2  = k_1k_2+k_1 k_3+ k_2k_3$, $K_3 = k_1k_2k_3$ and $K_1= \sum_{i}k_i$. The four-point function receives contributions from both exchange and contact diagrams, but the exchange diagrams contribute to ${\cal O}(q^2)$ in the soft limit and so can be ignored.\footnote{This can be verified by counting the number of derivatives in the external legs, see also explicit check in~\cite{Creminelli:2012ed}.} The relevant late-time four-point function computed from contact diagrams is 
\begin{align}
\vev{\pi_{{\vec k}_1} \pi_{{\vec k}_2} \pi_{{\vec k}_3} \pi_{{\vec k}_4} }'_{\rm contact} &=  3\bigg( \beta _1-\frac{9 \alpha_1^2}{2} \bigg) \frac{1}{c_s^9  K_4 K_1^5}   \\ \nonumber
&\quad\, +\frac{1}{2}\bigg(2 \beta_3- \alpha_2^2\bigg) \frac{  ({\vec k}_1\cdot {\vec k}_2)({\vec k}_3\cdot{\vec k}_4) }{4 c_s^{13} K_4^3 K_1^5} \big( 12 K_4+3 K_3 K_1+K_2 K_1^2+K_1^4\big) \\ \nonumber
&\quad\, +(\beta_2-3 \alpha_1 \alpha_2  ) \frac{k_1^2 k_2^2 ({\vec k}_3\cdot {\vec k}_4) }{4 c_s^{11} K_4^3 K_1^5} \bigg( 1 +3\frac{k_3+k_4}{K_1} + 12\frac{ k_3 k_4}{K_1^2}\bigg) + 23\,\mbox{perms.} 
\end{align}
where we have defined the following combinations of momentum magnitudes
\be
K_1 = k_1+k_2+k_3+k_4\,,  \qquad K_2  = \sum_{i<j}k_ik_j \, , \qquad K_3  = \sum_{i<j<l}k_ik_jk_l \, , \qquad  K_4 = k_1k_2k_3k_4 \, . 
\ee
For arbitrary $\alpha_i$ and $\beta_i$,
the soft theorem~\eqref{4to3EFToI} imposes the following relations\footnote{Note that the correlators in \eqref{4to3EFToI} still involve a momentum-conserving delta function. Acting with the differential operator $\hat{{\cal K}}^i$ on the right-hand side of~\eqref{4to3EFToI} is therefore slightly subtle.
One can think of two different---but in the end equivalent---ways of proceeding. One can first act  with the SCT operator $\hat{{\cal K}}^i $ on $ B_3(k_1,k_2,k_3) \equiv \vev{\pi_{{\vec k}_1} \pi_{{\vec k}_2} \pi_{{\vec k}_3} }'$, and only then afterward use the momentum conservation to set ${\vec k}_3=-{\vec k}_1-{\vec k}_2$:
\begin{equation*}
\left[(\hat{{\cal K}}^i_{\vec k_1}  +\hat{{\cal K}}^i_{\vec k_2} +\hat{{\cal K}}^i_{\vec k_3}    )B_3 \right]' = \left[ k_1^i\left( \frac{4}{k_1} \partial_{k_1} -  \frac{4}{k_3} \partial_{k_3} +\partial_{k_1}^2 - \partial_{k_3}^2 \right)B_3     - k_2^i\left(\frac{4}{k_2} \partial_{k_2} -  \frac{4}{k_3} \partial_{k_3}  +\partial_{k_2}^2- \partial_{k_3}^2\right) B_3 \right]' .
\end{equation*}
Alternatively,  one  can use momentum conservation first to make $B'_3$ a function of ${\vec k}_1,\, {\vec k}_2$ only, and then write  the SCT operator $\hat{{\cal K}}^i$ acting  on it as 
\begin{equation*}
(\hat{{\cal K}}^i_{\vec k_1}  +\hat{{\cal K}}^i_{\vec k_2}  )B_3' = k_1^i \left[\frac{4}{k_1}\partial_{k_1}B_3 + \partial_{k_1}^2 B_3+ \frac{10}{k_3} \partial_{k_3}B_3 + \partial_{k_3}^2 B_3  + 2 \frac{k_1}{k_3}\partial_{k_1}\partial_{k_3} B_3 + 2 \frac{k_2}{k_3}\partial_{k_2}\partial_{k_3} B_3   \right]' + ( 1 \leftrightarrow 2) \, .
\end{equation*}
These two very different-looking expressions are not equal for arbitrary $B_3$.  However, when evaluated explicitly on the actual three-point function computed in the EFT of inflation, they turn out to be identical.}
\begin{align}
\beta_2& = 3 \alpha _1 \left(\alpha _2-c_s^5\right) -2 \alpha _2 c_s^3 \, ,   \\
\beta _3 &= \frac{1}{2} \alpha _2 \left(\alpha _2-c_s^5\right)  \, ,
\end{align}
which fixes two of the parameters of the quartic interactions in terms of the cubic couplings (where the coupling $\alpha_2$ itself is fixed in terms of the sound speed by the soft theorem~\eqref{3to2EFToI}.

So, we see that the nonlinearly realized symmetries of $\pi$ reduce the number of free parameters in the EFT of inflation. It would be interesting to more systematically implement these constraints from the bootstrap point of view at higher derivative order and at higher points.

\section{Ultra-slow-roll with time-dependent sound speed}
\label{app:timedependentsoundspeed}

In this Appendix, we investigate another example of ultra-slow-roll evolution where the time dependence of the sound speed is not negligible. Our motivation is to investigate the physical mode condition in this context. As opposed to cases where $s\equiv\dot{c}_s/(H c_s)\ll1$, we will show that the physical mode condition is now not satisfied. This implies that, in such a class of models, it is not possible to extend, in a straightforward way, the soft theorem \eqref{softthd1} to late times.

(Equal-time) soft theorems in shift-symmetric ultra-slow-roll inflation  were first derived in \cite{Finelli:2017fml,Bravo:2017wyw}. These consistency conditions have been shown to arise from a combination of a large diffeomorphism and the internal shift symmetry~\cite{Finelli:2017fml}, and have been checked explicitly in the case of constant sound speed $c_s$ for the perturbations. (The existence of such equal-time identities  is guaranteed by  the physical mode condition~\cite{Hui:2018cag}, which is satisfied  when $c_s=\text{constant}$.)

To consider the case with a time-dependent sound speed, we start with a  shift-symmetric theory of the form ${\cal L}=P(X)$, where $X\equiv -(\partial_\mu\phi)^2$. In the unitary gauge language, this is equivalent to
\be
\begin{aligned}
{\cal L}= \frac{M_\text{Pl}^2}{2}R &- M_\text{Pl}^2 \left(3H^2 +2 \dot{H} \right) 
+ M_\text{Pl}^2\dot{H}\delta g^{00} + \frac{M_\text{Pl}^2\dot{H}}{4}\frac{c_s^2-1}{c_s^2}(\delta g^{00})^2
\\
&\hspace{2.15cm}+ \frac{M_\text{Pl}^2\dot{H}}{6c_s^2} \left[ c_s^2-1
-\frac{H\bar X}{\dot{\bar{X}}}\frac{ 2s + (1-c_s^2)(2\varepsilon-\eta) }{2}
\right](\delta g^{00})^3 
+ \cdots\,,
\label{pxthL}
\end{aligned}
\ee
where  we have defined the background quantities
\begin{equation}
\frac{\dot{\bar{X}}}{H \bar X} = -6 c_s^2 \, ,
\qquad
\bar X \equiv \dot{\bar{\phi}}^2 \, ,
\end{equation}
where $\bar \phi$ denotes the background profile of the scalar field $\phi$. In \eqref{pxthL} we introduced the slow-roll parameters $\varepsilon\equiv-\dot{H}/H^2$ and $\eta\equiv \dot{\varepsilon}/(H\varepsilon)$.

The lagrangian~\eqref{pxthL} is a particular case of the general EFT~\eqref{EFTofI}, where the coefficients are fixed by the underlying shift symmetry on $\phi$~\cite{Finelli:2018upr}. In particular, the slow-roll parameters  satisfy the following background equations of motion:
\begin{equation}
\dot{H } \left[ \eta - 2\varepsilon +3 (1+c_s^2) \right] =0 \, .
\label{usreoms}
\end{equation}
In the $\zeta$-gauge, defined by $\delta g_{ij}= a^2 \Exp^{2\zeta}\delta_{ij} $, the quadratic action for the scalar perturbation $\zeta$ is
\begin{equation}
S^{(2)} 
	= \frac{1}{2}\int\D^3x \, \D\tau \, z^2  \left(
	{\zeta'}^2 - c_s^2(\partial_i\zeta)^2\right) \, ,
\label{mgnobox}
\end{equation}
where we defined
\begin{equation}
z^2\equiv M_\text{Pl}^2\frac{2\varepsilon a^2}{c_s^2} 
\label{z}
\end{equation}
and where the prime $'$ denotes differentiation with respect to the conformal time $\tau$. In the following, we will  assume  $\varepsilon,\dot{\eta}/H,\dot{s}/H\ll 1$, while we will allow $\eta$, $c_s$ and $s$ to be $\mathcal{O}(1)$. It can be shown that the spectrum of perturbations is scale-invariant in the two cases~\cite{Akhshik:2015nfa}
\begin{equation}
\eta+s=0 
\qquad \text{or} \qquad
6+\eta-5s =0 \, .
\label{condsi}
\end{equation}
In the second case, i.e.,~$s=(6+\eta)/5$, the background dynamics is non-attractor, as in ultra-slow-roll.

\vspace{-10pt}
\paragraph{Violation of the physical mode condition:}

The lagrangian \eqref{pxthL} has a nonlinearly realized symmetry on $\zeta$ of the form \cite{Finelli:2017fml,Finelli:2018upr}
\begin{equation}
\delta\zeta = - \lambda(t)\left(1+x^i\partial_i \zeta\right)  - \xi^0 \left( H +\dot{\zeta} \right) \, ,
\label{zetatrans}
\end{equation}
where
\begin{equation}
\xi^0(t) =  \frac{c}{\dot{\bar{\phi}}(t)} \, ,
\qquad
\dot{\lambda}(t) = - \xi^0(t)\dot{H}(t) \, ,
\end{equation}
or equivalently, in momentum space,
\begin{equation}
\delta\zeta_{\bfk} = - \lambda(t) - \xi^0H + \lambda(t)\left(3+\bfk\cdot\partial_\bfk\right) \zeta_{\bfk} - \xi^0 \dot{\zeta}_{\bfk}  \, .
\label{zetatransFT}
\end{equation}
In the case $s\ll1$, one can use the symmetry \eqref{zetatrans} to derive a soft theorem  for late-time correlators \cite{Finelli:2017fml}. The soft theorem is guaranteed to hold thanks to the physical mode condition, which is satisfied if $s\ll1$ \cite{Hui:2018cag}.
Using the explicit form of the three-point function \cite{Akhshik:2015nfa} in the case where $s$ is instead non-negligible, in particular  $s=(6+\eta)/5$ (see Eq.~\eqref{condsi}), one can check that the late-time soft theorem ceases to hold. This can be attributed to the violation of the physical mode condition.   

To see this, consider the background equation of motion for the scalar field,
\begin{equation}
\partial_t\left( a^3 \dot{\bar{\phi}} P_X  \right) =0\, ,
\end{equation}
which can be solved for $\dot{\bar{\phi}}$ as
\begin{equation}
\dot{\bar{\phi}} \propto a^3 \dot{H} \, .
\label{app:solphid}
\end{equation}
Assuming that $\eta$ and $s$ do not vary significantly  during the non-attractor phase, we can write \cite{Akhshik:2015nfa}
\begin{equation}
\varepsilon=\varepsilon_\text{e}\left(\frac{\tau}{\tau_\text{e}}\right)^{-\eta} \, ,
\qquad
c_s = c_{s\text{e}}\left(\frac{\tau}{\tau_\text{e}}\right)^{-s} \, ,
\label{scaling}
\end{equation}
Then, from~\eqref{zetatransFT}, the nonlinear piece of the transformation~\eqref{zetatransFT} takes the form
\begin{equation}
\delta_\text{NL} \zeta = \int \D t \,  \xi^0\dot H - \xi^0H
\propto  \int \frac{\D t}{a^3}  - \frac{H}{a^3 \dot H} 
\approx \frac{1}{a^3 H \varepsilon} \, ,
\end{equation}
where we used the background solution \eqref{app:solphid}.
Assuming the scaling~\eqref{scaling}, this implies
\begin{equation}
\Delta\zeta_\text{NL} \propto   \tau^{3 +\eta}
 \, .
\label{e1}
\end{equation}
On the other hand, let us consider  the linearized equations of motion for $\zeta$---see, e.g., \eqref{mgnobox}---in the limit $\bfq\rightarrow0$,
\begin{equation}
\partial_\tau \left( \frac{a^2 \varepsilon }{c_s^2} \zeta_{\bfq\rightarrow0}'  \right) =0 \, ,
\end{equation}
which can be solved as
\begin{equation}
\zeta_{\bfq\rightarrow0} \propto \int \D \tau \frac{c_s^2}{a^2 \varepsilon}
\approx H^2 \int \D \tau \frac{\tau^2c_s^2}{ \varepsilon} \propto \tau^{3+\eta-2s} \, .
\label{e2}
\end{equation}
where we assumed again the scaling  \eqref{scaling}. Comparing \eqref{e1}  and \eqref{e2} it is clear that, under the assumption \eqref{scaling}, the physical mode condition is satisfied  only for $s\ll1$.

\section{Soft theorems and scattering amplitudes}
\label{sec:softampl}

In Section~\ref{sec:superfluidmain} we derived soft theorems for the correlation functions of a superfluid phonon that result from nonlinearly realized time translations and Lorentz boosts. These soft theorems relate certain couplings in the EFT at the level of observables. 
A natural question to ask is whether one can derive analogous soft theorems for scattering amplitudes. That is, do the correlator soft theorems have an analogue in terms of Ward identities relating $(N+1)$-point scattering amplitudes in the soft limit to $N$-point scattering amplitudes? This question was recently asked in~\cite{Green:2022slj} and explored in detail, so we only make some brief comments, and refer the reader there for a more complete discussion.

It is worth recalling that, for Poincar\'e-invariant scalar field theories, the presence of a shift symmetry is responsible for an Adler zero~\cite{Adler:1964um,Adler:1965ga}. This is nothing but the statement that the scalar's scattering amplitudes vanish in the limit where one of the external momenta is taken to be soft, i.e., they are schematically of the form 
\begin{equation}
\lim_{q \rightarrow 0}{\cal A} \sim q^\sigma \, ,
\qquad
\text{with } \sigma\geq1 \, ,
\end{equation}
where we can think of theories having larger values of $\sigma$ as having enhanced symmetries~\cite{Cheung:2014dqa,Cachazo:2014xea,Hinterbichler:2015pqa,Cheung:2016drk}.
This should be contrasted with the case of scattering amplitudes of fields with spins, which do not vanish in the soft limit, leading to the well-known Weinberg's theorems for the emission of low-energy photons and gravitons  \cite{Weinberg:1964ew,Weinberg:1965nx,Strominger:2013jfa, He:2014laa,He:2014cra,Strominger:2017zoo}. 
The  Adler zero for shift-symmetric, Poincar\'e-invariant scalar theories can also be understood as a consequence of the fact that there are no cubic vertices in the theory that can yield a nonzero $3$-point amplitude \cite{Weinberg:1996kr}.\footnote{All cubic operators
can be removed by suitable field redefinitions---the only exception is  $\phi^3$, which is forbidden by the shift symmetry.}

The story changes if (some of) the Poincar\'e symmetries are spontaneously broken. In this case, there can be cubic operators that are nontrivial on shell (and thus cannot be redefined away), resulting in nonzero $3$-point amplitudes and violating the Adler zero condition~\cite{Weinberg:1996kr,Cheung:2016drk,Grall:2020ibl,Green:2022slj}. This is precisely what happens with the superfluid. 
The lagrangian for the  perturbations on the Lorentz-breaking background is (see also Appendix~\ref{sec:superfluid})
\beq
\lagr = - \frac{1}{2}(\tilde{\prl}_\mu \pi)^2 + \tilde \a_1 \dot \pi^3 - \tilde \a_2 \dot \pi (\tilde\prl_\mu \pi)^2 + \tilde \b_1 \dot \pi^4 - \tilde \b_2 \dot \pi^2 (\tilde\prl_\mu \pi)^2 + \tilde \b_3 (\tilde\prl_\mu \pi)^4 \, ,
\label{superflfakelorlagr}
\eeq
where we have defined
\begin{align}
\tilde \a_1  &= \left(c_s^2+\frac{2 c_3}{3}-1  \right) \tilde \a_2\, , & \tilde \a_2  &= \frac{1-c_s^2}{2 c_s^{5/2}\sqrt{2 \mu^2 P'(\mu^2)}} \, ,  \\
\tilde \b_1  &= c_s^2 \left( c_s^2 + 4 c_3 + 2 c_4 -2 - \frac{4 c_3}{c_s^2}+\frac{1}{c_s^2} \right) \, , &
\tilde \b_2  &= 2 (c_s^2 + 2 c_3 -1) \tilde \b_3 \, , \\
  \tilde \b_3  &= \frac{1-c_s^2}{16 c_s^5 \mu^2 P'(\mu^2)} \, ,
\end{align}
and where we introduced for convenience the rescaled coordinates 
\beq
x^\mu = \left(t,\, \v x/ c_s \right), \qquad p^\mu = \left(\omega,\, c_s \v p \right)\, ,
\eeq
which make the quadratic Lagrangian invariant under ``fake" Lorentz transformations.
To cubic order, there are only two operators in \eqref{superflfakelorlagr}:  $\dot{\pi}^3$ and $\dot{\pi}(\tilde\partial_\mu\pi)^2$. After straightforward integrations by parts, the latter can be recast in the form $\pi^2\tilde \square \dot \pi$, which vanishes on the linearized equations of motion. As a result, only $\dot{\pi}^3$ yields a non-vanishing 3-point scattering amplitude, which takes the form
\beq
{\cal A}_3(\pi_{\v p_1} \rightarrow \pi_{\v p_2} \pi_{\v p_3}) = -6 i \tilde \a_1 \omega_1 \omega_2 \omega_3 \,  ,
\label{ampl1to2}
\eeq
with $\omega_k \equiv c_s p_k$. In particular, as a result of the  non-vanishing amplitude \eqref{ampl1to2},  one can check that the 2-to-2 scattering amplitude of $\pi$ is nonzero in the soft limit, and  factorizes into the product of \eqref{ampl1to2} and an overall momentum-dependent coefficient, 
\beq
\lim \limits_{\v p_4 \rightarrow 0}{\cal A}_4 (\pi_{\v p_1} \pi_{\v p_2} \rightarrow \pi_{\v p_3} \pi_{\v p_4})  =\left( \lim \limits_{\v p_4 \rightarrow 0}  \sum_{i=1}^3  \frac{i \eta_i}{2 c_s^2 p_i p_4} \frac{6 c_s^3 \tilde \a_1 p_i^2 p_4
}{1- \frac{(\v p_i \cdot \v p_4)}{p_i p_4}} \right) {\cal A}_3 (\pi_{\v p_1} \pi_{\v p_2} \rightarrow \pi_{\v p_3} ) \, ,
\label{weinbergtheorem}
\eeq
where $\eta_i =-1$ for the incoming phonons 1 and 2, and $\eta_i =+1$ for the outgoing phonon 3,
consistently with standard `polology' arguments for soft  amplitudes~\cite{Weinberg:1996kr}.

The soft theorem~\eqref{weinbergtheorem} can be  viewed as resulting from  the conservation of the $U(1)$ current~\cite{Hamada:2017atr,Green:2022slj}
\beq
j^\mu = \frac{\d \lagr}{\d (\prl_\mu \pi)} = - \prl^\mu \pi + 3 \tilde \a_1 \d^\mu_0 \dot \pi^2 - \tilde \a_2 \d ^\mu_0 (\prl_\nu \pi)^2 - 2 \tilde \a_2 \dot \pi \prl^\mu \pi\, + \cdots\, .
\label{superflcurrent}
\eeq
To derive \eqref{weinbergtheorem} from the conservation of  \eqref{superflcurrent}, one can use the Ward identity
\be
\begin{aligned}
\braket{\prl_\mu j^\mu (x) \pi(x_1) \pi (x_2) \pi (x_3)} = &-i \d^{(4)}(x-x_1) \braket{\pi (x_2) \pi (x_3)} \\
  &-i \d^{(4)}(x-x_2) \braket{\pi (x_1) \pi (x_3)} -i \d^{(4)}(x-x_3) \braket{\pi (x_1) \pi (x_2)}  \, .
\label{usualward}
\end{aligned}
\ee
After plugging the first term of  \eqref{superflcurrent} into \eqref{usualward}, we can take the  Fourier transform with respect to $x$ and use LSZ reduction on the fields $\pi(x_1),\, \pi (x_2),\, \pi (x_3)$ to get the $2 $-to-$ 2$ scattering amplitude. Insertion of the quadratic part of the operator \eqref{superflcurrent} in \eqref{usualward} results in a nearly on-shell propagator that diverges in $\v q \rightarrow 0$ limit, whereas cubic and higher terms in $j^\mu$ give subdominant contributions~\cite{Mirbabayi:2016xvc}. Taking into account that all but $\dot \pi^3$ vertices vanish on shell, this logic leads to the right-hand side of \eqref{weinbergtheorem}.

Similar statements to~\eqref{weinbergtheorem} can be obtained for higher-point amplitudes. Notably---in contrast to the analogous statements for correlators---they are somewhat model-dependent, in the sense that one must specify the cubic vertices present in the theory in order to write down a soft theorem. Alternatively, one can subtract off the contributions from the cubic vertices at tree level in order to obtain an Adler zero for the remaining contributions, as was done in~\cite{Green:2022slj}.

{\small
\renewcommand{\em}{}
\linespread{1.025}\selectfont
\bibliographystyle{utphys}
\addcontentsline{toc}{section}{References}
\bibliography{physicalmodesbib}

\providecommand{\href}[2]{#2}\begingroup\raggedright\begin{thebibliography}{100}

\bibitem{Adler:1964um}
S.~L. Adler, ``{Consistency conditions on the strong interactions implied by a
  partially conserved axial vector current},''
  \href{http://dx.doi.org/10.1103/PhysRev.137.B1022}{{\em Phys. Rev.} {\bf 137}
  (1965)  B1022--B1033}.

\bibitem{Adler:1965ga}
S.~L. Adler, ``{Consistency conditions on the strong interactions implied by a
  partially conserved axial-vector current. II},''
  \href{http://dx.doi.org/10.1103/PhysRev.139.B1638}{{\em Phys. Rev.} {\bf 139}
  (1965)  B1638--B1643}.

\bibitem{Weinberg:1964ew}
S.~Weinberg, ``{Photons and Gravitons in $S$-Matrix Theory: Derivation of
  Charge Conservation and Equality of Gravitational and Inertial Mass},''
  \href{http://dx.doi.org/10.1103/PhysRev.135.B1049}{{\em Phys. Rev.} {\bf 135}
  (1964)  B1049--B1056}.

\bibitem{Weinberg:1965nx}
S.~Weinberg, ``{Infrared photons and gravitons},''
  \href{http://dx.doi.org/10.1103/PhysRev.140.B516}{{\em Phys. Rev.} {\bf 140}
  (1965)  B516--B524}.

\bibitem{Strominger:2013jfa}
A.~Strominger, ``{On BMS Invariance of Gravitational Scattering},''
  \href{http://dx.doi.org/10.1007/JHEP07(2014)152}{{\em JHEP} {\bf 07} (2014)
  152}, \href{http://arxiv.org/abs/1312.2229}{{\tt arXiv:1312.2229 [hep-th]}}.

\bibitem{He:2014laa}
T.~He, V.~Lysov, P.~Mitra, and A.~Strominger, ``{BMS supertranslations and
  Weinberg\textquoteright{}s soft graviton theorem},''
  \href{http://dx.doi.org/10.1007/JHEP05(2015)151}{{\em JHEP} {\bf 05} (2015)
  151}, \href{http://arxiv.org/abs/1401.7026}{{\tt arXiv:1401.7026 [hep-th]}}.

\bibitem{He:2014cra}
T.~He, P.~Mitra, A.~P. Porfyriadis, and A.~Strominger, ``{New Symmetries of
  Massless QED},'' \href{http://dx.doi.org/10.1007/JHEP10(2014)112}{{\em JHEP}
  {\bf 10} (2014)  112}, \href{http://arxiv.org/abs/1407.3789}{{\tt
  arXiv:1407.3789 [hep-th]}}.

\bibitem{Strominger:2017zoo}
A.~Strominger, ``{Lectures on the Infrared Structure of Gravity and Gauge
  Theory},'' \href{http://arxiv.org/abs/1703.05448}{{\tt arXiv:1703.05448
  [hep-th]}}.

\bibitem{Mirbabayi:2016xvc}
M.~Mirbabayi and M.~Simonovi\'c, ``{Weinberg Soft Theorems from Weinberg
  Adiabatic Modes},''
\href{http://arxiv.org/abs/1602.05196}{{\tt arXiv:1602.05196 [hep-th]}}.

\bibitem{Cheung:2014dqa}
C.~Cheung, K.~Kampf, J.~Novotny, and J.~Trnka, ``{Effective Field Theories from
  Soft Limits of Scattering Amplitudes},''
  \href{http://dx.doi.org/10.1103/PhysRevLett.114.221602}{{\em Phys. Rev.
  Lett.} {\bf 114} (2015) no.~22, 221602},
\href{http://arxiv.org/abs/1412.4095}{{\tt arXiv:1412.4095 [hep-th]}}.

\bibitem{Cachazo:2014xea}
F.~Cachazo, S.~He, and E.~Y. Yuan, ``{Scattering Equations and Matrices: From
  Einstein To Yang-Mills, DBI and NLSM},''
  \href{http://dx.doi.org/10.1007/JHEP07(2015)149}{{\em JHEP} {\bf 07} (2015)
  149}, \href{http://arxiv.org/abs/1412.3479}{{\tt arXiv:1412.3479 [hep-th]}}.

\bibitem{Hinterbichler:2015pqa}
K.~Hinterbichler and A.~Joyce, ``{Hidden symmetry of the Galileon},''
  \href{http://dx.doi.org/10.1103/PhysRevD.92.023503}{{\em Phys. Rev. D} {\bf
  92} (2015) no.~2, 023503}, \href{http://arxiv.org/abs/1501.07600}{{\tt
  arXiv:1501.07600 [hep-th]}}.

\bibitem{Cheung:2016drk}
C.~Cheung, K.~Kampf, J.~Novotny, C.-H. Shen, and J.~Trnka, ``{A Periodic Table
  of Effective Field Theories},''
  \href{http://dx.doi.org/10.1007/JHEP02(2017)020}{{\em JHEP} {\bf 02} (2017)
  020},
\href{http://arxiv.org/abs/1611.03137}{{\tt arXiv:1611.03137 [hep-th]}}.

\bibitem{Bittermann:2022nfh}
N.~Bittermann and A.~Joyce, ``{Soft limits of the wavefunction in exceptional
  scalar theories},'' \href{http://arxiv.org/abs/2203.05576}{{\tt
  arXiv:2203.05576 [hep-th]}}.

\bibitem{Maldacena:2002vr}
J.~M. Maldacena, ``{Non-Gaussian features of primordial fluctuations in single
  field inflationary models},''
  \href{http://dx.doi.org/10.1088/1126-6708/2003/05/013}{{\em JHEP} {\bf 05}
  (2003)  013},
\href{http://arxiv.org/abs/astro-ph/0210603}{{\tt arXiv:astro-ph/0210603
  [astro-ph]}}.

\bibitem{Creminelli:2004yq}
P.~Creminelli and M.~Zaldarriaga, ``{Single field consistency relation for the
  3-point function},''
  \href{http://dx.doi.org/10.1088/1475-7516/2004/10/006}{{\em JCAP} {\bf 0410}
  (2004)  006},
\href{http://arxiv.org/abs/astro-ph/0407059}{{\tt arXiv:astro-ph/0407059
  [astro-ph]}}.

\bibitem{Creminelli:2012ed}
P.~Creminelli, J.~Nore\~na, and M.~Simonovi\'c, ``{Conformal consistency
  relations for single-field inflation},''
  \href{http://dx.doi.org/10.1088/1475-7516/2012/07/052}{{\em JCAP} {\bf 1207}
  (2012)  052},
\href{http://arxiv.org/abs/1203.4595}{{\tt arXiv:1203.4595 [hep-th]}}.

\bibitem{Assassi:2012zq}
V.~Assassi, D.~Baumann, and D.~Green, ``{On Soft Limits of Inflationary
  Correlation Functions},''
  \href{http://dx.doi.org/10.1088/1475-7516/2012/11/047}{{\em JCAP} {\bf 1211}
  (2012)  047},
\href{http://arxiv.org/abs/1204.4207}{{\tt arXiv:1204.4207 [hep-th]}}.

\bibitem{Hinterbichler:2012nm}
K.~Hinterbichler, L.~Hui, and J.~Khoury, ``{Conformal Symmetries of Adiabatic
  Modes in Cosmology},''
  \href{http://dx.doi.org/10.1088/1475-7516/2012/08/017}{{\em JCAP} {\bf 1208}
  (2012)  017},
\href{http://arxiv.org/abs/1203.6351}{{\tt arXiv:1203.6351 [hep-th]}}.

\bibitem{Flauger:2013hra}
R.~Flauger, D.~Green, and R.~A. Porto, ``{On squeezed limits in single-field
  inflation. Part I},'' \href{http://dx.doi.org/10.1088/1475-7516/2013/08/032,
  10.1088/1475-7516/2013/08/032/}{{\em JCAP} {\bf 1308} (2013)  032},
\href{http://arxiv.org/abs/1303.1430}{{\tt arXiv:1303.1430 [hep-th]}}.

\bibitem{Pimentel:2013gza}
G.~L. Pimentel, ``{Inflationary Consistency Conditions from a Wavefunctional
  Perspective},'' \href{http://dx.doi.org/10.1007/JHEP02(2014)124}{{\em JHEP}
  {\bf 02} (2014)  124},
\href{http://arxiv.org/abs/1309.1793}{{\tt arXiv:1309.1793 [hep-th]}}.

\bibitem{Goldberger:2013rsa}
W.~D. Goldberger, L.~Hui, and A.~Nicolis, ``{One-particle-irreducible
  consistency relations for cosmological perturbations},''
  \href{http://dx.doi.org/10.1103/PhysRevD.87.103520}{{\em Phys. Rev.} {\bf
  D87} (2013) no.~10, 103520},
\href{http://arxiv.org/abs/1303.1193}{{\tt arXiv:1303.1193 [hep-th]}}.

\bibitem{Hinterbichler:2013dpa}
K.~Hinterbichler, L.~Hui, and J.~Khoury, ``{An Infinite Set of Ward Identities
  for Adiabatic Modes in Cosmology},''
  \href{http://dx.doi.org/10.1088/1475-7516/2014/01/039}{{\em JCAP} {\bf 1401}
  (2014)  039},
\href{http://arxiv.org/abs/1304.5527}{{\tt arXiv:1304.5527 [hep-th]}}.

\bibitem{Berezhiani:2013ewa}
L.~Berezhiani and J.~Khoury, ``{Slavnov-Taylor Identities for Primordial
  Perturbations},'' \href{http://dx.doi.org/10.1088/1475-7516/2014/02/003}{{\em
  JCAP} {\bf 1402} (2014)  003},
\href{http://arxiv.org/abs/1309.4461}{{\tt arXiv:1309.4461 [hep-th]}}.

\bibitem{Mirbabayi:2014zpa}
M.~Mirbabayi and M.~Zaldarriaga, ``{Double Soft Limits of Cosmological
  Correlations},'' \href{http://dx.doi.org/10.1088/1475-7516/2015/03/025}{{\em
  JCAP} {\bf 1503} (2015) no.~03, 025},
\href{http://arxiv.org/abs/1409.6317}{{\tt arXiv:1409.6317 [hep-th]}}.

\bibitem{Joyce:2014aqa}
A.~Joyce, J.~Khoury, and M.~Simonovi\'c, ``{Multiple Soft Limits of
  Cosmological Correlation Functions},''
  \href{http://dx.doi.org/10.1088/1475-7516/2015/01/012}{{\em JCAP} {\bf 1501}
  (2015) no.~01, 012},
\href{http://arxiv.org/abs/1409.6318}{{\tt arXiv:1409.6318 [hep-th]}}.

\bibitem{Kundu:2015xta}
N.~Kundu, A.~Shukla, and S.~P. Trivedi, ``{Ward Identities for Scale and
  Special Conformal Transformations in Inflation},''
  \href{http://dx.doi.org/10.1007/JHEP01(2016)046}{{\em JHEP} {\bf 01} (2016)
  046},
\href{http://arxiv.org/abs/1507.06017}{{\tt arXiv:1507.06017 [hep-th]}}.

\bibitem{Pajer:2017hmb}
E.~Pajer and S.~Jazayeri, ``{Systematics of Adiabatic Modes: Flat Universes},''
  \href{http://dx.doi.org/10.1088/1475-7516/2018/03/013}{{\em JCAP} {\bf 1803}
  (2018) no.~03, 013},
\href{http://arxiv.org/abs/1710.02177}{{\tt arXiv:1710.02177 [astro-ph.CO]}}.

\bibitem{Finelli:2017fml}
B.~Finelli, G.~Goon, E.~Pajer, and L.~Santoni, ``{Soft Theorems For
  Shift-Symmetric Cosmologies},''
  \href{http://dx.doi.org/10.1103/PhysRevD.97.063531}{{\em Phys. Rev.} {\bf
  D97} (2018) no.~6, 063531},
\href{http://arxiv.org/abs/1711.03737}{{\tt arXiv:1711.03737 [hep-th]}}.

\bibitem{Bravo:2017wyw}
R.~Bravo, S.~Mooij, G.~A. Palma, and B.~Pradenas, ``{A generalized non-Gaussian
  consistency relation for single field inflation},''
  \href{http://dx.doi.org/10.1088/1475-7516/2018/05/024}{{\em JCAP} {\bf 1805}
  (2018) no.~05, 024},
\href{http://arxiv.org/abs/1711.02680}{{\tt arXiv:1711.02680 [astro-ph.CO]}}.

\bibitem{Bordin:2017ozj}
L.~Bordin, P.~Creminelli, M.~Mirbabayi, and J.~Nore\~na, ``{Solid
  Consistency},'' \href{http://dx.doi.org/10.1088/1475-7516/2017/03/004}{{\em
  JCAP} {\bf 03} (2017)  004}, \href{http://arxiv.org/abs/1701.04382}{{\tt
  arXiv:1701.04382 [astro-ph.CO]}}.

\bibitem{Jazayeri:2019nbi}
S.~Jazayeri, E.~Pajer, and D.~van~der Woude, ``{Solid Soft Theorems},''
  \href{http://dx.doi.org/10.1088/1475-7516/2019/06/011}{{\em JCAP} {\bf 06}
  (2019)  011}, \href{http://arxiv.org/abs/1902.09020}{{\tt arXiv:1902.09020
  [hep-th]}}.

\bibitem{Avis:2019eav}
G.~Avis, S.~Jazayeri, E.~Pajer, and J.~Supel, ``{Spatial Curvature at the Sound
  Horizon},'' \href{http://dx.doi.org/10.1088/1475-7516/2020/02/034}{{\em JCAP}
  {\bf 02} (2020)  034}, \href{http://arxiv.org/abs/1911.04454}{{\tt
  arXiv:1911.04454 [astro-ph.CO]}}.

\bibitem{Hui:2018cag}
L.~Hui, A.~Joyce, and S.~S.~C. Wong, ``{Inflationary soft theorems revisited: A
  generalized consistency relation},''
  \href{http://dx.doi.org/10.1088/1475-7516/2019/02/060}{{\em JCAP} {\bf 1902}
  (2019)  060},
\href{http://arxiv.org/abs/1811.05951}{{\tt arXiv:1811.05951 [hep-th]}}.

\bibitem{Kehagias:2013yd}
A.~Kehagias and A.~Riotto, ``{Symmetries and Consistency Relations in the Large
  Scale Structure of the Universe},''
  \href{http://dx.doi.org/10.1016/j.nuclphysb.2013.05.009}{{\em Nucl. Phys.}
  {\bf B873} (2013)  514--529},
\href{http://arxiv.org/abs/1302.0130}{{\tt arXiv:1302.0130 [astro-ph.CO]}}.

\bibitem{Peloso:2013zw}
M.~Peloso and M.~Pietroni, ``{Galilean invariance and the consistency relation
  for the nonlinear squeezed bispectrum of large scale structure},''
  \href{http://dx.doi.org/10.1088/1475-7516/2013/05/031}{{\em JCAP} {\bf 1305}
  (2013)  031},
\href{http://arxiv.org/abs/1302.0223}{{\tt arXiv:1302.0223 [astro-ph.CO]}}.

\bibitem{Creminelli:2013mca}
P.~Creminelli, J.~Nore\~na, M.~Simonovi\'c, and F.~Vernizzi, ``{Single-Field
  Consistency Relations of Large Scale Structure},''
  \href{http://dx.doi.org/10.1088/1475-7516/2013/12/025}{{\em JCAP} {\bf 1312}
  (2013)  025},
\href{http://arxiv.org/abs/1309.3557}{{\tt arXiv:1309.3557 [astro-ph.CO]}}.

\bibitem{Horn:2014rta}
B.~Horn, L.~Hui, and X.~Xiao, ``{Soft-Pion Theorems for Large Scale
  Structure},'' \href{http://dx.doi.org/10.1088/1475-7516/2014/09/044}{{\em
  JCAP} {\bf 1409} (2014) no.~09, 044},
\href{http://arxiv.org/abs/1406.0842}{{\tt arXiv:1406.0842 [hep-th]}}.

\bibitem{Esposito:2019jkb}
A.~Esposito, L.~Hui, and R.~Scoccimarro, ``{Nonperturbative test of consistency
  relations and their violation},''
  \href{http://dx.doi.org/10.1103/PhysRevD.100.043536}{{\em Phys. Rev. D} {\bf
  100} (2019) no.~4, 043536}, \href{http://arxiv.org/abs/1905.11423}{{\tt
  arXiv:1905.11423 [astro-ph.CO]}}.

\bibitem{Goldstein:2022hgr}
S.~Goldstein, A.~Esposito, O.~H.~E. Philcox, L.~Hui, J.~C. Hill,
  R.~Scoccimarro, and M.~H. Abitbol, ``{Squeezing ${\boldsymbol f_{\rm NL}}$
  out of the matter bispectrum with consistency relations},''
  \href{http://arxiv.org/abs/2209.06228}{{\tt arXiv:2209.06228 [astro-ph.CO]}}.

\bibitem{Cheung:2007st}
C.~Cheung, P.~Creminelli, A.~L. Fitzpatrick, J.~Kaplan, and L.~Senatore, ``{The
  Effective Field Theory of Inflation},''
  \href{http://dx.doi.org/10.1088/1126-6708/2008/03/014}{{\em JHEP} {\bf 03}
  (2008)  014},
\href{http://arxiv.org/abs/0709.0293}{{\tt arXiv:0709.0293 [hep-th]}}.

\bibitem{Dubovsky:2005xd}
S.~Dubovsky, T.~Gregoire, A.~Nicolis, and R.~Rattazzi, ``{Null energy condition
  and superluminal propagation},''
  \href{http://dx.doi.org/10.1088/1126-6708/2006/03/025}{{\em JHEP} {\bf 03}
  (2006)  025}, \href{http://arxiv.org/abs/hep-th/0512260}{{\tt
  arXiv:hep-th/0512260}}.

\bibitem{Dubovsky:2011sj}
S.~Dubovsky, L.~Hui, A.~Nicolis, and D.~T. Son, ``{Effective field theory for
  hydrodynamics: thermodynamics, and the derivative expansion},''
  \href{http://dx.doi.org/10.1103/PhysRevD.85.085029}{{\em Phys. Rev. D} {\bf
  85} (2012)  085029}, \href{http://arxiv.org/abs/1107.0731}{{\tt
  arXiv:1107.0731 [hep-th]}}.

\bibitem{Nicolis:2013lma}
A.~Nicolis, R.~Penco, and R.~A. Rosen, ``{Relativistic Fluids, Superfluids,
  Solids and Supersolids from a Coset Construction},''
  \href{http://dx.doi.org/10.1103/PhysRevD.89.045002}{{\em Phys. Rev. D} {\bf
  89} (2014) no.~4, 045002}, \href{http://arxiv.org/abs/1307.0517}{{\tt
  arXiv:1307.0517 [hep-th]}}.

\bibitem{Nicolis:2015sra}
A.~Nicolis, R.~Penco, F.~Piazza, and R.~Rattazzi, ``{Zoology of condensed
  matter: Framids, ordinary stuff, extra-ordinary stuff},''
  \href{http://dx.doi.org/10.1007/JHEP06(2015)155}{{\em JHEP} {\bf 06} (2015)
  155}, \href{http://arxiv.org/abs/1501.03845}{{\tt arXiv:1501.03845
  [hep-th]}}.

\bibitem{Alberte:2020eil}
L.~Alberte and A.~Nicolis, ``{Spontaneously broken boosts and the Goldstone
  continuum},'' \href{http://dx.doi.org/10.1007/JHEP07(2020)076}{{\em JHEP}
  {\bf 07} (2020)  076}, \href{http://arxiv.org/abs/2001.06024}{{\tt
  arXiv:2001.06024 [hep-th]}}.

\bibitem{Gaiotto:2014kfa}
D.~Gaiotto, A.~Kapustin, N.~Seiberg, and B.~Willett, ``{Generalized Global
  Symmetries},'' \href{http://dx.doi.org/10.1007/JHEP02(2015)172}{{\em JHEP}
  {\bf 02} (2015)  172}, \href{http://arxiv.org/abs/1412.5148}{{\tt
  arXiv:1412.5148 [hep-th]}}.

\bibitem{Lake:2018dqm}
E.~Lake, ``{Higher-form symmetries and spontaneous symmetry breaking},''
  \href{http://arxiv.org/abs/1802.07747}{{\tt arXiv:1802.07747 [hep-th]}}.

\bibitem{Hofman:2018lfz}
D.~M. Hofman and N.~Iqbal, ``{Goldstone modes and photonization for higher form
  symmetries},'' \href{http://dx.doi.org/10.21468/SciPostPhys.6.1.006}{{\em
  SciPost Phys.} {\bf 6} (2019) no.~1, 006},
  \href{http://arxiv.org/abs/1802.09512}{{\tt arXiv:1802.09512 [hep-th]}}.

\bibitem{Delacretaz:2019brr}
L.~V. Delacr\'etaz, D.~M. Hofman, and G.~Mathys, ``{Superfluids as Higher-form
  Anomalies},'' \href{http://dx.doi.org/10.21468/SciPostPhys.8.3.047}{{\em
  SciPost Phys.} {\bf 8} (2020)  047},
  \href{http://arxiv.org/abs/1908.06977}{{\tt arXiv:1908.06977 [hep-th]}}.

\bibitem{Benedetti:2021lxj}
V.~Benedetti, H.~Casini, and J.~M. Magan, ``{Generalized symmetries of the
  graviton},'' \href{http://dx.doi.org/10.1007/JHEP05(2022)045}{{\em JHEP} {\bf
  05} (2022)  045}, \href{http://arxiv.org/abs/2111.12089}{{\tt
  arXiv:2111.12089 [hep-th]}}.

\bibitem{Hinterbichler:2022agn}
K.~Hinterbichler, D.~M. Hofman, A.~Joyce, and G.~Mathys, ``{Gravity as a
  gapless phase and biform symmetries},''
  \href{http://arxiv.org/abs/2205.12272}{{\tt arXiv:2205.12272 [hep-th]}}.

\bibitem{McGreevy:2022oyu}
J.~McGreevy, ``{Generalized Symmetries in Condensed Matter},''
  \href{http://arxiv.org/abs/2204.03045}{{\tt arXiv:2204.03045
  [cond-mat.str-el]}}.

\bibitem{Stokes}
T.~S. van~den Bremer and {\O}.~Breivik, ``Stokes drift,'' {\em Philosophical
  Transactions of the Royal Society A: Mathematical, Physical and Engineering
  Sciences} {\bf 376} (2017)  .

\bibitem{Maldacena:2011nz}
J.~M. Maldacena and G.~L. Pimentel, ``{On graviton non-Gaussianities during
  inflation},'' \href{http://dx.doi.org/10.1007/JHEP09(2011)045}{{\em JHEP}
  {\bf 09} (2011)  045},
\href{http://arxiv.org/abs/1104.2846}{{\tt arXiv:1104.2846 [hep-th]}}.

\bibitem{Creminelli:2011mw}
P.~Creminelli, ``{Conformal invariance of scalar perturbations in inflation},''
  \href{http://dx.doi.org/10.1103/PhysRevD.85.041302}{{\em Phys. Rev.} {\bf
  D85} (2012)  041302},
\href{http://arxiv.org/abs/1108.0874}{{\tt arXiv:1108.0874 [hep-th]}}.

\bibitem{Mata:2012bx}
I.~Mata, S.~Raju, and S.~Trivedi, ``{CMB from CFT},''
  \href{http://dx.doi.org/10.1007/JHEP07(2013)015}{{\em JHEP} {\bf 07} (2013)
  015},
\href{http://arxiv.org/abs/1211.5482}{{\tt arXiv:1211.5482 [hep-th]}}.

\bibitem{Bzowski:2013sza}
A.~Bzowski, P.~McFadden, and K.~Skenderis, ``{Implications of conformal
  invariance in momentum space},''
  \href{http://dx.doi.org/10.1007/JHEP03(2014)111}{{\em JHEP} {\bf 03} (2014)
  111}, \href{http://arxiv.org/abs/1304.7760}{{\tt arXiv:1304.7760 [hep-th]}}.

\bibitem{Ghosh:2014kba}
A.~Ghosh, N.~Kundu, S.~Raju, and S.~P. Trivedi, ``{Conformal Invariance and the
  Four Point Scalar Correlator in Slow-Roll Inflation},''
  \href{http://dx.doi.org/10.1007/JHEP07(2014)011}{{\em JHEP} {\bf 07} (2014)
  011},
\href{http://arxiv.org/abs/1401.1426}{{\tt arXiv:1401.1426 [hep-th]}}.

\bibitem{Arkani-Hamed:2015bza}
N.~Arkani-Hamed and J.~Maldacena, ``{Cosmological Collider Physics},''
\href{http://arxiv.org/abs/1503.08043}{{\tt arXiv:1503.08043 [hep-th]}}.

\bibitem{Pajer:2016ieg}
E.~Pajer, G.~L. Pimentel, and J.~V.~S. Van~Wijck, ``{The Conformal Limit of
  Inflation in the Era of CMB Polarimetry},''
  \href{http://dx.doi.org/10.1088/1475-7516/2017/06/009}{{\em JCAP} {\bf 06}
  (2017)  009}, \href{http://arxiv.org/abs/1609.06993}{{\tt arXiv:1609.06993
  [hep-th]}}.

\bibitem{Arkani-Hamed:2017fdk}
N.~Arkani-Hamed, P.~Benincasa, and A.~Postnikov, ``{Cosmological Polytopes and
  the Wavefunction of the Universe},''
  \href{http://arxiv.org/abs/1709.02813}{{\tt arXiv:1709.02813 [hep-th]}}.

\bibitem{Arkani-Hamed:2018kmz}
N.~Arkani-Hamed, D.~Baumann, H.~Lee, and G.~L. Pimentel, ``{The Cosmological
  Bootstrap: Inflationary Correlators from Symmetries and Singularities},''
\href{http://arxiv.org/abs/1811.00024}{{\tt arXiv:1811.00024 [hep-th]}}.

\bibitem{Goon:2018fyu}
G.~Goon, K.~Hinterbichler, A.~Joyce, and M.~Trodden, ``{Shapes of gravity:
  Tensor non-Gaussianity and massive spin-2 fields},''
  \href{http://dx.doi.org/10.1007/JHEP10(2019)182}{{\em JHEP} {\bf 10} (2019)
  182}, \href{http://arxiv.org/abs/1812.07571}{{\tt arXiv:1812.07571
  [hep-th]}}.

\bibitem{Benincasa:2019vqr}
P.~Benincasa, ``{Cosmological Polytopes and the Wavefuncton of the Universe for
  Light States},'' \href{http://arxiv.org/abs/1909.02517}{{\tt arXiv:1909.02517
  [hep-th]}}.

\bibitem{Bzowski:2019kwd}
A.~Bzowski, P.~McFadden, and K.~Skenderis, ``{Conformal $n$-point functions in
  momentum space},''
  \href{http://dx.doi.org/10.1103/PhysRevLett.124.131602}{{\em Phys. Rev.
  Lett.} {\bf 124} (2020) no.~13, 131602},
  \href{http://arxiv.org/abs/1910.10162}{{\tt arXiv:1910.10162 [hep-th]}}.

\bibitem{Hillman:2019wgh}
A.~Hillman, ``{Symbol Recursion for the dS Wave Function},''
  \href{http://arxiv.org/abs/1912.09450}{{\tt arXiv:1912.09450 [hep-th]}}.

\bibitem{Baumann:2019oyu}
D.~Baumann, C.~Duaso~Pueyo, A.~Joyce, H.~Lee, and G.~L. Pimentel, ``{The
  cosmological bootstrap: weight-shifting operators and scalar seeds},''
  \href{http://dx.doi.org/10.1007/JHEP12(2020)204}{{\em JHEP} {\bf 12} (2020)
  204}, \href{http://arxiv.org/abs/1910.14051}{{\tt arXiv:1910.14051
  [hep-th]}}.

\bibitem{Baumann:2020dch}
D.~Baumann, C.~Duaso~Pueyo, A.~Joyce, H.~Lee, and G.~L. Pimentel, ``{The
  Cosmological Bootstrap: Spinning Correlators from Symmetries and
  Factorization},'' \href{http://dx.doi.org/10.21468/SciPostPhys.11.3.071}{{\em
  SciPost Phys.} {\bf 11} (2021)  071},
  \href{http://arxiv.org/abs/2005.04234}{{\tt arXiv:2005.04234 [hep-th]}}.

\bibitem{Goodhew:2020hob}
H.~Goodhew, S.~Jazayeri, and E.~Pajer, ``{The Cosmological Optical Theorem},''
  \href{http://dx.doi.org/10.1088/1475-7516/2021/04/021}{{\em JCAP} {\bf 04}
  (2021)  021}, \href{http://arxiv.org/abs/2009.02898}{{\tt arXiv:2009.02898
  [hep-th]}}.

\bibitem{Melville:2021lst}
S.~Melville and E.~Pajer, ``{Cosmological Cutting Rules},''
  \href{http://dx.doi.org/10.1007/JHEP05(2021)249}{{\em JHEP} {\bf 05} (2021)
  249}, \href{http://arxiv.org/abs/2103.09832}{{\tt arXiv:2103.09832
  [hep-th]}}.

\bibitem{Cabass:2021fnw}
G.~Cabass, E.~Pajer, D.~Stefanyszyn, and J.~Supel, ``{Bootstrapping large
  graviton non-Gaussianities},''
  \href{http://dx.doi.org/10.1007/JHEP05(2022)077}{{\em JHEP} {\bf 05} (2022)
  077}, \href{http://arxiv.org/abs/2109.10189}{{\tt arXiv:2109.10189
  [hep-th]}}.

\bibitem{Goodhew:2021oqg}
H.~Goodhew, S.~Jazayeri, M.~H. Gordon~Lee, and E.~Pajer, ``{Cutting
  cosmological correlators},''
  \href{http://dx.doi.org/10.1088/1475-7516/2021/08/003}{{\em JCAP} {\bf 08}
  (2021)  003}, \href{http://arxiv.org/abs/2104.06587}{{\tt arXiv:2104.06587
  [hep-th]}}.

\bibitem{Cabass:2022jda}
G.~Cabass, D.~Stefanyszyn, J.~Supel, and A.~Thavanesan, ``{On Graviton
  non-Gaussianities in the Effective Field Theory of Inflation},''
  \href{http://arxiv.org/abs/2209.00677}{{\tt arXiv:2209.00677 [hep-th]}}.

\bibitem{Benincasa:2022gtd}
P.~Benincasa, ``{Amplitudes meet Cosmology: A (Scalar) Primer},''
  \href{http://arxiv.org/abs/2203.15330}{{\tt arXiv:2203.15330 [hep-th]}}.

\bibitem{Baumann:2022jpr}
D.~Baumann, D.~Green, A.~Joyce, E.~Pajer, G.~L. Pimentel, C.~Sleight, and
  M.~Taronna, ``{Snowmass White Paper: The Cosmological Bootstrap},'' in {\em
  {2022 Snowmass Summer Study}}.
\newblock 3, 2022.
\newblock \href{http://arxiv.org/abs/2203.08121}{{\tt arXiv:2203.08121
  [hep-th]}}.

\bibitem{Pajer:2020wxk}
E.~Pajer, ``{Building a Boostless Bootstrap for the Bispectrum},''
  \href{http://dx.doi.org/10.1088/1475-7516/2021/01/023}{{\em JCAP} {\bf 01}
  (2021)  023}, \href{http://arxiv.org/abs/2010.12818}{{\tt arXiv:2010.12818
  [hep-th]}}.

\bibitem{Jazayeri:2021fvk}
S.~Jazayeri, E.~Pajer, and D.~Stefanyszyn, ``{From locality and unitarity to
  cosmological correlators},''
  \href{http://dx.doi.org/10.1007/JHEP10(2021)065}{{\em JHEP} {\bf 10} (2021)
  065}, \href{http://arxiv.org/abs/2103.08649}{{\tt arXiv:2103.08649
  [hep-th]}}.

\bibitem{Baumann:2021fxj}
D.~Baumann, W.-M. Chen, C.~Duaso~Pueyo, A.~Joyce, H.~Lee, and G.~L. Pimentel,
  ``{Linking the singularities of cosmological correlators},''
  \href{http://dx.doi.org/10.1007/JHEP09(2022)010}{{\em JHEP} {\bf 09} (2022)
  010}, \href{http://arxiv.org/abs/2106.05294}{{\tt arXiv:2106.05294
  [hep-th]}}.

\bibitem{Hillman:2021bnk}
A.~Hillman and E.~Pajer, ``{A differential representation of cosmological
  wavefunctions},'' \href{http://dx.doi.org/10.1007/JHEP04(2022)012}{{\em JHEP}
  {\bf 04} (2022)  012}, \href{http://arxiv.org/abs/2112.01619}{{\tt
  arXiv:2112.01619 [hep-th]}}.

\bibitem{Bonifacio:2021azc}
J.~Bonifacio, E.~Pajer, and D.-G. Wang, ``{From amplitudes to contact
  cosmological correlators},''
  \href{http://dx.doi.org/10.1007/JHEP10(2021)001}{{\em JHEP} {\bf 10} (2021)
  001}, \href{http://arxiv.org/abs/2106.15468}{{\tt arXiv:2106.15468
  [hep-th]}}.

\bibitem{Pimentel:2022fsc}
G.~L. Pimentel and D.-G. Wang, ``{Boostless Cosmological Collider Bootstrap},''
  \href{http://arxiv.org/abs/2205.00013}{{\tt arXiv:2205.00013 [hep-th]}}.

\bibitem{Jazayeri:2022kjy}
S.~Jazayeri and S.~Renaux-Petel, ``{Cosmological Bootstrap in Slow Motion},''
  \href{http://arxiv.org/abs/2205.10340}{{\tt arXiv:2205.10340 [hep-th]}}.

\bibitem{Itzykson:1980rh}
C.~Itzykson and J.~B. Zuber, {\em {Quantum Field Theory}}.
\newblock International Series In Pure and Applied Physics. McGraw-Hill, New
  York, 1980.

\bibitem{Brauner:2010wm}
T.~Brauner, ``{Spontaneous Symmetry Breaking and Nambu-Goldstone Bosons in
  Quantum Many-Body Systems},''
  \href{http://dx.doi.org/10.3390/sym2020609}{{\em Symmetry} {\bf 2} (2010)
  609--657}, \href{http://arxiv.org/abs/1001.5212}{{\tt arXiv:1001.5212
  [hep-th]}}.

\bibitem{Son:2002zn}
D.~T. Son, ``{Low-energy quantum effective action for relativistic
  superfluids},'' \href{http://arxiv.org/abs/hep-ph/0204199}{{\tt
  arXiv:hep-ph/0204199}}.

\bibitem{Assassi:2012et}
V.~Assassi, D.~Baumann, and D.~Green, ``{Symmetries and Loops in Inflation},''
  \href{http://dx.doi.org/10.1007/JHEP02(2013)151}{{\em JHEP} {\bf 02} (2013)
  151},
\href{http://arxiv.org/abs/1210.7792}{{\tt arXiv:1210.7792 [hep-th]}}.

\bibitem{Pimentel:2012tw}
G.~L. Pimentel, L.~Senatore, and M.~Zaldarriaga, ``{On Loops in Inflation III:
  Time Independence of zeta in Single Clock Inflation},''
  \href{http://dx.doi.org/10.1007/JHEP07(2012)166}{{\em JHEP} {\bf 07} (2012)
  166}, \href{http://arxiv.org/abs/1203.6651}{{\tt arXiv:1203.6651 [hep-th]}}.

\bibitem{Tsamis:2003px}
N.~C. Tsamis and R.~P. Woodard, ``{Improved estimates of cosmological
  perturbations},'' \href{http://dx.doi.org/10.1103/PhysRevD.69.084005}{{\em
  Phys. Rev.} {\bf D69} (2004)  084005},
\href{http://arxiv.org/abs/astro-ph/0307463}{{\tt arXiv:astro-ph/0307463
  [astro-ph]}}.

\bibitem{Kinney:2005vj}
W.~H. Kinney, ``{Horizon crossing and inflation with large eta},''
  \href{http://dx.doi.org/10.1103/PhysRevD.72.023515}{{\em Phys. Rev.} {\bf
  D72} (2005)  023515},
\href{http://arxiv.org/abs/gr-qc/0503017}{{\tt arXiv:gr-qc/0503017 [gr-qc]}}.

\bibitem{Nicolis:2011pv}
A.~Nicolis and F.~Piazza, ``{Spontaneous Symmetry Probing},''
  \href{http://dx.doi.org/10.1007/JHEP06(2012)025}{{\em JHEP} {\bf 06} (2012)
  025}, \href{http://arxiv.org/abs/1112.5174}{{\tt arXiv:1112.5174 [hep-th]}}.

\bibitem{Green:2022slj}
D.~Green, Y.~Huang, and C.-H. Shen, ``{Inflationary Adler Conditions},''
  \href{http://arxiv.org/abs/2208.14544}{{\tt arXiv:2208.14544 [hep-th]}}.

\bibitem{Finelli:2018upr}
B.~Finelli, G.~Goon, E.~Pajer, and L.~Santoni, ``{The Effective Theory of
  Shift-Symmetric Cosmologies},''
  \href{http://dx.doi.org/10.1088/1475-7516/2018/05/060}{{\em JCAP} {\bf 1805}
  (2018) no.~05, 060},
\href{http://arxiv.org/abs/1802.01580}{{\tt arXiv:1802.01580 [hep-th]}}.

\bibitem{Floess:2018ths}
T.~Fl\"{o}ss, ``{Inflationary Consistency Conditions and Shift-Symmetric
  Cosmologies}.'' \url{https://studenttheses.uu.nl/handle/20.500.12932/29309},
  Master Thesis, Utrecht University, 2018.

\bibitem{Creminelli:2006xe}
P.~Creminelli, M.~A. Luty, A.~Nicolis, and L.~Senatore, ``{Starting the
  Universe: Stable Violation of the Null Energy Condition and Non-standard
  Cosmologies},'' \href{http://dx.doi.org/10.1088/1126-6708/2006/12/080}{{\em
  JHEP} {\bf 12} (2006)  080},
\href{http://arxiv.org/abs/hep-th/0606090}{{\tt arXiv:hep-th/0606090
  [hep-th]}}.

\bibitem{Creminelli:2016zwa}
P.~Creminelli, D.~Pirtskhalava, L.~Santoni, and E.~Trincherini, ``{Stability of
  Geodesically Complete Cosmologies},''
  \href{http://dx.doi.org/10.1088/1475-7516/2016/11/047}{{\em JCAP} {\bf 11}
  (2016)  047}, \href{http://arxiv.org/abs/1610.04207}{{\tt arXiv:1610.04207
  [hep-th]}}.

\bibitem{Kehagias:2012pd}
A.~Kehagias and A.~Riotto, ``{Operator Product Expansion of Inflationary
  Correlators and Conformal Symmetry of de Sitter},''
  \href{http://dx.doi.org/10.1016/j.nuclphysb.2012.07.004}{{\em Nucl. Phys.}
  {\bf B864} (2012)  492--529},
\href{http://arxiv.org/abs/1205.1523}{{\tt arXiv:1205.1523 [hep-th]}}.

\bibitem{Berezhiani:2014tda}
L.~Berezhiani, J.~Khoury, and J.~Wang, ``{Non-Trivial Checks of Novel
  Consistency Relations},''
  \href{http://dx.doi.org/10.1088/1475-7516/2014/06/056}{{\em JCAP} {\bf 1406}
  (2014)  056},
\href{http://arxiv.org/abs/1401.7991}{{\tt arXiv:1401.7991 [hep-th]}}.

\bibitem{Avery:2015rga}
S.~G. Avery and B.~U.~W. Schwab, ``{Noether's second theorem and Ward
  identities for gauge symmetries},''
  \href{http://dx.doi.org/10.1007/JHEP02(2016)031}{{\em JHEP} {\bf 02} (2016)
  031},
\href{http://arxiv.org/abs/1510.07038}{{\tt arXiv:1510.07038 [hep-th]}}.

\bibitem{Weinberg:2005vy}
S.~Weinberg, ``{Quantum contributions to cosmological correlations},''
  \href{http://dx.doi.org/10.1103/PhysRevD.72.043514}{{\em Phys. Rev.} {\bf
  D72} (2005)  043514},
\href{http://arxiv.org/abs/hep-th/0506236}{{\tt arXiv:hep-th/0506236
  [hep-th]}}.

\bibitem{Pajer:2018egx}
E.~Pajer and D.~Stefanyszyn, ``{Symmetric Superfluids},''
  \href{http://dx.doi.org/10.1007/JHEP06(2019)008}{{\em JHEP} {\bf 06} (2019)
  008}, \href{http://arxiv.org/abs/1812.05133}{{\tt arXiv:1812.05133
  [hep-th]}}.

\bibitem{Bardeen:1983qw}
J.~M. Bardeen, P.~J. Steinhardt, and M.~S. Turner, ``{Spontaneous Creation of
  Almost Scale - Free Density Perturbations in an Inflationary Universe},''
  \href{http://dx.doi.org/10.1103/PhysRevD.28.679}{{\em Phys. Rev. D} {\bf 28}
  (1983)  679}.

\bibitem{Salopek:1990jq}
D.~S. Salopek and J.~R. Bond, ``{Nonlinear evolution of long wavelength metric
  fluctuations in inflationary models},''
  \href{http://dx.doi.org/10.1103/PhysRevD.42.3936}{{\em Phys. Rev. D} {\bf 42}
  (1990)  3936--3962}.

\bibitem{Weinberg:2003sw}
S.~Weinberg, ``{Adiabatic modes in cosmology},''
  \href{http://dx.doi.org/10.1103/PhysRevD.67.123504}{{\em Phys. Rev.} {\bf
  D67} (2003)  123504},
\href{http://arxiv.org/abs/astro-ph/0302326}{{\tt arXiv:astro-ph/0302326
  [astro-ph]}}.

\bibitem{Weinberg:2008zzc}
S.~Weinberg, {\em {Cosmology}}.
\newblock 2008.

\bibitem{Cheung:2007sv}
C.~Cheung, A.~L. Fitzpatrick, J.~Kaplan, and L.~Senatore, ``{On the consistency
  relation of the 3-point function in single field inflation},''
  \href{http://dx.doi.org/10.1088/1475-7516/2008/02/021}{{\em JCAP} {\bf 0802}
  (2008)  021},
\href{http://arxiv.org/abs/0709.0295}{{\tt arXiv:0709.0295 [hep-th]}}.

\bibitem{Akhshik:2015nfa}
M.~Akhshik, H.~Firouzjahi, and S.~Jazayeri, ``{Effective Field Theory of
  non-Attractor Inflation},''
  \href{http://dx.doi.org/10.1088/1475-7516/2015/07/048}{{\em JCAP} {\bf 1507}
  (2015) no.~07, 048},
\href{http://arxiv.org/abs/1501.01099}{{\tt arXiv:1501.01099 [hep-th]}}.

\bibitem{Weinberg:1996kr}
S.~Weinberg, {\em {The quantum theory of fields. Vol. 2: Modern applications}}.
\newblock Cambridge University Press,
2013.
\newblock

\bibitem{Grall:2020ibl}
T.~Grall, S.~Jazayeri, and D.~Stefanyszyn, ``{The cosmological phonon:
  symmetries and amplitudes on sub-horizon scales},''
  \href{http://dx.doi.org/10.1007/JHEP11(2020)097}{{\em JHEP} {\bf 11} (2020)
  097}, \href{http://arxiv.org/abs/2005.12937}{{\tt arXiv:2005.12937
  [hep-th]}}.

\bibitem{Hamada:2017atr}
Y.~Hamada and S.~Sugishita, ``{Soft pion theorem, asymptotic symmetry and new
  memory effect},'' \href{http://dx.doi.org/10.1007/JHEP11(2017)203}{{\em JHEP}
  {\bf 11} (2017)  203}, \href{http://arxiv.org/abs/1709.05018}{{\tt
  arXiv:1709.05018 [hep-th]}}.

\end{thebibliography}\endgroup
}

\end{document}